\documentclass[a4paper,11pt]{article}
\pdfoutput=1
\usepackage{amsmath,amsfonts,amstext,amssymb,mathrsfs,amsthm}
\usepackage{jheppubnew}
\usepackage{bm}
\usepackage{comment}
\usepackage[numbers,sort&compress]{natbib}
\usepackage{tikz}
\usepackage{caption,subcaption}
\usepackage{titlesec}

\usetikzlibrary{decorations.pathmorphing}
\tikzset{snake it/.style={decorate, decoration=snake}}

\titleformat*{\section}{\Large\bfseries}
\titleformat*{\subsection}{\large\bfseries}
\titleformat*{\subsubsection}{\bfseries}
\titlespacing*{\section}{0pt}{20pt}{14pt}
\titlespacing*{\subsection}{0pt}{16pt}{10pt}
\titlespacing*{\subsubsection}{0pt}{12pt}{6pt}

\newcommand{\I}{\mathrm{i}}
\newcommand{\D}{\mathrm{d}}
\newcommand{\C}{\mathbb{C}}
\newcommand{\R}{\mathbb{R}}
\newcommand{\Z}{\mathbb{Z}}

\renewcommand{\j}{\varphi}
\renewcommand{\O}{\mathcal{O}}
\newcommand{\<}{\langle}
\renewcommand{\>}{\rangle}
\newcommand{\bs}[1]{\boldsymbol{#1}}
\newcommand{\nn}{\nonumber}

\renewcommand{\a}{\alpha}
\renewcommand{\b}{\beta}

\newcommand{\p}{\rho_h}

\newcommand{\limR}{\lim_R}

\newcommand{\limRL}{\lim_{R,L}}

\newcommand{\Mre}{\mathcal{M}^{\R}}
\newcommand{\Mcx}{\mathcal{M}^{\C}}
\renewcommand{\H}{\mathcal{H}}
\newcommand{\Hone}{\mathcal{H}^{(1)}}
\newcommand{\Hsemi}{\mathcal{H}_{\text{semi}}}
\newcommand{\Hg}{\H_g}
\newcommand{\Ire}{\mathcal{I}^{\R}}
\newcommand{\Icx}{\mathcal{I}^{\C}}
\newcommand{\Ineg}{\mathcal{I}^{-}}
\newcommand{\Ipos}{\mathcal{I}^{+}}

\newcommand{\Bcx}{\mathfrak{B}^{\C}}
\newcommand{\Bneg}{\mathfrak{B}^{-}}
\newcommand{\Bpos}{\mathfrak{B}^{+}}

\newcommand{\Dcx}{\mathfrak{D}^{\C}}

\DeclareMathOperator{\Tr}{Tr}

\DeclareMathOperator{\sech}{sech}

\newcommand{\op}[1]{\hat{#1}}

\colorlet{colorR}{blue!50!white}
\colorlet{colorL}{orange!50!white}

\begin{document}

\title{\LARGE Wormholes, geons, and the illusion of the tensor product}

\author[a]{Adam Bzowski,}
\affiliation[a]{Faculty of Physics, 
	University of Warsaw, 
	Pasteura 5,
	02-093 Warsaw, 
	Poland}

\emailAdd{abzowski@fuw.edu.pl}

\abstract{In this paper I argue that the Hilbert space of states of a holographic, traversable wormhole does not factorize into the tensor product of the boundary Hilbert spaces. After presenting the general argument I analyze two examples: the scalar sectors of the BTZ geon and the AdS$_2$ eternal wormhole. Utilizing real-time holography I derive the Hilbert spaces, identify the dual states and evaluate correlation functions. I show that the number of peculiarities associated with the wormhole and black hole physics emerges once the factorization is \textit{a priori} assumed. This includes null states and null operators, highly entangled vacuum states and the cross-boundary interactions all emerging as avatars of non-factorization.
}

\maketitle

\section{Introduction}

The most persistent assumption regarding holographic, traversable wormholes, permeating perhaps hundreds of papers, is the idea that the Hilbert space of states $\H$ splits into the tensor product of their boundary Hilbert spaces, $\H \cong \H_L \otimes \H_R$. The reasoning behind this statement is simple: with the vacuum state $| 0 \>_L \otimes | 0 \>_R$ dual to two disconnected AdS spaces, the thermofield double dual to a semiclassical black hole, the wormhole is believed to be dual to a highly entangled state in the tensor product. The strong entanglement within the wormhole state should then be responsible for the information transfer between the two asymptotic regions.

In this paper I want to argue that the Hilbert space of states $\H$ associated with a holographic, traversable wormhole does \emph{not} factorize into the tensor product of the boundary Hilbert spaces, $\H \neq \H_L \otimes \H_R$. I present here the extension of the results of \cite{Bzowski:2021vno}. Furthermore, I want to argue that a number of peculiar features associated with quantum black hole or wormholes can be naturally explained by non-factorization. This includes the emergence of \emph{null states}, effective \emph{interaction between the two boundary theories}, the \emph{vacuum state resembling a highly entangled, Bell-like state}, the \emph{operatorial relation between the boundary Hamiltonians} and more. 

I will present two examples of wormholes and the associated phenomena: the geon-wormhole in section \ref{sec:geon} and the Jackiw–Teitelboim (JT) wormhole, \textit{i.e.}, the eternal AdS$_2$ wormhole in section \ref{sec:jt}. While these systems were extensively analyzed before (see the subsequent section for the discussion of previous results), it is usually assumed that the Hilbert space of the system factorizes. In this paper I will show how the non-factorization enters the picture and explain a number of peculiar features listed above.

\subsection{Review of the literature}

\paragraph{Non-factorization.} The fact that the entanglement carried by the state dual to the wormhole is not enough for the transfer of information between the two boundaries was argued almost a decade ago in several papers, \textit{e.g.}, \cite{Chowdhury:2013tza,Maldacena:2013xja,Susskind:2014moa,Balasubramanian:2014gla}. In \cite{Shenker:2013pqa,Shenker:2013yza} it was shown how a wormhole can be opened by emitting shock waves from the boundaries. In \cite{Gao:2016bin,Maldacena:2017axo} this procedure was reinterpreted as the introduction of cross-boundary couplings between the dual boundary field theories.

More recently, however, it became more apparent that even the interaction between the left and right boundary systems is not sufficient for transferring the information through the wormhole. In \cite{Baez:2014bka} the authors argue that in the wormhole geometries certain entangled degrees of freedom turn out to be the same and must be identified between the two boundaries. This is the hallmark of non-factorization: states that seem to be naively distinct in the two boundaries represent in fact the same state. In particular in \cite{Harlow:2015lma,Guica:2015zpf} the authors identified specific charged bulk operators in higher dimensional black holes, whose zero modes represent such states.

More recently the problem of non-factorization was approached from a few different angles. In \cite{Bzowski:2018aiq} a model was built, where the factorization of the otherwise non-factorizable Hilbert space emerges at low energies. In the context of holographic theories a comprehensive analysis was carried out in the series of papers \cite{Chowdhury:2020hse,Geng:2021hlu,Raju:2021lwh,Chowdhury:2021nxw,Chakraborty:2021rvy,Banerjee:2022jnv}. It was shown that in the presence of gravity the holographic theory `sees' more degrees of freedom than naively expected, which suggests that the bulk Hilbert space is `smaller' than naively expected. In particular in \cite{Raju:2021lwh} it was explicitly argued that the factorization property between the two sides of a black hole fails. Finally, based on the results of \cite{Leutheusser:2021qhd,Leutheusser:2021frk}, in \cite{Witten:2021unn} it was explicitly stated that the Hilbert space associated with the excitations on top of the BTZ black hole does not factorize into the tensor product of the boundary spaces.

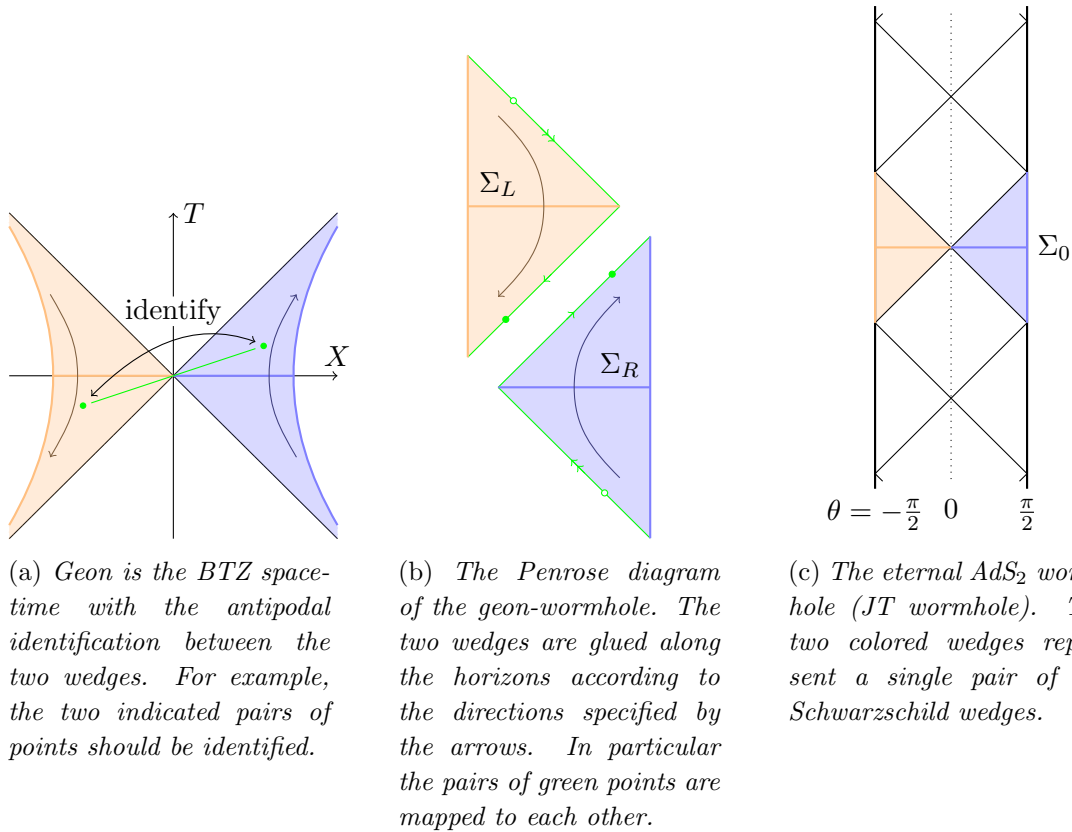
\begin{figure}[t]
\centering
\begin{subfigure}[t]{0.28\textwidth}
\begin{tikzpicture}[scale=1.8]
\draw (-1.2,-1.2) -- ( 1.2,1.2);
\draw ( 1.2,-1.2) -- (-1.2,1.2);
\draw[->] (-1.2,0) -- (1.2,0);
\draw[->] (0,-1.2) -- (0,1.2);
\node[above] at (1.2,0) {$X$};
\node[right] at (0,1.2) {$T$};
\draw[dotted, colorL, fill=colorL, opacity=0.3] (-1.2,-1.1) to [out=60,in=-60] (-1.2,1.1) -- (-1.2,1.2) -- (0,0) -- (-1.2,-1.2) -- cycle;
\draw[dotted, colorR, fill=colorR, opacity=0.3] (1.2,1.2) -- (1.2,1.1) to [out=-120,in=120] (1.2,-1.1) -- (1.2,-1.2) -- (0,0) -- cycle;
\draw[thick, colorL] (-1.2,1.1) to [out=-60,in=60] (-1.2,-1.1);
\draw[thick, colorR] (1.2,1.1) to [out=-120,in=120] (1.2,-1.1);
\draw[thick, colorL] (-0.88,0) -- (0,0);
\draw[thick, colorR] (0.88,0) -- (0,0);
\draw[green, fill=green] (0.66,0.22) circle (0.02);
\draw[green, fill=green] (-0.66,-0.22) circle (0.02);
\draw[green] (-0.6,-0.2) -- (0.6,0.2);
\draw[<->] (-0.6,-0.15) to [out=45, in=160] (0.6,0.25);
\draw[white, fill=white] (-0.32,0.58) rectangle (0.32,0.37);
\node[above] at (0,0.3) {identify};
\draw[colorR!50!black, ->] (0.9,-0.6) to [out=120,in=-90] ( 0.7,0) to[out=90,in=-120] (0.9,0.6);
\draw[colorL!50!black, <-] (-0.9,-0.6) to [out=60,in=-90] (-0.7,0) to[out=90,in=-60] (-0.9,0.6);
\end{tikzpicture}
\centering
\caption{Geon is the BTZ spacetime with the antipodal identification between the two wedges. For example, the two indicated pairs of points should be identified.\label{fig:intro_geon}}
\end{subfigure}
\qquad
\begin{subfigure}[t]{0.28\textwidth}
\begin{tikzpicture}[scale=2.0]
\draw[white, colorL, fill=colorL, opacity=0.3] (-1,-1) -- (-1,1) -- (0,0) -- cycle;
\draw[white, colorR, fill=colorR, opacity=0.3] (0.2,-0.2) -- (0.2,-2.2) -- (-0.8,-1.2) -- cycle;
\draw[green] (-1,-1) -- ( 0,0);
\draw[green] (-0.8,-1.2) -- (0.2,-0.2);
\draw[green] (-1,1) -- ( 0,0);
\draw[green] (-0.8,-1.2) -- (0.2,-2.2);
\draw[->, green] (-0.45,-0.45) -- (-0.5,-0.5);
\draw[->, green] (-0.35,-0.75) -- (-0.3,-0.7);
\draw[->, green] (-0.52,0.52) -- (-0.47,0.47);
\draw[->, green] (-0.48,0.48) -- (-0.43,0.43);
\draw[->, green] (-0.27,-1.73) -- (-0.32,-1.68);
\draw[->, green] (-0.23,-1.77) -- (-0.28,-1.72);
\draw[green, fill=green] (-0.75,-0.75) circle (0.02);
\draw[green, fill=green] (-0.05,-0.45) circle (0.02);
\draw[green, fill=white] (-0.7,0.7) circle (0.02);
\draw[green, fill=white] (-0.1,-1.9) circle (0.02);
\draw[thick, colorL] (-1,1) -- (-1,-1);
\draw[thick, colorR] (0.2,-0.2) -- (0.2,-2.2);
\draw[thick, colorL] (-1,0) -- (0,0);
\draw[thick, colorR] (-0.8,-1.2) -- (0.2,-1.2);
\draw[colorR!50!black, ->] (0,-1.8) to [out=135,in=-90] (-0.3,-1.2) to[out=90,in=-135] (0,-0.6);
\draw[colorL!50!black, <-] (-0.8,-0.6) to [out=45,in=-90] (-0.5,0) to[out=90,in=-45] (-0.8,0.6);
\node[above] at (0,-1.2) {$\Sigma_R$};
\node[above] at (-0.8,0) {$\Sigma_L$};
\end{tikzpicture}
\centering
\caption{The Penrose diagram of the geon-wormhole. The two wedges are glued along the horizons according to the directions specified by the arrows. In particular the pairs of green points are mapped to each other.\label{fig:intro_geonwh}}
\end{subfigure}
\qquad
\begin{subfigure}[t]{0.28\textwidth}
\begin{tikzpicture}[scale=1.0]
\draw[dotted, colorL, fill=colorL, opacity=0.3] (-1,1) -- (0,0) -- (-1,-1) -- cycle;
\draw[dotted, colorR, fill=colorR, opacity=0.3] (1,1) -- (0,0) -- (1,-1) -- cycle;
\draw[thick] (-1,3.2) -- (-1,-3.2);
\draw[thick] ( 1,3.2) -- ( 1,-3.2);
\draw (0.9,3.1) -- (1,3) -- (-1,1) -- (1,-1) -- (-1,-3) -- (-0.9,-3.1);
\draw (-0.9,3.1) -- (-1,3) -- (1,1) -- (-1,-1) -- (1,-3) -- (0.9,-3.1);
\draw[thick,colorL] (-1,0) -- (0,0);
\draw[thick,colorR] (1,0) -- (0,0);
\node[below] at (1,-3.2) {$\frac{\pi}{2}$};
\node[below] at (-1,-3.2) {$\theta = -\frac{\pi}{2}$};
\draw[dotted] (0,3.1) -- (0,-3.1);
\node[below] at (0,-3.2) {$0$};
\node[right] at (1,0) {$\Sigma_0$};
\draw[thick, colorR] (1,1) -- (1,-1);
\draw[thick, colorL] (-1,1) -- (-1,-1);
\end{tikzpicture}
\centering
\caption{The eternal AdS$_2$ wormhole (JT wormhole). The two colored wedges represent a single pair of the Schwarzschild wedges.\label{fig:intro_jt}}
\end{subfigure}
\caption{The Penrose diagrams of two wormhole geometries considered in this paper: the geon-wormhole on the left and the AdS$_2$ eternal wormhole (JT wormhole) on the right.\label{fig:intro_wh}}
\end{figure}

\paragraph{The geon.} In this paper I present the detailed analysis of the matter field in the backgrounds of two wormholes: the geon-wormhole in section \ref{sec:geon} and the Jackiw–Teitelboim (JT) wormhole, \textit{i.e.}, the eternal AdS$_2$ wormhole in section \ref{sec:jt}. Originally, \cite{Louko:1998hc}, the $\R P^2$ geon, or just \emph{the geon}, is the quotient space of the BTZ black hole with points identified by the antipodal map, see figure \ref{fig:intro_geon}. Such a spacetime is smooth, but contains a single asymptotic boundary. The geon then can be regarded as a toy model of a unitary, radiating black  hole and was extensively studied in \cite{Louko:1998hc,tHooft:2016qoo,tHooft:2016rrl,Betzios:2016yaq,Betzios:2020xuj}.

Here we will `unfold' the geon and treat it as the wormhole where the underlying geometry is that of the full BTZ black hole, but the scalar field on top of it is parity-even under the antipodal map. The Penrose diagram of the resulting wormhole is presented in figure \ref{fig:intro_geonwh}. The original geon formulation and our geon-wormhole are equivalent. 

The analysis of the Hilbert space of the geon was first carried out in \cite{Sanchez:1986qn}, where it was pointed out that the standard Fock quantization fails due to zero-norm states. In the context of holography the analysis was carried in \cite{Louko:1998hc,Guica:2014dfa}. All three papers assume that the Hilbert space is that of the BTZ black hole and thus it splits into the tensor product of the boundary Hilbert spaces. Nevertheless, the authors reach different conclusions regarding the geon state as well as the structure of the correlation functions.

\paragraph{The JT wormhole.} The JT wormhole is the AdS$_2$ spacetime, which possesses two asymptotic boundaries, as shown in figure \ref{fig:intro_jt}. In the context of AdS$_2$ holography, \cite{Maldacena:2016hyu,Maldacena:2016upp}, in \cite{Arias:2010xg} it was shown that the Hilbert space of the matter sector does not factorize into the tensor product of the boundary Hilbert spaces. In \cite{Harlow:2018tqv} the authors argued that the Hilbert space structure of the gravitational sector of the JT gravity does not factorize into the tensor product either. Regardless of these results, the standard starting point in the analysis of the Hilbert space of the matter sector of the JT gravity is the factorized Hilbert space. It is then believed that knitting of the wormhole requires the construction of a highly entangled, infinite temperature state, \cite{Kourkoulou:2017zaj,Maldacena:2018lmt,Su:2020zgc,Lin:2022rbf,Antonini:2022lmg}. While such highly entangled states appear in the context of the geon as well, \cite{Guica:2014dfa}, in case of the AdS$_2$ wormhole the state is known as the Kourkoulou-Maldacena state or the SYK thermofield double.

\subsection{Outline of the results}

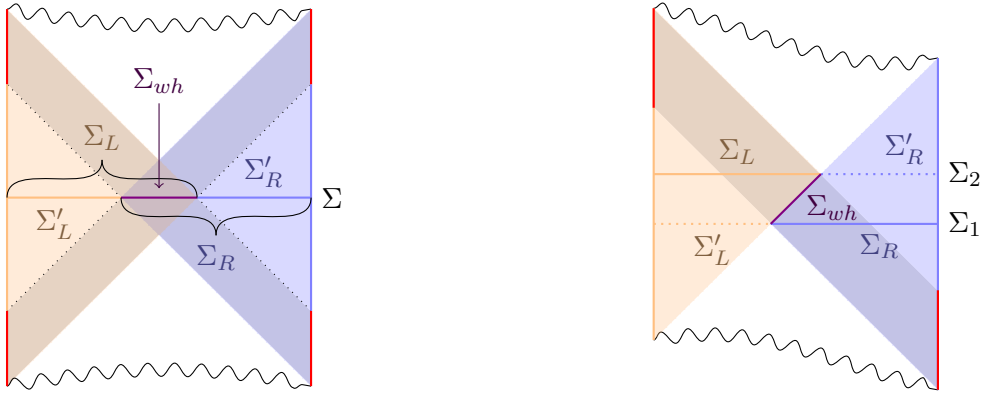
\begin{figure}[t]
\centering
\begin{subfigure}[t]{0.4\textwidth}
\begin{tikzpicture}[scale=2.5]
\draw[dotted, colorR, fill=colorR, opacity=0.3] (0.8,1) -- (0.8,-1) -- (-0.2,0) -- cycle;
\draw[dotted, colorL, fill=colorL, opacity=0.3] (-0.8,1) -- (-0.8,-1) -- (0.2,0) -- cycle;
\draw[white, fill=black, opacity=0.1] (-0.8,1) -- (0,0.2) -- (0.8,1) -- (0.8,0.6) -- (0.2,0) -- (0.8,-0.6) -- (0.8,-1) -- (0,-0.2) -- (-0.8,-1) -- (-0.8,-0.6) -- (-0.2,0) -- (-0.8,0.6) -- cycle;
\draw[snake it] (-0.8,1) to [out=-10,in=-170] (0.8,1);
\draw[snake it] (-0.8,-1) to [out=10,in=170] (0.8,-1);
\draw[dotted] (0.8,-0.6) -- (0.2,0) -- (0.8, 0.6);
\draw[dotted] (-0.8,-0.6) -- (-0.2,0) -- (-0.8, 0.6);
\draw[thick, red] (-0.8,0.6) -- (-0.8,1);
\draw[thick, red] (-0.8,-0.6) -- (-0.8,-1);
\draw[thick, red] (0.8,0.6) -- (0.8,1);
\draw[thick, red] (0.8,-0.6) -- (0.8,-1);
\draw[thick, colorL] (-0.8,0.6) -- (-0.8,-0.6);
\draw[thick, colorR] ( 0.8,0.6) -- ( 0.8,-0.6);
\draw[thick, colorL] (-0.8,0) -- (-0.2,0);
\draw[thick, colorR] (0.8,0) -- (0.2,0);
\draw[thick, violet] (-0.2,0) -- (0.2,0);
\draw (-0.2,0) to [out=-90,in=90] (0.3,-0.2);
\draw (0.8,0) to [out=-90,in=90] (0.3,-0.2);
\node[below, colorR!50!black] at (0.3,-0.2) {$\Sigma_R$};
\node[above, colorR!50!black] at (0.55,0) {$\Sigma'_R$};
\draw (-0.8,0) to [out=90,in=-90] (-0.3,0.2);
\draw (0.2,0) to [out=90,in=-90] (-0.3,0.2);
\node[above, colorL!50!black] at (-0.3,0.2) {$\Sigma_L$};
\node[below, colorL!50!black] at (-0.55,0) {$\Sigma'_L$};
\draw[->, violet!50!black] (0,0.5) -- (0,0.05);
\node[above, violet!50!black] at (0,0.5) {$\Sigma_{wh}$};
\node[right] at (0.8,0) {$\Sigma$};
\end{tikzpicture}
\caption{A symmetric wormhole, where the wormhole modes traversing between the two boundaries are in 1-to-1 correspondence with the initial data on the portion $\Sigma_{wh} \subseteq \Sigma$ of the complete Cauchy slice.\label{fig:wh_sym}}
\centering
\end{subfigure}
\qquad\qquad
\begin{subfigure}[t]{0.4\textwidth}
\begin{tikzpicture}[scale=2.2]
\draw[snake it] (-1,1) to [out=-10,in=-170] (0.7,0.7);
\draw[snake it] (-1,-1) to [out=10,in=170] (0.7,-1.3);
\draw[dotted, colorL, fill=colorL, opacity=0.3] (-1,-1) -- (-1,1) -- (0,0) -- cycle;
\draw[dotted, colorR, fill=colorR, opacity=0.3] (0.7,0.7) -- (0.7,-1.3) -- (-0.3,-0.3) -- cycle;
\draw[white, fill=black, opacity=0.1] (-1,1) -- (0.7,-0.7) -- (0.7,-1.3) -- (-1,0.4) -- cycle;
\draw[thick, colorL] (-1,1) -- (-1,-1);
\draw[thick, colorR] ( 0.7,0.7) -- ( 0.7,-1.3);
\draw[thick, red] (-1,0.4) -- (-1,1);
\draw[thick, red] (0.7,-0.7) -- (0.7,-1.3);
\draw[thick, colorL] (-1,0) -- (0,0);
\draw[thick, colorR] (-0.3,-0.3) -- (0.7,-0.3);
\draw[thick, dotted, colorL] (-1,-0.3) -- (-0.3,-0.3);
\draw[thick, dotted, colorR] (0,0) -- (0.7,0);
\draw[thick, violet] (-0.3,-0.3) -- (0,0);
\node[colorL!50!black, above] at (-0.5,0) {$\Sigma_L$};
\node[colorL!50!black, below] at (-0.65,-0.3) {$\Sigma'_L$};
\node[colorR!50!black, above] at (0.5,0) {$\Sigma'_R$};
\node[colorR!50!black, below] at (0.35,-0.3) {$\Sigma_R$};
\node[violet!50!black, right] at (-0.15,-0.20) {$\Sigma_{wh}$};
\node[right] at (0.7,0) {$\Sigma_2$};
\node[right] at (0.7,-0.3) {$\Sigma_1$};
\end{tikzpicture}
\centering
\caption{The wormhole can be regarded as two wedges glued together along the violet line segment $\Sigma_{wh}$.\label{fig:wh_asym}}
\end{subfigure}
\caption{An example of a wormhole traversable from the right boundary to the left.\label{fig:wh1}}
\end{figure}

\paragraph{Non-factorization.} Consider asymptotically AdS, traversable wormholes as shown in figures \ref{fig:wh_sym} and \ref{fig:wh_asym}. Both wormholes consist of two wedges, the left orange wedge and the right blue wedge, and the wormhole geometry determines how the two wedges are stitched together. Red pieces of the opposite boundaries indicate the regions between which the information can be exchanged.

The Hilbert space of states $\H$ associated with the excitations on top of the wormhole geometry can be associated with states living on any complete Cauchy surface and obeying suitable asymptotic boundary conditions. For example, we can associate the Hilbert space $\H$ to the Cauchy surface $\Sigma$ presented in figure \ref{fig:wh_sym}. Each of the subregions of $\Sigma$ gives rise to a tensor factor inside $\H$. For example, as $\Sigma = \Sigma'_L \cup \Sigma_R$ with the intersection $\Sigma'_L \cap \Sigma_R$ of measure zero, we have the split $\H \cong \H'_L \otimes \H_R$. On the other hand, since $\Sigma_L$ and $\Sigma_R$ have a non-trivial intersection, $\Sigma_L \cap \Sigma_R = \Sigma_{wh}$, the initial data on $\Sigma_L$ and $\Sigma_R$ are not independent: they must agree along the intersection $\Sigma_{wh}$. Thus, we arrive at the \emph{non-factorization} property,
\begin{align} \label{nonfact}
\H_L \otimes \H_R \cong \H \otimes \H_{wh}.
\end{align}
This simply expresses the fact that the initial data on $\Sigma_L$ and $\Sigma_R$ are not independent due to the \emph{wormhole modes} propagating between the boundaries. 

The surfaces $\Sigma_L$ and $\Sigma_R$ in figure \ref{fig:wh_sym} are the cuts through the middle of the causal developments of the boundaries in the bulk. Thus, we can utilize the real-time holography \cite{Skenderis:2008dg,Skenderis:2008dh,KaplansNotes} to identify $\H_L$ and $\H_R$ as the actual dual Hilbert spaces. By matching the bulk modes with the boundary data as in \cite{Papadodimas:2012aq,Kim:2016ipt,Bahiru:2022ukn} we find a 1-to-1 map between the initial data on, say, $\Sigma_R$ and the boundary data on the right boundary.

The situation is analogous for the asymmetric wormhole presented in figure \ref{fig:wh_asym} with the complete Cauchy slice $\Sigma$ corresponding to $\Sigma = \Sigma'_L \cup \Sigma_{wh} \cup \Sigma'_R$. Although now $\Sigma_L$ and $\Sigma_R$ are not subregions of $\Sigma$, one can find a suitable foliation of the wormhole such that $\Sigma'_{L} \cup \Sigma_{wh}$ evolves to $\Sigma_L$ and $\Sigma_R$ to $\Sigma'_{R} \cup \Sigma_{wh}$. Whenever we can evolve between two Cauchy surfaces, the associated Hilbert spaces of states living on these surfaces are isomorphic and the isomorphism is given by the unitary evolution operator. Thus, for example,
\begin{align} \label{fact}
& \H_L \cong \H_{L'} \otimes \H_{wh}, && \H_R \cong \H_{R'} \otimes \H_{wh}.
\end{align}
Consequently, the non-factorization property \eqref{nonfact} holds.

\paragraph{Consequences.} In the paper we want to argue that a number of peculiar features and illusions associated with the wormhole and black hole physics naturally stems from the non-factorization property \eqref{nonfact}. Since the assumption of factorization of the Hilbert space can only ever be an approximation to the physical setup, we will refer to such a situation as \emph{semiclassical approximation}. We work out two examples: the geon-wormhole, as shown in figure \ref{fig:intro_geonwh} as well as the AdS$_2$ wormhole presented in figure \ref{fig:intro_jt} on page \pageref{fig:intro_jt}.

\begin{enumerate}
\item \textbf{Illusion of null states.}  In \cite{Baez:2014bka,Harlow:2018tqv,Gardiner:2022zlx} it was argued that certain states must be identified between the two boundary theories. By combining \eqref{fact} and \eqref{nonfact} we see that $\H_L \otimes \H_R$ contains two copies of $\H_{wh}$, one copy from each $\Sigma_L$ and $\Sigma_R$, while the physical Hilbert space, $\H$, contains only a single copy of $\H_{wh}$. Thus, from the point of view of the semiclassical approximation the corresponding wormhole states in $\H_L \otimes \H_R$ must be identified. In other words any state having its part contained in $\H_{wh}$ in \eqref{nonfact} is \emph{null}. 

\item \textbf{Illusion of null operators.} With the physical Hilbert space $\H$ `smaller' than the tensor product $\Hsemi = \H_L \otimes \H_R$ the algebra of observables does not factorize either. In general operators of the form $\bs{1} \otimes \O_R$, which are supposed to act within a single boundary, are ill-defined. From the point of view of the semiclassical approximations such operators must be declared unphysical.

\item \textbf{Illusion of entanglement.} Both in the context of the geon in \cite{Guica:2014dfa} as well as the AdS$_2$ wormhole, \cite{Kourkoulou:2017zaj,Maldacena:2018lmt,Su:2020zgc,Lin:2022rbf,Antonini:2022lmg}, it is believed that to construct the wormhole holographically, a highly entangled, infinite temperature state is required. Such states satisfy relations of the form 
\begin{align} \label{inftemp}
& (\op{a}^R_{wh} - \op{a}^{L \dagger}_{wh}) | \psi \> = 0, &&  (\op{a}^L_{wh} - \op{a}^{R \dagger}_{wh}) | \psi \> = 0,
\end{align}
where $\op{a}^{L, R \dagger}_{wh}, \op{a}^{L, R}_{wh}$ stand for some creation-annihilation operators associated with the wormhole modes in the two boundaries. We will see that such relations are the consequences of the matching of the wormhole modes along $\Sigma_{wh}$. The relations are naturally embedded in the full, physical theory, but are very difficult to impose once the factorization had been assumed.

\item \textbf{Illusion of interactions.} Since the bulk system is a free system, it is governed by a free Hamiltonian $\op{H}$ once the time direction is selected. The cross-boundary correlation functions emerge due to the non-factorization of the Hilbert space rather than entanglement or interactions. From the point of view of the semiclassical approximation, however, in order to impose the relations \eqref{inftemp}, a suitable cross-boundary interaction can be added. Effectively, the interaction acts as the approximate projector onto the states satisfying \eqref{inftemp}. In the context of the matter sector of the JT gravity such interactions were introduced in \cite{Kourkoulou:2017zaj,Maldacena:2018lmt}.

\end{enumerate}

\paragraph{Examples.} In order to present the peculiarities and illusions of the wormholes in a simple set-up, in section \ref{sec:geon} we analyze the structure of the geon, while in section \ref{sec:jt} the structure of the AdS$_2$ wormhole. Since the systems we analyze have fixed backgrounds, the results can be regarded as the $G_N = 0$ or the leading $1/N$ statements in holography. However, with the gravity turned on one should expect that the non-factorization of the Hilbert space becomes even more severe. Indeed, without gravity present the Hilbert space of the Minkowski spacetime or that of an eternal black hole splits into the tensor product of the Hilbert spaces associated with the two wedges. However, in the presence of gravity the arguments of \cite{Raju:2021lwh,Witten:2021unn} show that the factorization must fail. In this sense one can also treat the results of this paper as toy models of non-factorization. 

In particular in the paper we derive the following results regarding the structure of the geon and the AdS$_2$ wormhole:
\begin{enumerate}
\item \textbf{The Hilbert spaces.} The two examples of the wormholes, as shown in figure \ref{fig:intro_wh} on page \pageref{fig:intro_wh}, are very extreme in the sense that all bulk modes are wormhole modes propagating between the boundaries. Hence, in both cases, we find that the total Hilbert space $\H$ is isomorphic to the left and right boundary Hilbert spaces, $\H \cong \H_L \cong \H_R$ \emph{separately}. From the point of view of the semiclassical approximation this result can only be achieved correctly within the framework of constrained quantization.
\item \textbf{The dual geon state.} We find that the state dual to the geon is different than those advocated in \cite{Louko:1998hc,Guica:2014dfa}. In particular, our state lives in the physical Hilbert space isomorphic to each boundary Hilbert space separately,  $\H \cong \H_L \cong \H_R$. It is a `thermofield single' state: a 1-particle squeezed state exhibiting thermal properties when limited to a single wedge.
\item \textbf{Bogoliubov coefficients for the JT wormhole.} We explicitly calculate the Bogoliubov coefficients between the global modes and the Schwarzschild modes and show that the state dual to the eternal AdS$_2$ wormhole is the thermofield double state from the point of view of the Schwarzschild wedge. The difference between the eternal wormhole and two disconnected boundaries is hidden in the structure of the Hilbert space rather than the state.
\end{enumerate}

\section{Quantization and holography} \label{sec:qh}

\subsection{Canonical quantization}

We consider the free real field $\Phi$ governed by the standard action
\begin{align} \label{Sfree}
S = - \frac{1}{2} \int \D^{d+1} x \sqrt{-g} \left[ g^{\mu\nu} \partial_\mu \Phi \partial_\nu \Phi + m^2 \Phi^2 \right].
\end{align}
The $(d+1)$-dimensional background metric $g_{\mu\nu}$ is fixed. The field obeys Klein-Gordon equation $(-\Box_g + m^2) \Phi = 0$. If the spacetime possesses boundaries or asymptotic boundaries, we assume enough boundary conditions were specified so that the problem becomes globally hyperbolic.

Let $\Mre$ denote the space of real bulk solutions to the Klein-Gordon equation obeying the specified boundary conditions on all (asymptotic) boundaries and by $\Mcx$ denote its complexification. Let us choose a time foliation $\{ \Sigma_{t} \}_t$ by complete Cauchy slices and consider the initial value problem at the surface $\Sigma_t$ of constant time. The canonical form on $\Mre$ induces the non-degenerate Klein-Gordon product on $\Mcx$,
\begin{align} \label{KG}
( \Phi, \Psi ) = - \I \int_{\Sigma_t} \D^d x \sqrt{\gamma} \, n^{\mu} \left[  \Phi \partial_\mu \Psi^{\ast} - \partial_\mu \Phi \, \Psi^{\ast} \right],
\end{align}
where $n^{\mu}$ is a time-like unit vector orthogonal to $\Sigma_t$ and $\gamma_{\mu\nu}$ denotes the metric induced on $\Sigma_t$. The Klein-Gordon product does not depend on the choice of the leaf $\Sigma_t$ of the foliation.

The Klein-Gordon product is a non-degenerate sesquilinear form, but it is not positive-definite; in fact $(\Phi^{\ast}, \Phi^{\ast}) = - (\Phi, \Phi)$. Thus, one can choose a polarization on the space of the complexified solutions, 
\begin{align} \label{polarization}
\Mcx \cong \Hone \oplus (\Hone)^{\ast},
\end{align}
such that the Klein-Gordon product is positive-definite on $\Hone$ and negative-definite on $(\Hone)^{\ast}$. The solutions $\phi \in \Hone$ will be called \emph{negative frequency}, while those in $(\Hone)^{\ast}$ \emph{positive frequency}. At this point the split \eqref{polarization} is arbitrary. After a suitable completion, see \textit{e.g.}, \cite{Wald:1995yp} for details, $\Hone$ becomes the Hilbert space of \emph{1-particle states}.

The total Hilbert space $\H$ is the Fock space, \textit{i.e.} it is obtained as a symmetric power (for a bosonic field) of the 1-particle space,
\begin{align}
\H = Sym(\Hone) = \C \oplus \Hone \oplus ( \Hone \circledS \Hone ) \oplus \ldots,
\end{align}
where $\circledS$ denotes the symmetrized tensor product. The element $1 \in \C$ is known as \emph{the vacuum state} and denoted by $| 0 \>$.

Assume a complete set of negative frequency modes $\{ \phi_n \}_n$ was selected, orthonormal with respect to the Klein-Gordon product \eqref{KG}, $(\phi_m, \phi_n) = \delta_{mn}$. A 1-particle state $\phi_n \in \Hone$ is denoted as $|1 \>_n$ while a $j$-particle state $\phi_n \circledS \ldots \circledS \phi_n$ obtained as the $j$-fold symmetric tensor product of $\phi_n$ is denoted by $| j \>_n \in Sym^j(\Hone)$. 

The Hilbert space $\H$ carries the representation of the canonical commutation relations. The \emph{creation} and \emph{annihilation operators} $\op{a}^{\dagger}_{n}$ and $\op{a}_{n}$ act on multi-particle states as
\begin{align}
& \op{a}^{\dagger}_{n} | j \>_n = \sqrt{j+1} | j+1 \>_n, && \op{a}_{n} | j \>_n = \sqrt{j} | j-1 \>_n, && \op{a}_n | 0 \> = 0.
\end{align}
This results in the creation-annihilation operators being Hermitian conjugates of each other and to satisfy the canonical commutation relations, $\left[ \op{a}_{m}, \op{a}^{\dagger}_{n} \right] = \delta_{mn}$.

\paragraph{Smeared operators.} It will be convenient for our analysis to work with smeared creation-annihilation operators $\op{a}^{\dagger}_{\phi}$ and $\op{a}_{\phi}$. Any $\phi \in \Hone$ can be decomposed in the chosen basis using its Fourier coefficients $c_n = ( \phi, \phi_n )$ so that $\phi = \sum_n c_n \phi_n$. Thus we can define smeared operators
\begin{align}
& \op{a}_\phi = \sum_n c_n \op{a}_n, && \op{a}^{\dagger}_\phi = \sum_n c^{\ast}_n \op{a}^{\dagger}_n,
\end{align}
In particular we can identify a negative frequency mode $\phi \in \Hone$ with the 1-particle state $| 1_{\phi} \> = \op{a}^{\dagger}_\phi | 0 \>  = \sum_n c^{\ast}_n | 1_n \>$ obtained by the action of the smeared creation operator. In terms of the smeared operators the canonical commutation relations can be conveniently expressed in terms of the Klein-Gordon product as
\begin{align} \label{ccr_smeared}
& \left[ \op{a}_{\phi}, \op{a}_{\psi}^{\dagger} \right] = (\phi, \psi) \, \bs{1}, && \< 1_{\phi} | 1_{\psi} \> = (\phi, \psi).
\end{align}

\subsection{The polarization and the foliation} \label{sec:polar}

The choice of the polarization in \eqref{polarization} was arbitrary and devoid of any physical meaning. In practice one prefers to think about the states of quantum theory as living on constant time slices. Given a foliation $\{ \Sigma_t \}_t$ there exists a natural split into positive and negative frequency modes on each leaf. A negative frequency mode behaves as $e^{-\I \omega t}$ in the vicinity of a given time slice, where $t$ is the time direction perpendicular to that leaf. To be more precise consider the gauge in which the metric $g_{\mu\nu}$ takes form
\begin{align} \label{g-split}
\D s^2 = - \D t^2 + \gamma_{ij}(x, t) \D x^i \D x^j
\end{align}
at least locally in the neighborhood of a single leaf. We can choose this leaf to be $\Sigma_0$. The unit normal vector to $\Sigma_0$ is $\partial_t$ and we assume that the metric $\gamma_{ij}$ induced on the leaf $\Sigma_0$ is smooth as $t \rightarrow 0$. This means that in the vicinity of $\Sigma_0$ the Klein-Gordon equation becomes
\begin{align} \label{KG-split}
0 = \left( - \frac{\partial^2}{\partial t^2} + \Delta_x - m^2 + O(t) \right) \Phi,
\end{align}
up to small corrections of order $O(t)$. Here $\Delta_x$ denotes the Laplacian of $\gamma_{ij}$ on $\Sigma_0$.

\paragraph{Time-independent backgrounds.} Consider first the situation where the induced metric $\gamma_{ij}$ in \eqref{g-split} is time-independent, $\gamma_{ij}(x, t) = \gamma_{ij}(x)$, \textit{i.e.}, the geometry of each slice $\Sigma_t$ is identical. In such a case the Klein-Gordon equation \eqref{KG-split} can be Fourier transformed in $t$ and split into positive and negative frequency solutions. \emph{Negative frequency} modes $\phi_{\omega \ell}$ with respect to the foliation take form
\begin{align} \label{Phi-negative}
\phi_{\omega \ell}(t, x) = \frac{e^{- \I \omega t}}{\sqrt{2 \omega}} f_{\omega \ell}(x), \quad \omega > 0,
\end{align}
where $f_{\omega \ell}$ is a time-independent \emph{wave function} satisfying 
\begin{align}
\Delta_x f_{\omega \ell} = (m^2 - \omega^2) f_{\omega \ell}.
\end{align}
The precise range of $\omega > 0$ depends on the background geometry and the boundary conditions imposed. The index $\ell$ denotes collectively different solutions to this equation for a fixed $\omega$. The wave function $f_{\omega \ell}$ lives entirely on a given time slice $\Sigma_0$ and can be chosen to be real, $f_{\omega \ell}^{\ast} = f_{\omega \ell}$. Positive frequency modes are complex conjugates $\phi_{\omega \ell}^{\ast}$ and they behave as $\phi_{\omega \ell}^{\ast} \sim e^{\I \omega t}$ for $\omega > 0$. In this way we obtain the natural polarization $\Mcx \cong \Hone \oplus (\Hone)^{\ast}$ into negative and positive frequency modes. The field operator $\op{\Phi}$ takes form
\begin{align} \label{gensol}
\op{\Phi}(t, x) = \sum_{\ell} \int \frac{\D \omega}{2 \pi} \left[ \phi_{\omega \ell} \op{a}_{\omega \ell} + \phi^{\ast}_{\omega \ell} \op{a}^{\dagger}_{\omega \ell} \right],
\end{align}
where $\op{a}^{\dagger}_{\omega \ell}$ and $\op{a}_{\omega \ell}$ denote the creation-annihilation operators. The vacuum state $|0\>$ is annihilated by all annihilation operators, $\op{a}_{\omega \ell} | 0 \> = 0$, while 1-particle states are identified with the action of the creation operators on the vacuum, $| 1 \>_{\omega \ell} = \op{a}^{\dagger}_{\omega \ell} | 0 \>$.

With the induced metric $\gamma_{ij}$ time-independent, the designation of a mode as positive or negative frequency is global in the sense that the modes \eqref{Phi-negative} are negative frequency for all surfaces $\Sigma_t$. Furthermore, negative frequency modes on one Cauchy surface $\Sigma_t$ evolve into negative frequency modes on another surface $\Sigma_s$, while the positive frequency modes evolve into positive frequency modes. Consequently, there is the global unique vacuum state $| 0 \>$, which is the lowest-energy state with respect to the Hamiltonian defined in the standard way,
\begin{align} \label{genH}
H & = \int_{\Sigma_t} \D^d x \sqrt{\gamma} \left[ \frac{1}{2} \Pi^2  + \frac{1}{2} \gamma^{ij} \partial_i \Phi \partial_j \Phi + \frac{1}{2} m^2 \Phi^2 \right],
\end{align}
where $\Pi = \partial_t \Phi$ is the canonical momentum. Since $\gamma_{ij}$ is time-independent, so is the Hamiltonian. After substituting the modes and normal ordering the creation-annihilation operators, the quantum Hamiltonian becomes
\begin{align} \label{Ham}
\op{H} & = \sum_{\ell} \int \frac{\D \omega}{2 \pi} \omega \op{a}^{\dagger}_{\omega \ell} \op{a}_{\omega \ell}.
\end{align}
Clearly, $\op{H} | 0 \> = 0$ and no particles are produced as the system evolves.

\paragraph{Time-dependent backgrounds.} Let us now return to the situation where the induced metric $\gamma_{ij}$ on each time-slice in \eqref{g-split} remains time-dependent. In the vicinity of $\Sigma_0$ the modes \eqref{Phi-negative} approximate the solutions to the Klein-Gordon equation. Thus, among solutions in $\Mcx$ one can identify those, which are negative frequency with respect to the given time slice $\Sigma_t$. However, as $t$ varies, the split into positive and negative frequency modes would vary as well. 

Given the foliation $\{ \Sigma_t\}_t$ each leaf induces its own polarization
\begin{align} \label{polarization_t}
\Mcx \cong \Hone_t \oplus (\Hone_t)^{\ast}
\end{align}
into negative and positive frequency modes. Instead of a single Hilbert space we obtain the family of Hilbert spaces $\H_t = Sym(\Hone_t)$ with each $\H_t$ associated with the constant time slice $\Sigma_t$. Each Hilbert space $\H_t$ is associated with its own set of creation-annihilation operators $\op{a}^{\Sigma_t \dagger}_{\omega \ell}, \op{a}^{\Sigma_t}_{\omega \ell}$. The field can be written as in \eqref{gensol}, but the creation-annihilation operators are now associated with a given time-slice and change from leaf to leaf. Each leaf has its own \emph{instantaneous vacuum} $| 0 \>_{\Sigma_t}$ annihilated by all $\op{a}^{\Sigma_t}_{\omega \ell}$ and their own notion of particles as created by $\op{a}^{\Sigma_t \dagger}_{\omega \ell}$. The Hamiltonian \eqref{Ham} still retains its form, but now it contains instantaneous creation-annihilation operators.  The instantaneous vacuum $| 0 \>_{\Sigma_t}$ is the lowest energy state with respect to $\op{H}_t$.

\paragraph{Bogoliubov transformations.} With the family of polarizations \eqref{polarization_t} associated with every leaf $\Sigma_t$ of the foliation, we obtained the family of Hilbert spaces $\H_t$, each with its own set of creation-annihilation operators and the instantaneous vacua $| 0 \>_{\Sigma_t}$. All the Hilbert spaces are unitairly isomorphic with the isomorphism $U_{ts} : \H_t \rightarrow \H_s$ induced by the evolution. Let $\phi_t \in \Hone_t$ be a negative frequency mode with respect to the leaf $\Sigma_t$ and normalized in the Klein-Gordon product \eqref{KG}. Since the polarization \eqref{polarization_t} associated with $\Sigma_s$ may be a different one, $\phi_t$ can be decomposed as
\begin{align}
& \phi_t = \mu_{ts} \phi_s + \nu_{ts} \phi_s^{\ast}, && \mu_{ts} = (\phi_t, \phi_s), & \nu_{ts} = - (\phi_t, \phi_s^{\ast}),
\end{align}
where $\phi_s$ and $\phi_s^{\ast}$ are normalized negative and positive frequency modes with respect to $\Sigma_s$. The \emph{Bogoliubov coefficients} $\mu_{ts}, \nu_{ts} \in \C$ satisfy $| \mu_{ts} |^2 - | \nu_{ts} |^2 = 1$.

The isomorphism between $\H_t$ and $\H_s$ is obtained by mapping the operators $\op{a}_{\phi_s}^{\Sigma_s}$ and $\op{a}_{\phi_s}^{\Sigma_s \dagger}$, which act on $\H_s$, to the operators $\op{b}_{\phi_s}^{\Sigma_t}$ and $\op{b}_{\phi_s}^{\Sigma_t \dagger}$, which act on $\H_t$, by
\begin{align} \label{bogo}
\op{a}_{\phi_s}^{\Sigma_s} \: \longmapsto \: \op{b}_{\phi_s}^{\Sigma_t} = \mu_{ts} \op{a}^{\Sigma_t}_{\phi_t} + \nu_{ts}^{\ast} \op{a}_{\phi_t}^{\Sigma_t \dagger}.
\end{align}
This means, for example, that the instantaneous vacuum $|0\>_s \in \H_s$, annihilated by $\op{a}_{\phi_s}^{\Sigma_s}$, is the image of the excited state $| \psi \>_t \in \H_t$ annihilated by $\op{b}_{\phi_s}^{\Sigma_t}$ in $\H_t$. As this \emph{Bogoliubov transformation} induces the unitary isomorphism between $\H_t$ and $\H_s$, one usually does not treat these spaces as completely different entities. Instead, we choose one of them, say $\H_0$, and designate it as \emph{the} Hilbert space of the system $\H = \H_0$. The isomorphism $U_{0s}$ becomes the automorphism of $\H$ with each Hilbert space $\H_s$ being the image of $U_{0s}$. What this means is that we effectively identify operators $\op{a}_{\phi_s}^{\Sigma_s \dagger}, \op{a}_{\phi_s}^{\Sigma_s}$ with $\op{b}_{\phi_s}^{\Sigma_0 \dagger}, \op{b}_{\phi_s}^{\Sigma_0}$. The exact form of the operator $U_{ts}$ and the image of the vacuum state $U_{ts} | 0 \>_t$ is presented in appendix \ref{sec:2-particle_squeeze}.

\paragraph{Isomorphisms.} Every two separable, infinitely-dimensional Hilbert spaces are isomorphic. The Fock spaces, however, carry additional structure, namely the representation of the commutation relations. Thus, we will say that two Fock spaces, $\H_1$ and $\H_2$, are \emph{isomorphic}, $\H_1 \cong \H_2$, if there exists a unitary isomorphism $U : \H_1 \rightarrow \H_2$ of the representations of the canonical commutation relations. Up to some technical assumptions, see \cite{Wald:1995yp}, $\H_1 \cong \H_2$ if and only if there exists a Bogoliubov transformation between their creation-annihilation operators.

Two isomorphic Fock spaces in general have different vacua, due to the non-trivial mixing of the creation and annihilation operators in the Bogoliubov transformation \eqref{bogo}. If, however, there is no mixing, we say that the two Fock spaces are \emph{equal}. This means that the annihilation operators on $\H_1$ are mapped to annihilation operators on $\H_2$, the vacuum of $\H_1$ is mapped to the vacuum of $\H_2$ and 1-particle states $\Hone_1$ are mapped to 1-particle states $\Hone_2$.

\paragraph{Initial data and Hilbert spaces.} In the construction of the family of Hilbert spaces $\H_t$ we have used the family of polarizations \eqref{polarization_t} on the space of global solutions $\Mcx$. The elements of the 1-particle spaces $\Hone_t$ are global solutions $\phi \in \Mcx$. On the other hand it is natural to think about Hilbert space $\H_t$ and its states as living on the slice $\Sigma_t$ and the evolution mapping states on one slice to the other.

Consider a leaf $\Sigma_t$ of the selected foliation $\{ \Sigma_t \}_t$ and consider the system described by \eqref{Sfree} as the initial value problem on $\Sigma_t$. By $\Ire_t$ and $\Icx_t$ we denote the set of real and complex initial data on $\Sigma_t$ respectively. In general, $\Ire_t$ consists of the values of the bulk field $\Phi |_{\Sigma_t}$ and its conjugate momentum $n^\mu \partial_{\mu} \Phi|_{\Sigma_t}$ specified on $\Sigma_t$. Here $n^{\mu}$ is a timelike unit vector orthogonal to the surface $\Sigma_t$. Additional (asymptotic) boundary conditions may decrease the number of independent degrees of freedom and thus the space of initial conditions $\Ire_t$ may be smaller.

We assume that enough data have been specified so that the problem becomes globally hyperbolic. This means that the evolution defines a 1-to-1 map between the initial data and the full bulk solutions. Thus, $\Mcx \cong \Icx_t$ for all $t$. In particular any polarization \eqref{polarization_t} induces the polarization of the complex initial data $\Icx_t$ on each leaf $\Sigma_t$,
\begin{align}
& \Icx_t \cong \Ineg_t \oplus \Ipos_t, && \Ipos_t = (\Ineg_t)^{\ast}.
\end{align}
Any initial data $i_t \in \Ineg_t$ gives rise to a negative frequency mode, $\phi_t \in \Hone_t$, while $i_t^{\ast} \in \Ipos_t$ evolves into a positive frequency mode $\phi_t^{\ast} \in (\Hone_t)^{\ast}$. In this way we obtain the 1-to-1 map between the space of 1-particle states $\Hone_t$ and the set of negative frequency initial data $\Ineg_t$ for each leaf $\Sigma_t$,
\begin{align}
\Hone_t = \{ \phi_{\omega \ell}^{\Sigma_t} \}_{\omega \ell} \cong \{ f_{\omega \ell}^{\Sigma_t} \}_{\omega \ell} \cong \Ineg_t.
\end{align}
We can identify the elements of $\Ineg_t$, and by extension the elements of $\Hone_t$, with the wave functions $f_{\omega \ell}$. These wave functions live entirely within a given time-slice and thus fulfill the requirement that the 1-particle states live on constant time slices.

Using the Klein-Gordon product \eqref{KG} we can endow the space $\Ineg_t$ of the wave functions $f_{\omega \ell}$ with the scalar product. We define the scalar product on $\Ineg_t$ to be equal the Klein-Gordon product for the corresponding bulk fields in \eqref{Phi-negative}
\begin{align} \label{f_L2}
( f_{\omega \ell}, f_{\omega' \ell'})_{\Ineg_{t}} = ( \phi_{\omega \ell}, \phi_{\omega' \ell'} )_{\Hone_t} = \int_{\Sigma_t} \D^d x \sqrt{\gamma} f_{\omega \ell} f^{\ast}_{\omega' \ell'},
\end{align}
which means that the wave functions $f_{\omega \ell}$ are square-integrable on each time slice.

\subsection{Factorization properties} \label{sec:factor}

Consider a complete Cauchy slice $\Sigma$ and assume it splits into two disconnected regions, $\Sigma = \Sigma_L \cup \Sigma_R$ and $\Sigma_L \cap \Sigma_R = \emptyset$. Clearly, the wave function $f \in \Ineg$ on $\Sigma$ splits into the sum of two wave functions, $f = f_L + f_R$ on $\Sigma_L$ and $\Sigma_R$ respectively. Thus $\Ineg \cong \Ineg_L \oplus \Ineg_R$ and the Hilbert space splits into the tensor product, $\H \cong \H_L \otimes \H_R$. Here and in the remainder of the section we write $\Ineg$ for $\Ineg_{\Sigma}$, $\Ineg_L$ for $\Ineg_{\Sigma_L}$ and so on.

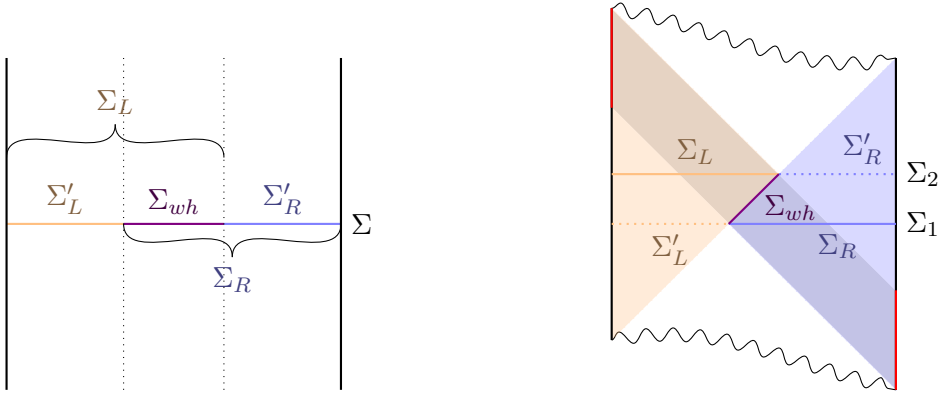
\begin{figure}[ht]
\centering
\begin{subfigure}[t]{0.4\textwidth}
\begin{tikzpicture}[scale=2.2]
\draw[thick, colorL] (-1,0) -- (-0.3,0);
\draw[thick, colorR] ( 1,0) -- ( 0.3,0);
\draw[thick, violet] (-0.3,0) -- (0.3,0);
\draw[thick, black] (-1,-1) -- (-1,1);
\draw[thick, black] ( 1,-1) -- ( 1,1);
\draw[dotted] (-0.3,-1) -- (-0.3,1);
\draw[dotted] ( 0.3,-1) -- ( 0.3,1);
\node[colorL!50!black, above] at (-0.65,0) {$\Sigma'_L$};
\node[colorR!50!black, above] at ( 0.65,0) {$\Sigma'_R$};
\node[violet!50!black, above] at (0,0) {$\Sigma_{wh}$};
\draw (-0.3,0) to [out=-90,in=90] (0.35,-0.2);
\draw (1,0) to [out=-90,in=90] (0.35,-0.2);
\node[below, colorR!50!black] at (0.35,-0.2) {$\Sigma_R$};
\draw (-1,0.4) to [out=90,in=-90] (-0.35,0.6);
\draw (0.3,0.4) to [out=90,in=-90] (-0.35,0.6);
\node[above, colorL!50!black] at (-0.35,0.6) {$\Sigma_L$};
\node[right] at (1,0) {$\Sigma$};
\end{tikzpicture}
\centering
\caption{Three disjoints regions, $\Sigma'_L$, $\Sigma_{wh}$, $\Sigma'_R$, on a constant time slice $\Sigma$ of some foliation. The three regions combine into two sets, $\Sigma_L = \Sigma'_L \cup \Sigma_{wh}$ and $\Sigma_R = \Sigma'_R \cup \Sigma_{wh}$. These two sets have a non-trivial intersection, $\Sigma_L \cap \Sigma_R = \Sigma_{wh} \neq \emptyset$.\label{fig:splits}}
\end{subfigure}
\qquad\qquad
\begin{subfigure}[t]{0.4\textwidth}
\begin{tikzpicture}[scale=2.2]
\draw[snake it] (-1,1) to [out=-10,in=-170] (0.7,0.7);
\draw[snake it] (-1,-1) to [out=10,in=170] (0.7,-1.3);
\draw[dotted, colorL, fill=colorL, opacity=0.3] (-1,-1) -- (-1,1) -- (0,0) -- cycle;
\draw[dotted, colorR, fill=colorR, opacity=0.3] (0.7,0.7) -- (0.7,-1.3) -- (-0.3,-0.3) -- cycle;
\draw[white, fill=black, opacity=0.1] (-1,1) -- (0.7,-0.7) -- (0.7,-1.3) -- (-1,0.4) -- cycle;
\draw[thick] (-1,1) -- (-1,-1);
\draw[thick] ( 0.7,0.7) -- ( 0.7,-1.3);
\draw[thick, red] (-1,0.4) -- (-1,1);
\draw[thick, red] (0.7,-0.7) -- (0.7,-1.3);
\draw[thick, colorL] (-1,0) -- (0,0);
\draw[thick, colorR] (-0.3,-0.3) -- (0.7,-0.3);
\draw[thick, dotted, colorL] (-1,-0.3) -- (-0.3,-0.3);
\draw[thick, dotted, colorR] (0,0) -- (0.7,0);
\draw[thick, violet] (-0.3,-0.3) -- (0,0);
\node[colorL!50!black, above] at (-0.5,0) {$\Sigma_L$};
\node[colorL!50!black, below] at (-0.65,-0.3) {$\Sigma'_L$};
\node[colorR!50!black, above] at (0.5,0) {$\Sigma'_R$};
\node[colorR!50!black, below] at (0.35,-0.3) {$\Sigma_R$};
\node[violet!50!black, right] at (-0.15,-0.20) {$\Sigma_{wh}$};
\node[right] at (0.7,0) {$\Sigma_2$};
\node[right] at (0.7,-0.3) {$\Sigma_1$};
\end{tikzpicture}
\centering
\caption{The same picture as in the left panel, but with various surfaces evolved in time. For example, the initial data on $\Sigma_R$ and $\Sigma'_R \cup \Sigma_{wh}$ are equivalent as they both specify the same portions of the bulk fields.\label{fig:splitsWh}}
\end{subfigure}
\caption{Inital data for wormholes and the factorization property.\label{fig:wh3}}
\end{figure}

On the other hand consider now the two regions $\Sigma_L$ and $\Sigma_R$ with a non-empty overlap, $\Sigma_{wh} = \Sigma_L \cap \Sigma_R \neq \emptyset$. Such a situation is presented in figure \ref{fig:splits}. The initial conditions $\Ineg_{L}$ and $\Ineg_{R}$ restricted to $\Sigma_L$ and $\Sigma_R$ are now dependent. Given a pair of wave functions $f_L \oplus f_R \in \Ineg_{L} \oplus \Ineg_{R}$ one can stitch them together into a single $f \in \Ineg$ only if they satisfy $f_R - f_L = 0$ on the overlap $\Sigma_{wh}$. We can write the map
\begin{align} \label{isoLR}
& F : \Ineg_{L} \oplus \Ineg_{R} \overset{\cong}{\longrightarrow}  \Ineg \oplus \Ineg_{wh}, && F = (f_L, f_R) \longmapsto (f, f_{wh}),
\end{align} 
where 
\begin{align}
f & = f_L |_{\Sigma'_L} + f_R |_{\Sigma'_L} + \frac{1}{\sqrt{2}} ( f_R + f_L )|_{\Sigma_{wh}}, \\
f_{wh} & = \frac{1}{\sqrt{2}} ( f_R - f_L )|_{\Sigma_{wh}}.
\end{align} 
Using the polarization identity one can check that this is indeed the isomorphism. The subscript stands for \emph{wormhole}, as we will identify this factor as the 1-particle space of wormhole modes traveling through the wormhole geometry. For the full Hilbert spaces this means
\begin{align} \label{non-factor}
\H_L \otimes \H_R \cong \H \otimes \H_{wh}.
\end{align}
The isomorphism is given by applying the functor of the symmetric power to \eqref{isoLR}. For the system presented in figure \ref{fig:splits} this is in fact equality of Fock spaces in the sense of section \ref{sec:polar} as the two sides share the same vacuum. We will refer to this equation as the \emph{non-factorization property} of the total Hilbert space $\H$.

If the two regions $\Sigma_L$ and $\Sigma_R$ intersect along their common boundary only, the situation may be subtle. However, with the wave functions being square-integrable  according to \eqref{f_L2}, the measure-zero boundaries are invisible to the initial data. This means, in particular, that we can split $\Ineg_L$ and $\Ineg_R$ into the wave functions supported on $\Sigma_{wh}$ and those supported on $\Sigma'_L$ and $\Sigma'_R$ respectively,
\begin{align} \label{split_ILR}
& \Ineg_L \cong \Ineg_{L'} \oplus \Ineg_{wh}, && \Ineg_R \cong \Ineg_{R'} \oplus \Ineg_{wh},
\end{align}
which means that
\begin{align} \label{split_HLR}
& \H_L \cong \H_{L'} \otimes \H_{wh}, && \H_R \cong \H_{R'} \otimes \H_{wh}.
\end{align}
Finally, we can substitute \eqref{split_ILR} to \eqref{isoLR} and split $\Ineg$ into wave functions associated with the three regions $\Sigma_{L'}$, $\Sigma_{R'}$ and $\Sigma_{wh}$. Thus, we obtain the split
\begin{align} \label{split_HLRwh}
\H \cong \H_{L'} \otimes \H_{wh} \otimes \H_{R'}.
\end{align}
For the system presented in figure \ref{fig:splits} both this isomorphism and those in \eqref{split_HLR} are equalities of Fock spaces.

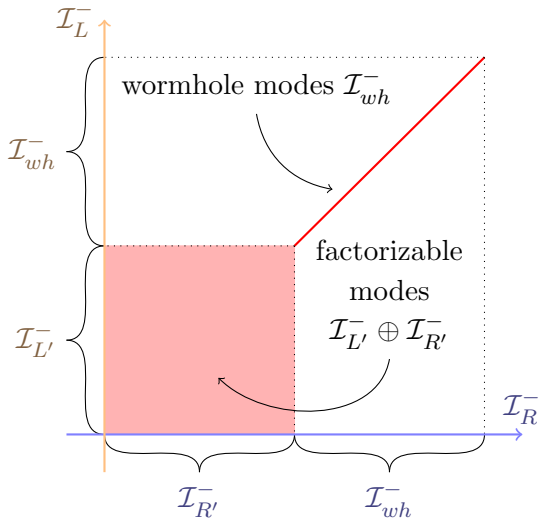
\begin{figure}[ht]
\begin{tikzpicture}[scale=2.5]
\node[above, colorR!50!black] at (2.2,0) {$\Ineg_R$};
\node[left, colorL!50!black] at (0,2.2) {$\Ineg_L$};
\draw[red, fill=red, opacity=0.3] (0,0) rectangle (1,1);
\draw[thick, red] (1,1) -- (2,2);
\draw[dotted] (2,2) -- (2,0);
\draw[dotted] (2,2) -- (0,2);
\draw[dotted] (1,1) -- (1,0);
\draw[dotted] (1,1) -- (0,1);
\draw (0,0) to [out=-90,in=90] (0.5,-0.2);
\draw (1,0) to [out=-90,in=90] (0.5,-0.2);
\draw (1,0) to [out=-90,in=90] (1.5,-0.2);
\draw (2,0) to [out=-90,in=90] (1.5,-0.2);
\node[below, colorR!50!black] at (0.5,-0.2) {$\Ineg_{R'}$};
\node[below, colorR!50!black] at (1.5,-0.2) {$\Ineg_{wh}$};
\draw (0,0) to [out=180,in=0] (-0.2,0.5);
\draw (0,1) to [out=180,in=0] (-0.2,0.5);
\draw (0,1) to [out=180,in=0] (-0.2,1.5);
\draw (0,2) to [out=180,in=0] (-0.2,1.5);
\node[left, colorL!50!black] at (-0.2,0.5) {$\Ineg_{L'}$};
\node[left, colorL!50!black] at (-0.2,1.5) {$\Ineg_{wh}$};
\draw[->] (0.8,1.7) to [out=-80,in=170] (1.2,1.3);
\node[above] at (0.8,1.7) {wormhole modes $\Ineg_{wh}$};
\draw[->] (1.5,0.4) to [out=-100,in=-45] (0.6,0.3);
\node[above, text width=2.0cm, align=center] at (1.5,0.4) {factorizable modes $\Ineg_{L'} \oplus \Ineg_{R'}$};
\draw[thick, colorR, ->] (-0.2,0) -- (2.2,0);
\draw[thick, colorL, ->] (0,-0.2) -- (0,2.2);
\end{tikzpicture}
\centering
\caption{Schematic structure of the initial data $\Ineg$ (in red) for the wormholes from figure \ref{fig:wh1}. The initial data $\Ineg$ contains the factorizable modes $\Ineg_{L'} \oplus \Ineg_{R'}$, which can be chosen independently in each factor $\Ineg_L$ and $\Ineg_R$, as well as the wormhole modes $\Ineg_{wh}$, which must match between the two factors along $\Sigma_{wh}$. \label{fig:scheme}}
\end{figure}

\paragraph{Wormholes.} Notice that the conclusions of the discussion above, expressed in the non-factorization statement \eqref{non-factor}, remain valid after the evolution is taken into account. Consider the geometry presented in figure \ref{fig:splitsWh}. We can identify the Cauchy surface $\Sigma$ from figure \ref{fig:splits} with the surface $\Sigma = \Sigma'_L \cup \Sigma_{wh} \cup \Sigma'_R$ in figure \ref{fig:splitsWh}. While $\Sigma_{wh}$ is formally null, we can still regard $\Sigma$ as the complete Cauchy surface. Indeed, $\Ineg_{wh}$ provides the initial conditions for the left moving wormhole modes, which propagate between the red segments of the boundaries.

Unlike in figure \ref{fig:splits}, the surface $\Sigma_L$ is not equal $\Sigma'_L \cup \Sigma_{wh}$. However, there exists a foliation and the associated Hamiltonian, which brings $\Sigma_L$ to $\Sigma'_L \cup \Sigma_{wh}$. Thus, $\Icx_L \cong \Icx_{L'} \oplus \Icx_{wh}$ on the level of the complex initial data. The positive and negative frequency modes do not have to agree necessarily between $\Icx_L$ and $\Icx_{L'} \oplus \Icx_{wh}$, so the isomorphisms \eqref{split_ILR} in general fail. However, as discussed in section \ref{sec:polar}, the choice of the foliation induces the Bogoliubov transformations between Hilbert spaces associated with different leaves. Thus the isomorphisms \eqref{split_HLR} and \eqref{split_HLRwh} hold. In this case, however, the isomorphisms involve the Bogoliubov transformation induced by the evolution from $\Sigma_L$ to $\Sigma'_L \cup \Sigma_{wh}$. In particular the non-factorization statement \eqref{non-factor} remains valid for the wormhole geometry in figure \ref{fig:splitsWh}.

\subsection{Holography} \label{sec:holo}

\begin{figure}[ht]
\centering
\begin{subfigure}[t]{0.25\textwidth}
\begin{tikzpicture}[scale=1.5]
\draw[dotted, colorR, fill=colorR, opacity=0.3] (1.2,1) -- (1.2,-1) -- (0.2,0) -- cycle;
\draw[dotted, colorL, fill=colorL, opacity=0.3] (-1.2,1) -- (-1.2,-1) -- (-0.2,0) -- cycle;
\draw[dotted, black, fill=black, opacity=0.3] (0.2,0) -- (0,0.2) -- (-0.2,0) -- (0,-0.2) -- cycle;
\draw[thick, colorL] (-1.2,1) -- (-1.2,-1);
\draw[thick, colorR] ( 1.2,1) -- ( 1.2,-1);
\draw[snake it] (-1.2,1) to [out=-10,in=-170] (1.2,1);
\draw[snake it] (-1.2,-1) to [out=10,in=170] (1.2,-1);
\draw[dotted, black] (-0.7,0.9) -- (0,0.2) -- (0.7,0.9);
\draw[dotted, black] (-0.7,-0.9) -- (0,-0.2) -- (0.7,-0.9);
\draw[thick, colorL] (-1.2, 0) -- (-0.2,0);
\draw[thick, colorR] ( 1.2, 0) -- ( 0.2,0);
\draw[thick, black] ( -0.2, 0) -- ( 0.2,0);
\end{tikzpicture}
\centering
\caption{The region marked in gray cannot be probed from the boundaries: the geometry is non-holographic.\label{fig:holoL}}
\end{subfigure}
\qquad
\begin{subfigure}[t]{0.25\textwidth}
\begin{tikzpicture}[scale=1.5]
\draw[dotted, colorR, fill=colorR, opacity=0.3] (1,1) -- (1,-1) -- (0,0) -- cycle;
\draw[dotted, colorL, fill=colorL, opacity=0.3] (-1,1) -- (-1,-1) -- (0,0) -- cycle;
\draw[snake it] (-1,1) to [out=-10,in=-170] (1,1);
\draw[snake it] (-1,-1) to [out=10,in=170] (1,-1);
\draw[thick, colorL] (-1,1) -- (-1,-1);
\draw[thick, colorR] ( 1,1) -- ( 1,-1);
\draw[thick, colorL] (-1,0) -- (0,0);
\draw[thick, colorR] (0,0) -- (1,0);
\end{tikzpicture}
\centering
\caption{A typical semiclassical black hole, where the bulk degrees of freedom are in 1-to-1 correspondence with the boundary states.\label{fig:holoM}}
\end{subfigure}
\qquad
\begin{subfigure}[t]{0.25\textwidth}
\begin{tikzpicture}[scale=1.5]
\draw[dotted, colorR, fill=colorR, opacity=0.3] (0.8,1) -- (0.8,-1) -- (-0.2,0) -- cycle;
\draw[dotted, colorL, fill=colorL, opacity=0.3] (-0.8,1) -- (-0.8,-1) -- (0.2,0) -- cycle;
\draw[snake it] (-0.8,1) to [out=-10,in=-170] (0.8,1);
\draw[snake it] (-0.8,-1) to [out=10,in=170] (0.8,-1);
\draw[dotted] (0.8,-0.6) -- (0.2,0) -- (0.8, 0.6);
\draw[dotted] (-0.8,-0.6) -- (-0.2,0) -- (-0.8, 0.6);
\draw[thick, colorL] (-0.8,0.6) -- (-0.8,-0.6);
\draw[thick, colorR] ( 0.8,0.6) -- ( 0.8,-0.6);
\draw[thick, red] (-0.8,0.6) -- (-0.8,1);
\draw[thick, red] (0.8,-0.6) -- (0.8,-1);
\draw[thick, red] (-0.8,-0.6) -- (-0.8,-1);
\draw[thick, red] (0.8,0.6) -- (0.8,1);
\draw[thick, colorL] (-0.8,0) -- (-0.2,0);
\draw[thick, colorR] (0.8,0) -- (0.2,0);
\draw[thick, violet] (-0.2,0) -- (0.2,0);
\end{tikzpicture}
\centering
\caption{A wormhole, where the boundary data on the red line segments is not independent due to the wormhole modes traversing the wormhole.\label{fig:holoR}}
\end{subfigure}
\caption{Penrose diagrams of non-holographic and holographic spacetimes, assuming no quantum gravity in the bulk.\label{fig:holo1}}
\end{figure}
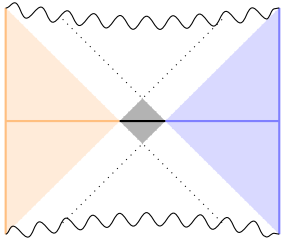
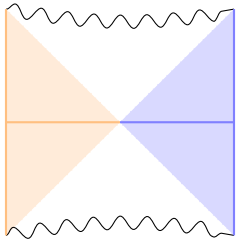
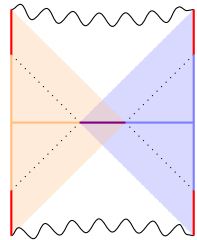

A bulk system is holographic if it is completely encoded in the boundary data. How exactly the bulk data is encoded in the boundary data may be a complicated question, particularly in the presence of gravity. In our simple toy models the situation is straightforward and pictorially shown in figure \ref{fig:holo1}. The spacetime on the left is not holographic, since the shaded region cannot be probed by waves emitted from any boundary. The two remaining spacetimes are holographic, since the entirety of the bulk is reached from the boundaries. However, in the case of a wormhole presented on the right, the boundary data on the two boundaries is not independent. The evolution relates the boundary data on the corresponding red segments.

\paragraph{Classical analysis.} Let us consider free scalar field \eqref{Sfree} on the wormhole backgrounds described by the metric $g_{\mu\nu}$ with two asymptotic boundaries, left and right. We assume that the geometry near each boundary is that of AdS of radius $L_I$ for $I = L,R$. If $z_I$ denotes a radial variable near the $I$-th boundary component at $z_I = 0$, the classical field $\Phi$ exhibits the fall-off $\Phi \sim \phi_I z^{d - \Delta_I} + \pi_I z^{\Delta_I}$ near that boundary. Here $d$ is the dimension of the boundary and $m^2 L_I^2 = \Delta_I (\Delta_I - d)$. Holographically, $\phi_I$ is related to the deformation of the boundary theory, while $\pi_I$ to the state of the theory. In this paper we are interested in the undeformed theory and thus we impose Dirichlet boundary conditions, $\phi_I = 0$, on all boundaries. 

\begin{figure}[t]
\centering
\begin{subfigure}[t]{0.4\textwidth}
\begin{tikzpicture}[scale=2.5]
\draw[dotted, colorR, fill=colorR, opacity=0.3] (0.8,1) -- (0.8,-1) -- (-0.2,0) -- cycle;
\draw[dotted, colorL, fill=colorL, opacity=0.3] (-0.8,1) -- (-0.8,-1) -- (0.2,0) -- cycle;
\draw[snake it] (-0.8,1) to [out=-10,in=-170] (0.8,1);
\draw[snake it] (-0.8,-1) to [out=10,in=170] (0.8,-1);
\draw[dotted] (0.8,-0.6) -- (0.2,0) -- (0.8, 0.6);
\draw[dotted] (-0.8,-0.6) -- (-0.2,0) -- (-0.8, 0.6);
\draw[thick, red] (-0.8,0.6) -- (-0.8,1);
\draw[thick, red] (-0.8,-0.6) -- (-0.8,-1);
\draw[thick, red] (0.8,0.6) -- (0.8,1);
\draw[thick, red] (0.8,-0.6) -- (0.8,-1);
\draw[thick, colorL] (-0.8,0.6) -- (-0.8,-0.6);
\draw[thick, colorR] ( 0.8,0.6) -- ( 0.8,-0.6);
\draw[thick, colorL] (-0.8,0) -- (-0.2,0);
\draw[thick, colorR] (0.8,0) -- (0.2,0);
\draw[thick, violet] (-0.2,0) -- (0.2,0);
\draw[<->] (0.75,0.5) to [out=180,in=90] (0.3,0.05);
\draw[<->] (-0.75,0.5) to [out=0,in=90] (-0.3,0.05);
\node[below] at (0.4,0) {$\Icx_R$};
\node[below] at (-0.4,0) {$\Icx_L$};
\node[right] at (0.8,0.5) {$\Bcx_R$};
\node[left] at (-0.8,0.5) {$\Bcx_L$};
\end{tikzpicture}
\caption{In the `symmetric' wormhole from figure \ref{fig:holoR} the boundary data on the left and right boundary corresponds to the initial data on $\Sigma_L$ and $\Sigma_R$ from figure \ref{fig:splits}.\label{fig:bnd_to_iniL}}
\centering
\end{subfigure}
\qquad\qquad
\begin{subfigure}[t]{0.4\textwidth}
\begin{tikzpicture}[scale=2.2]
\draw[snake it] (-1,1) to [out=-10,in=-170] (0.7,0.7);
\draw[snake it] (-1,-1) to [out=10,in=170] (0.7,-1.3);
\draw[dotted, colorL, fill=colorL, opacity=0.3] (-1,-1) -- (-1,1) -- (0,0) -- cycle;
\draw[dotted, colorR, fill=colorR, opacity=0.3] (0.7,0.7) -- (0.7,-1.3) -- (-0.3,-0.3) -- cycle;
\draw[white, fill=black, opacity=0.1] (-1,1) -- (0.7,-0.7) -- (0.7,-1.3) -- (-1,0.4) -- cycle;
\draw[thick, red] (-1,0.4) -- (-1,1);
\draw[thick, red] ( 0.7,-1.3) -- ( 0.7,-0.7);
\draw[thick, colorL] (-1,0.4) -- (-1,-1);
\draw[thick, colorR] ( 0.7,0.7) -- ( 0.7,-0.7);
\draw[thick, colorL] (-1,0) -- (0,0);
\draw[thick, colorR] (-0.3,-0.3) -- (0.7,-0.3);
\draw[thick, violet] (-0.3,-0.3) -- (0,0);
\node[below] at (-0.5,0) {$\Icx_L$};
\node[below] at (0.35,-0.3) {$\Icx_R$};
\node[left] at (-1,0) {$\Bcx_L$};
\node[right] at (0.7,-0.3) {$\Bcx_R$};
\draw[<->] (0.65,0.2) to [out=180,in=90] (0.2,-0.25);
\draw[<->] (-0.95,0.5) to [out=0,in=90] (-0.5,0.05);
\end{tikzpicture}
\centering
\caption{The relation between the boundary and initial data for the wormhole presented in figure \ref{fig:splitsWh}.\label{fig:bnd_to_iniR}}
\end{subfigure}
\caption{The 1-to-1 map between the boundary data and initial data for wormholes.\label{fig:bnd_to_ini}}
\end{figure}
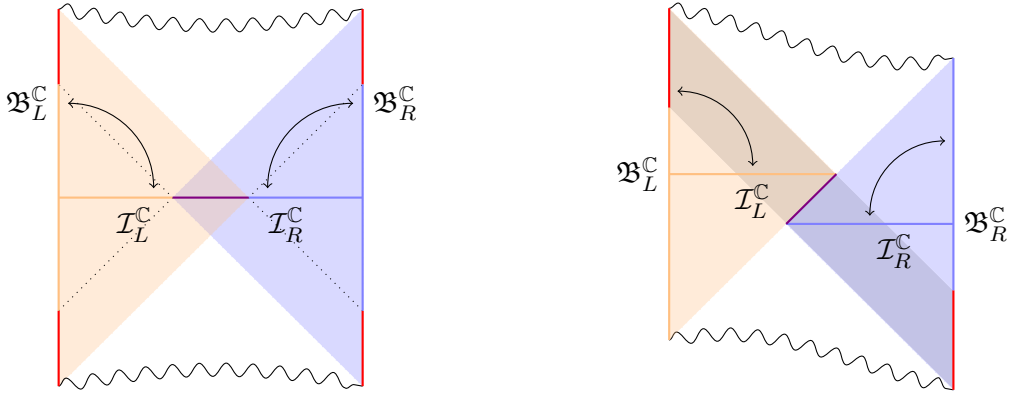

Given a bulk field $\Phi$ we can take its \emph{boundary values}
\begin{align} \label{pi}
\varphi_I = \lim_{z \rightarrow 0} z^{-\Delta_I} \Phi, \quad I = L,R.
\end{align}
Let us concentrate on the complex fields and by $\Bcx_L$ and $\Bcx_R$ denote the set of all complex boundary values on the left and right boundary component. As we vary $\Phi_{\C}$ over $\Mcx$, the space of all solutions to the Klein-Gordon equation, we obtain the set
\begin{align}
\Dcx = \{ (\varphi_L, \varphi_R) \, : \, \Phi_{\C} \in \Mcx \}.
\end{align}
Note that this set is in principle only a subset of
\begin{align}
\Bcx = \Bcx_L \oplus \Bcx_R.
\end{align}
If the boundary data on the two boundaries are not independent, $\Dcx$ is a proper subset of $\Bcx$. For example, consider the wormhole geometry presented in figure \ref{fig:bnd_to_iniR}. Let $\Phi_{\C} \in \Mcx$ be a bulk field with the right boundary value $\varphi_R$ non-vanishing on the red segment of the right boundary. The bulk evolution implies that $\Phi_{\C}$ must have a non-vanishing boundary value on the red portion of the left boundary as well. In particular the boundary data $(0, \varphi_R) \notin \Dcx$ does not represent any bulk field.

Consider now the converse: the bulk reconstruction from the boundary data. For a given pair $(\varphi_L, \varphi_R) \in \Dcx$ of the boundary data there exists the bulk field 
\begin{align} \label{Phi}
\Phi_{\C}[\varphi_L, \varphi_R] \in \Mcx
\end{align}
with the specified boundary values. In general holography implies the following:
\begin{enumerate}
\item If $\Phi_{\C}[\varphi_L, \varphi_R]$ exists for some $(\varphi_L, \varphi_R) \in \Bcx$, it is unique.
\item For a given $\varphi_R \in \Bcx_R$ at least one $\varphi_L \in \Bcx_L$ exists such that $(\varphi_L, \varphi_R) \in \Dcx$ is a valid boundary data. (And analogously for $\varphi_L \in \Bcx_L$.)
\end{enumerate}
Indeed, in holography, the boundary data must determine the bulk. If there existed more than one $\Phi_{\C}[\varphi_L, \varphi_R]$ for a given $(\varphi_L, \varphi_R)$, they would be indistinguishable from the point of view of the boundary theory. Similarly, if point 2 was false, then we would be able to remove  $\varphi_R$ from $\Bcx_R$ entirely, as no bulk field would be able to probe the associated state.

\paragraph{Boundary Hilbert spaces in the bulk.} Since we are interested in looking at the system from the bulk perspective, we want to identify the boundary Hilbert spaces $\H_L^{\partial}$ and $\H_R^{\partial}$ in the bulk Hilbert space $\H$. To do it, fix a complete Cauchy slice $\Sigma$ in the bulk. For a given boundary, say right, we want to determine the subset $\Icx_R \subseteq \Icx$ of the complex initial data on $\Sigma$, which corresponds to the boundary data in $\Bcx_R$ on the right boundary. To do it, first define the set 
\begin{align}
\bar{\mathcal{I}}^{\C}_R = \{ \left. \Phi_{\C}[\varphi_L, 0] \right|_{\Sigma} \: : \: (\varphi_L, 0) \in \Dcx \} \subseteq \Icx,
\end{align}
\textit{i.e.}, the set of the initial data on $\Sigma$ that is \emph{independent} of the boundary data in $\Bcx_R$. The complement of $\bar{\mathcal{I}}^{\C}_R$ in the Klein-Gordon product \eqref{KG} defines the subspace $\Icx_R \subseteq \Icx$ such that
\begin{align}
\Icx \cong \Icx_R \oplus \bar{\mathcal{I}}^{\C}_R.
\end{align}
The space $\Icx_R$ contains the initial data that uniquely determines and is determined by the boundary data $\Bcx_R$ on the right boundary. In the same way we define $\Icx_L$, the space of initial data associated with the left boundary. Note that $\Icx_L \cap \Icx_R = \Icx_{wh}$, the space of the initial data giving rise to the wormhole modes propagating between the two boundaries. In principle the subspaces $\Icx_L, \Icx_R$ and $\Icx_{wh}$ do not have to be related to any subregions of the Cauchy surface $\Sigma$.

In this paper we only consider the boundary geometry of the Minkowski spacetime or possibly of the form $M_I^{\partial} = I \times N_I$, where $I \subseteq \R$ represents time and $N_I$ some spacelike manifold. We assume that the near-boundary bulk coordinates match the boundary coordinates, meaning that when the radial coordinate $z_I$ is sent to zero, the remaining bulk coordinates $(t, x_j)$ become the boundary coordinates. Thus, we can think about each boundary Hilbert space $\H^{\partial}_I$ as having a global, time-independent Hamiltonian $\op{H}_I$ and a global, time-independent ground state $|0\>_I$. Positive and negative frequency modes in each boundary are globally defined and have the form $\varphi_I(t, x_j) \sim e^{\pm \I \omega t} Y_{\omega \ell}(x_j)$, where $\omega$ are suitable boundary frequencies and $Y_{\omega \ell}$ denote a set of suitable orthogonal space-like modes on $N_I$. This leads to the global split of each boundary value space, $\Bcx_I = \Bneg_I \oplus \Bpos_I$.

Finally, we can look for the subspaces $\Ineg_L \subseteq \Icx_L$ and $\Ineg_R \subseteq \Icx_R$ of the modes that correspond to the negative frequency modes $\Bneg_L$ and $\Bneg_R$ respectively. If $\Ineg_L$ and $\Ineg_R$ match when restricted to $\Icx_{wh}$, they can be glued together to $\Ineg$ according to \eqref{isoLR}. We obtain the specific quantization $\H$, where the boundary Hilbert spaces are equal to $\H_L^{\partial} = \H_L = Sym(\Ineg_L)$ and $\H_R^{\partial} = \H_R = Sym(\Ineg_R)$ and constitute tensor factors in $\H$. In particular the vacua of $\H_L$ and $\H_R$ are the vacua $| 0 \>_L \in \H_L^{\partial}$ and $| 0 \>_R \in \H_R^{\partial}$. 

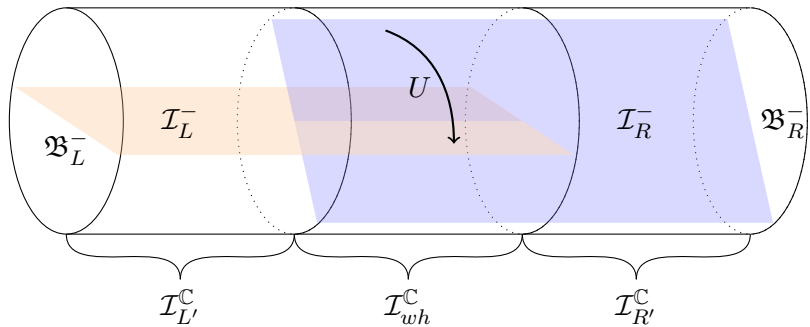
\begin{figure}[ht]
\begin{tikzpicture}[scale=1.5]
\draw (-3,0) ellipse [x radius=0.5, y radius=1.0];
\draw ( 3,0) ellipse [x radius=0.5, y radius=1.0];
\draw (-1,0) ellipse [x radius=0.5, y radius=1.0];
\draw ( 1,0) ellipse [x radius=0.5, y radius=1.0];
\draw[white, fill=white] (3,1.1) rectangle (2,-1.1);
\draw[white, fill=white] (1,1.1) rectangle (0,-1.1);
\draw[white, fill=white] (-1,1.1) rectangle (-2,-1.1);
\draw[dotted] ( 3,0) ellipse [x radius=0.5, y radius=1.0];
\draw[dotted] ( 1,0) ellipse [x radius=0.5, y radius=1.0];
\draw[dotted] (-1,0) ellipse [x radius=0.5, y radius=1.0];
\draw (-3,1) -- (3,1);
\draw (-3,-1) -- (3,-1);
\fill[colorR, opacity=0.2] (-1,0) -- (1,0) -- (1.45,-0.3) -- (-0.933,-0.3) -- cycle;
\fill[colorL, opacity=0.3] (-3.45,0.3) -- (0.55,0.3) -- (1.45,-0.3) -- (-2.55,-0.3) -- cycle;
\fill[colorR, opacity=0.3] (-1.2,0.9) -- (2.8,0.9) -- (3.2,-0.9) -- (-0.8,-0.9) -- (-0.933,-0.3) -- (1.45,-0.3) -- (1,0) -- (-1,0) -- cycle;
\draw[thick, ->] (-0.2,0.8) to [out=-20,in=90] (0.4,-0.2);
\node at (0.1,0.3) {$U$};
\draw (-3,-1) to [out=-90,in=90] (-2,-1.4);
\draw (-1,-1) to [out=-90,in=90] (-2,-1.4);
\draw (-1,-1) to [out=-90,in=90] ( 0,-1.4);
\draw ( 1,-1) to [out=-90,in=90] ( 0,-1.4);
\draw ( 3,-1) to [out=-90,in=90] ( 2,-1.4);
\draw ( 1,-1) to [out=-90,in=90] ( 2,-1.4);
\node[below] at (-2,-1.4) {$\Icx_{L'}$};
\node[below] at ( 0,-1.4) {$\Icx_{wh}$};
\node[below] at ( 2,-1.4) {$\Icx_{R'}$};
\node at (-2,0) {$\Ineg_{L}$};
\node at ( 2,0) {$\Ineg_{R}$};
\node[below] at (-3,0) {$\Bneg_L$};
\node[right] at (3,0) {$\Bneg_R$};
\end{tikzpicture}
\centering
\caption{Relation between various spaces of the initial data on a chosen Cauchy slice. The left and right boundary data $\Bneg_L$ and $\Bneg_R$ determines the subspaces $\Ineg_L$ and $\Ineg_R$ of the space of initial data $\Icx$ on $\Sigma$. These can be used to construct the Hilbert space $\H$ if the two subspaces agree in $\Icx_{wh}$. Otherwise, a suitable Bogoliubov transformation $U$ is required, denoted schematically by the arrow.\label{fig:bogo}}
\end{figure}

If, on the other hand, the two subspaces $\Ineg_L, \Ineg_R$ are different when restricted to $\Icx_{wh}$, a suitable Bogoliubov transformation is required to align the negative frequency modes in $\Icx_L$ with the negative frequency modes in $\Icx_R$. The matching becomes problematic for the wormhole modes contained in $\Icx_{wh}$ only, so we can split $\Ineg_L, \Ineg_R$ analogously to \eqref{split_ILR}, 
\begin{align} \label{split_match}
& \Ineg_L \cong \Ineg_{L'} \oplus \Ineg_{L wh}, && \Ineg_R \cong \Ineg_{R'} \oplus \Ineg_{R wh},
\end{align}
except that now $\Ineg_{L wh}$ and $\Ineg_{R wh}$ are different subspaces of $\Icx_{wh}$. See figure \ref{fig:bogo} for a graphic representation of the various spaces. Being right boundary-centric, we will declare $\Ineg_{wh} = \Ineg_{R wh}$ and identify the right Hilbert space $\H_R$ with $\H_R^{\partial}$. We can also treat $\Ineg_{L'}$ as negative frequency modes for the bulk quantization and thus identify $\H_{L'} = \H_{L'}^{\partial}$. For the remaining wormhole modes let $U$ be the Bogoliubov transformation between $\H_{wh}$ quantized with respect to $\Ineg_{R wh}$ and $\H_{wh}^{\partial}$ quantized with respect to $\Ineg_{Lwh}$. The isomorphism between $\H_L$ and $\H_L^{\partial}$ reads
\begin{align}
\bs{1} \otimes U \: : \: \H_L = \H_{L'} \otimes \H_{wh} \overset{\cong}{\longrightarrow} \H_L^{\partial} = \H_{L'}^{\partial} \otimes \H_{wh}^{\partial}.
\end{align}
Let us note that while formally the Bogoliubov transformation is unitary, in general $U$ can become anti-unitary. This can happen if the negative frequency modes in $\Ineg_{Rwh}$ turn out to be mostly positive frequency from the point of view of $\Ineg_{Lwh}$. It means that the time directions between the two boundaries mismatch and thus $U$ will now contain the time reversal operator making it anti-unitary.

\begin{figure}[ht]
\begin{subfigure}[t]{0.28\textwidth}
\begin{tikzpicture}[scale=1.8]
\draw[dotted, colorR, fill=colorR, opacity=0.3] (0,0) -- (1,1) -- (1,-1) -- cycle;
\draw[snake it] (-1,1) to [out=-10,in=-170] (1,1);
\draw[snake it] (-1,-1) to [out=10,in=170] (1,-1);
\draw[dotted] (-1, 1) -- (1,-1);
\draw[dotted] (-1, -1) -- (1,1);
\draw[thick] (-1,1) -- (-1,-1);
\draw[thick, colorR] ( 1,1) -- ( 1,-1);
\draw[thick, black] (-1, 0) -- (1, 0);
\draw[thick, colorR!50!black] (0,0) -- (1,0);
\draw (-1,0) to[out=-90,in=90] (0,-0.4);
\draw (1,0) to [out=-90,in=90] (0,-0.4);
\node[below] at (0,-0.4) {$\H$};
\node[above, colorR!50!black] at (0.5,0) {$\H_R$};
\node[left] at (-1,0) {$\Sigma$};
\end{tikzpicture}
\centering
\caption{The causal diamond and the corresponding boundary Hilbert space $\H_R$ as the subspace of the total Hilbert space for an eternal black hole.\label{fig:commentL}}
\end{subfigure}
\qquad
\begin{subfigure}[t]{0.28\textwidth}
\begin{tikzpicture}[scale=1.8]
\draw[dotted, colorR, fill=colorR, opacity=0.3] (0,0) -- (1,1) -- (1,-1) -- cycle;
\draw[thick, black] (1, -1.3) -- (1, 1.3);
\draw[thick, colorR] (1, -1) -- (1, 1);
\draw[thick, black] (-0.3, 0) -- (1, 0);
\draw[thick, colorR] (0,0) -- (1,0);
\draw (-0.3,-0.2) to [out=0,in=90] (0,-0.4);
\draw (1,0) to [out=-90,in=90] (0,-0.4);
\node[below] at (0,-0.4) {$\H$};
\node[above, colorR!50!black] at (0.5,0) {$\H_1$};
\node[right, colorR!50!black] at (1,0.3) {$U_1$};
\end{tikzpicture}
\centering
\caption{The causal diamond and the corresponding Hilbert space $\H_1$ as the subspace of the total Hilbert space for a subregion of a boundary.\label{fig:commentM}}
\end{subfigure}
\qquad
\begin{subfigure}[t]{0.28\textwidth}
\begin{tikzpicture}[scale=1.8]
\draw[dotted, colorR, fill=colorR, opacity=0.3] (0,0) -- (1,1) -- (1,-1) -- cycle;
\draw[dotted, colorL, fill=colorL, opacity=0.3] (0.5,0) -- (1, 0.5) -- (1,-0.5) -- cycle;
\draw[thick, black] (1, -1.3) -- (1, 1.3);
\draw[thick, colorR] (1, -1) -- (1, 1);
\draw[thick, colorL] (1, -0.5) -- (1, 0.5);
\draw[thick, black] (-0.3, 0) -- (1, 0);
\draw[thick, colorR] (0,0) -- (1,0);
\draw[thick, colorL] (0.5,0) -- (1,0);
\draw (0.5,0) to [out=90,in=-90] (0.75,0.1);
\draw (1,0) to [out=90,in=-90] (0.75,0.1);
\node[above, colorL!50!black] at (0.75,0.08) {$\H_2$};
\draw (0,0) to [out=-90,in=90] (0.5,-0.2);
\draw (1,0) to [out=-90,in=90] (0.5,-0.2);
\node[below, colorR!50!black] at (0.5,-0.18) {$\H_1$};
\node[right, colorL!50!black] at (0.98,0.25) {$U_2$};
\draw (1.2,1) to [out=0,in=180] (1.6, 0);
\draw (1.2,-1) to [out=0,in=180] (1.6, 0);
\node[right, colorR!50!black] at (1.6,0) {$U_1$};
\end{tikzpicture}
\centering
\caption{Two causal diamonds and the corresponding Hilbert spaces for two boundary subregions $U_2 \subseteq U_1$.\label{fig:commentR}}
\end{subfigure}
\centering
\caption{Various cases of the causal wedge reconstruction from the point of view of the bulk.\label{fig:wedges1}}
\end{figure}
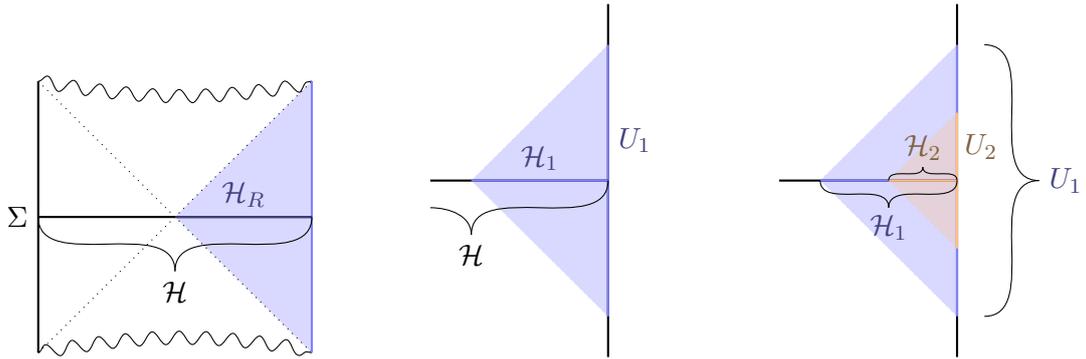

\paragraph{Comments and remarks.} The boundary limit \eqref{pi} can be applied to the quantized bulk operators as well. This is known as the BDHM dictionary, \cite{Banks:1998dd}, also \cite{Balasubramanian:1998de,Bena:1999jv}, the early version of real-time holography. The procedure takes the quantized bulk operator $\op{\Phi}$ and maps it to a pre-boundary operator $\op{\Phi}_I$ by taking the limit $\op{\Phi}_I = \lim_{z \rightarrow 0} z^{-\Delta_I} \op{\Phi}$, for $I = L,R$. We call it the \emph{pre-boundary operator} since $\op{\Phi}_I$ still acts on the total Hilbert space $\H$ rather than any `smaller' Hilbert space that we would associate with the boundary. 

In principle one can retrieve the actual boundary Hilbert space $\H^{\partial}_I$ by following the prescription of the real-time holography \cite{Skenderis:2008dg,Skenderis:2008dh}. In \cite{Botta-Cantcheff:2015sav,Christodoulou:2016nej,Marolf:2017kvq} the explicit correspondence between Lorentzian bulk modes and boundary excited states was provided. The prescription was also established in the presence of interactions in \cite{Botta-Cantcheff:2017qir}. As far as the AdS black holes are concerned, the prescription was worked out in \cite{Botta-Cantcheff:2018brv,Botta-Cantcheff:2019apr}.

What we described in the previous paragraph is a certain version of the converse problem known as the bulk reconstruction or the causal wedge reconstruction, see figure \ref{fig:commentL}. Given the boundary operators one can reconstruct the bulk operators by means of the HKLL reconstruction, \cite{Hamilton:2005ju,Hamilton:2006az,Heemskerk:2012np}. One smears the boundary operators with suitable smearing functions, which impose equations of motion and locality in the bulk. In the context of the scalar field on fixed background a detailed discussion can be found in \cite{Kim:2016ipt,Terashima:2020uqu,Bahiru:2022ukn}.

Much effort has been put into the analysis of how much bulk geometry can be actually recovered from the point of view of the boundary. In particular, if $U_1$ is a subset of, say, right boundary, as presented in figure \ref{fig:commentM}, how much of the bulk geometry can be reconstructed? Today this problem is known as the entanglement wedge reconstruction after \cite{Dong:2016eik}. In the context of AdS and BTZ backgrounds the problem was thoroughly analyzed in a number of papers, \cite{Czech:2012be,Parikh:2012kg,Leichenauer:2013kaa,Botta-Cantcheff:2018brv}.

Consider two regions $U_2 \subseteq U_1$ of the boundary as presented in figure \ref{fig:commentR}. By considering the causal developments of $U_1$ and $U_2$ into the bulk one can find the set of initial data $\Icx_2 \subseteq \Icx_1$. This means that the Hilbert space $\H_2$ describing the excitations in $U_2$ is a proper tensor factor of $\H_1$, the Hilbert space associated with $U_1$. On the other hand the Reeh-Schlieder theorem, \cite{Morrison:2014jha,Banerjee:2016mhh,Chowdhury:2020hse}, implies that the states obtained by the action on the vacuum $|0\>_1 \in \H_1$ of the operators from $\H_2$, supported on the smaller, orange portion of the Cauchy surface span a dense subspace of $\H_1$. There is no contradiction between these two statements, since the two Hilbert spaces do not share the vacuum states. For the operators of $\H_2$ to span a dense subspace of $\H_1$, they must act on a sufficiently regular state, whose precise definition is given in \cite{Morrison:2014jha}. From the point of view of $\H_1$ the vacuum state $|0\>_2$ is an irregular state and $| 0 \>_1$ certainly does not split into a simple tensor product involving $|0\>_2$.

More recently, in \cite{Leutheusser:2021frk,Leutheusser:2021qhd}, the match between the boundary Hilbert spaces and the subspaces of the bulk system was analyzed in case of the BTZ black hole. The authors carried out the analysis within the framework of the axiomatic field theory.

\subsection{Summary}

In section \ref{sec:factor} we argued a simple statement regarding the independence of the initial data on a selected Cauchy slice $\Sigma$. We considered the subregions $\Sigma_L$ and $\Sigma_R$ of the Cauchy slice in figure \ref{fig:splits} on page \pageref{fig:splits}. Since both regions contain the intersection $\Sigma_{wh} = \Sigma_L \cap \Sigma_R$, only those initial conditions in $\Icx_L \oplus \Icx_R$ which agree on $\Sigma_{wh}$ give rise to an initial condition on the full Cauchy surface $\Sigma$. On the level of the Hilbert spaces this leads to the non-factorization property \eqref{non-factor},
\begin{align} \label{sum_non-factor}
\H_L \otimes \H_R \cong \H \otimes \H_{wh}.
\end{align}
The bulk modes whose initial data is supported on $\Sigma_{wh}$ are the wormhole modes, which propagate between the two boundaries.

The same logic applies to the wormhole presented in figure \ref{fig:splitsWh} on page \pageref{fig:splitsWh}. In this case one has to find a suitable foliation and the associated Hamiltonian in order to relate initial data on various surfaces. This means that the isomorphisms such as the non-factorization property \eqref{sum_non-factor} holds up to the Bogoliubov transformation induced by the unitary evolution between the surfaces.

The Hilbert spaces $\H_L$ and $\H_R$ in \eqref{sum_non-factor} are defined from the point of view of the bulk. In the previous section we used real-time holography to argue that they can be identified with the actual dual boundary Hilbert spaces $\H_L^{\partial}$ and $\H_R^{\partial}$ describing the scalar sector of the theory. We argued that there exists the quantization of the bulk theory in such a way that $\H_R = \H_R^{\partial}$, while $\H_L \cong \H_L^{\partial}$ up to a Bogoliubov transformation, which may also include discreet symmetries such as time reversal. 

Putting these result together we conclude that the Hilbert space $\H$ factorizes into the tensor product of the boundary Hilbert spaces if and only if there are no wormhole modes propagating between the two asymptotic boundaries.

\subsection{\texorpdfstring{Example: AdS$_3$}{Example: AdS3}} \label{sec:AdS3}

As a simple example of the real-time holography let us consider a free, real scalar field $\Phi_{AdS}$ of mass $m$ in the empty AdS in the global coordinates. This is the textbook material and the detailed analysis can be found in the lecture notes \cite{KaplansNotes}. We will concentrate here on $3$-dimensional AdS, whose metric is given by,
\begin{align} \label{global_AdS3}
\D s^2 = \frac{L^2}{\cos^2 \theta} \left( - \D \tau^2 + \D \theta^2 + \sin^2 \theta \D \varphi^2 \right),
\end{align}
where $\varphi \in (0,2\pi)$ parameterizes a circle. The field satisfies the Klein-Gordon equation $(-\Box + m^2) \Phi_{AdS} = 0$ and we look for normalizable modes only and we parameterize the mass $L^2 m^2 = \Delta (\Delta - 2)$ as usual.

The solution can be decomposed into modes,
\begin{align}
\Phi_{AdS}(\tau, \theta, \Omega) = \sum_{k=0}^{\infty} \sum_{\ell = - \infty}^{\infty} \left( \phi_{k \ell} \a_{k \ell} + \phi^{\ast}_{k \ell} \a^{\ast}_{k \ell} \right),
\end{align}
where
\begin{align} \label{PhiAdSd_text}
\phi_{k \ell}(\tau, \theta, \varphi) =  c_{k \ell} e^{-\I \omega_{k \ell} \tau + \I \ell \varphi} \cos^\Delta \theta \sin^{|\ell|} \theta \: P_k^{(|\ell|, \Delta - 1)}(\cos(2 \theta)).
\end{align}
The solution is written in terms of Jacobi polynomials $P^{(\alpha, \beta)}_k$ and the frequencies are quantized as
\begin{align} \label{omegaQ}
\omega_{k \ell} = \Delta + |\ell| + 2 k, \quad k = 0,1,2,\ldots
\end{align}
The modes \eqref{PhiAdSd_text} represent standing waves in AdS and they are regular at $\theta = 0$. The normalization coefficients $c_{k \ell}$ are chosen in such a way that the modes are orthonormal in the Klein-Gordon norm, \textit{i.e.}, $(\phi_{k \ell}, \phi_{k' \ell'}) = \delta_{k k'} \delta_{\ell \ell'}$. More details on the Klein-Gordon equation and properties of the Jacobi polynomials are listed in Appendices \ref{app:gAdS} and \ref{app:Jacobi}.

Next, the coefficients in \eqref{PhiAdSd_text} are promoted to creation-annihilation operators and they obey the canonical commutation relations,
\begin{align} \label{AdS_ccr}
\left[ \op{\a}_{k \ell}, \op{\a}^{\dagger}_{k' \ell'} \right] = \delta_{k k'} \delta_{\ell \ell'}.
\end{align}
In this way we obtain the standard quantization of the free field with $\phi_{k \ell}$ designated as negative frequency modes. The vacuum state $| \Omega \>$ is defined by the condition $\op{\a}_{k \ell} | \Omega \> = 0$ for all $k, \ell$. The space of 1-particle states, $\H_{AdS}^{(1)}$, is defined as a span of all states obtained from the vacuum by the action of a single creation operator, $|1_{k \ell} \> = \op{\a}^{\dagger}_{k \ell} | \Omega \>$. The Klein-Gordon product is positive definite on $\H_{AdS}^{(1)}$ and thus, after completion, turns it into a Hilbert space. The full Hilbert space $\H_{AdS}$ is the symmetric Fock space built on top of $\H_{AdS}^{(1)}$, \textit{i.e.}, $\H_{AdS} = Sym(\H_{AdS}^{(1)})$, where $Sym$ denotes the universal symmetric algebra.

\paragraph{Going to the boundary.} We can now take the boundary limits in order to obtain the boundary data. To do it, one must pick up a defining function $F$, so that the boundary limit is well-defined. We choose $F = \cos \theta / L$ and the negative frequency boundary modes are defined as
\begin{align} \label{AdS_phi}
\varphi_{k \ell}(\tau, \varphi) = \lim_{\theta \rightarrow \pi/2} F^{-\Delta}(\theta) \phi_{k \ell}(\tau, \theta, \varphi)
\end{align}
and read
\begin{align} \label{gAdS_mode}
& \varphi_{k \ell}(\tau, \varphi) = \tilde{c}_{k \ell} e^{-\I \omega_{k \ell} \tau + \I \ell \varphi}, && \tilde{c}_{k \ell} = \frac{(-1)^k L^{\Delta - \frac{1}{2}}}{\Gamma(\Delta)} \sqrt{\frac{\Gamma(\Delta + k) \Gamma(\Delta + k + |\ell|)}{k! (k + | \ell |)!}}
\end{align}
which represent the standard waves of frequencies $\omega_{k \ell}$ on the cylinder with the non-standard normalization. When applied to the quantized bulk operator $\op{\Phi}_{AdS}$ we obtain the boundary operator,
\begin{align} \label{Obnd}
\O(\tau, \varphi) & = \lim_{\theta \rightarrow \frac{\pi}{2}} \left[ F^{-\Delta}(\theta) \op{\Phi}(\tau, \theta, \varphi) \right] \nn\\
& = \sum_{k=0}^{\infty} \sum_{\ell = -\infty}^{\infty} \left( \op{\a}_{k \ell} \varphi_{k \ell} + \op{\a}^{\dagger}_{k \ell} \varphi^{\ast}_{k \ell} \right).
\end{align}
Note that while the boundary operator depends on the boundary coordinates $\tau$ and $\Omega$ only, it acts on $\H_{AdS}$ and thus  $\H_{AdS}$ \emph{is} the boundary Hilbert space. This is the essence of holography for the empty AdS spacetime. In particular, the bulk creation-annihilation operators $\op{\a}^{\dagger}_{k \ell}, \op{\a}_{k \ell}$ are now interpreted as the boundary operators and the vacuum state $| \Omega \>$ is the boundary vacuum state.

\paragraph{Identification of the quantization scheme.} The advantage of the quantization scheme in global AdS coordinates is the fact that we can identify it as the radial quantization of the dual CFT. The bulk Hamiltonian is $H = \sum_{k \ell} \omega_{k \ell} \a_{k \ell}^{\dagger} \a_{k \ell}$, which corresponds to the dilatation operator on the boundary. The boundary cylinder can be Wick rotated, $\tau = - \I t$, and then mapped to the plane by $r = e^t$. Using \eqref{Obnd}, one defines a more familiar Euclidean operator
\begin{equation}
\O^{Eu}(t, \varphi) = e^{-\Delta t} \O(\tau = -\I t, \varphi),
\end{equation}
whose mode decomposition reads
\begin{align} \label{OEu}
\O^{Eu}(r,\varphi) = \sum_{\substack{k=0\\ \ell=-\infty}}^{\infty} \left( \op{\a}_{k \ell} \tilde{c}_{k \ell} r^{-2 \Delta - |\ell| - 2k} e^{\I \ell \varphi} + \op{\a}^{\dagger}_{k \ell} \tilde{c}_{k \ell} r^{|\ell| + 2k} e^{- \I \ell \varphi} \right).
\end{align}
Thus, we have obtained the Euclidean mode decomposition of a conformal primary operator of dimension $\Delta$ radially quantized. The operator insertion at $r = 0$ creates the state usually denoted by $| \Delta \>$,
\begin{equation}
| \Delta \> = \O^{Eu}(0) | \Omega \> = \frac{a_{00}^{\dagger}}{L^{\Delta - \frac{1}{2}}}| \Omega \>.
\end{equation}
Using the decomposition \eqref{OEu} it is now easy to check that the Euclidean 2-point function correctly reads
\begin{align}
\< \Omega | \O^{Eu}(z, \bar{z}) \O^{Eu}(0) | \Omega \> = \frac{1}{L^{2\Delta-1}} \frac{1}{|z|^{2 \Delta}}.
\end{align}
Further analysis for the higher order modes can be found in the lecture notes \cite{KaplansNotes}.

\paragraph{Rindler-AdS.} Let us now look at the 3-dimensional AdS spacetime in Rindler coordinates. The transformation between global and Rindler coordinates can be found in \cite{Parikh:2012kg} and the AdS metric takes form
\begin{equation} \label{Rindler_AdS}
\D s^2 = - \rho^2 \D t^2 + \frac{L^2 \D \rho^2}{L^2 + \rho^2} + ( L^2 + \rho^2 ) \D \varphi^2,
\end{equation}
with $\rho$ ranging from $0$ to $\infty$. The scalar field can be decomposed as
\begin{align}
\Phi_{RAdS}(t, \rho, \varphi) = \int_0^{\infty} \frac{\D \omega}{2 \pi} \sum_{n=-\infty}^{\infty} \left( \phi_{\omega n} \a_{\omega n} + \phi^{\ast}_{\omega n} \a^{\dagger}_{\omega n} \right).
\end{align}
We use the same symbols $\phi$ and $\a$ for the modes and the creation-annihilation operators as in the global AdS case as it should not cause any confusion. The negative frequency modes are then
\begin{align} \label{Rind_phi}
& \phi^{RAdS}_{\omega n}(t, \rho, \varphi) = c^{RAdS}_{\omega n} e^{-\I \omega t + \I n \varphi} R_{\omega n}(\rho),
\end{align}
where the normalizable radial functions $R_{\omega n}$ can be expressed in terms of the hypergeometric function, see appendix \ref{app:RAdS} for details. The normalization is such that near the boundary $\rho \rightarrow \infty$ we have
\begin{align}
R_{\omega n} = \left( \frac{\rho}{L} \right)^{-\Delta} \left[ 1 + O(\rho^{-2}) \right].
\end{align}
The normalization constant $c^{RAdS}_{\omega n}$ is chosen in such a way that the canonical commutation relations hold,
\begin{align}
\left[ \op{\a}_{\omega n}, \op{\a}_{\omega' n'}^{\dagger} \right] = ( \phi^{RAdS}_{\omega n}, \phi^{RAdS}_{\omega' n'} ) = 2 \pi  \delta(\omega - \omega') \delta_{nn'}.
\end{align}
The Hilbert space $\H_{RAdS}$ of the dual QFT is now built on top of the vacuum state $| 0 \>$ satisfying $\op{\a}_{\omega n} | 0 \> = 0$ for all $\omega > 0$ and $n \in \Z$. We denote this vacuum by $| 0 \>$ to distinguish it from the global vacuum $| \Omega \>$.

Finally, let us take the boundary limit and define the boundary operator $\O$ as
\begin{align}
\O(t, \varphi) & = \lim_{\rho \rightarrow \infty} \left( \frac{\rho}{L} \right)^{-\Delta} \phi^{RAdS}_{\omega n}(t, \rho, \varphi) \nn\\
& = \int_0^{\infty} \frac{\D \omega}{2 \pi} \sum_{n=-\infty}^{\infty} \left( \varphi^{RAdS}_{\omega n} \op{\a}_{\omega n} + \varphi^{RAdS \ast}_{\omega n} \op{\a}^{\dagger}_{\omega n} \right),
\end{align}
where
\begin{align} \label{varphiRAdS}
\varphi^{RAdS}_{\omega n} = c_{\omega n}^{RAdS} e^{-\I \omega t + \I n \varphi}.	
\end{align}

A relation between the Hilbert spaces $\H_{AdS}$ and $\H_{RAdS}$ obtained from the quantization in global and Schwarzschild coordinates was extensively studied in several papers, \textit{e.g.},  \cite{Czech:2012be,Parikh:2012kg,Leichenauer:2013kaa,Botta-Cantcheff:2018brv}. The difficulty in establishing the equivalence comes from the fact that the Schwarzschild modes \eqref{Rind_phi} have the distributional nature from the point of view of the global AdS. The careful analysis can be found in \cite{Morrison:2014jha}.

\section{The geon} \label{sec:geon}

\begin{figure}[ht]
\centering
\begin{subfigure}[t]{0.4\textwidth}
\begin{tikzpicture}[scale=2.0]
\draw (-1.2,-1.2) -- ( 1.2,1.2);
\draw ( 1.2,-1.2) -- (-1.2,1.2);
\draw[->] (-1.2,-1.2) -- ( 1.2,1.2);
\draw[->] ( 1.2,-1.2) -- (-1.2,1.2);
\node[above] at (1.2,1.2) {$U$};
\node[above] at (-1.2,1.2) {$V$};
\draw[->] (-1.2,0) -- (1.2,0);
\draw[->] (0,-1.2) -- (0,1.2);
\node[above] at (1.2,0) {$X$};
\node[right] at (0,1.2) {$T$};
\draw[snake it] (-1.1,1.2) to [out=-30,in=-150] (1.1,1.2);
\draw[snake it] (-1.1,-1.2) to [out=30,in=150] (1.1,-1.2);
\draw[dotted, colorL, fill=colorL, opacity=0.3] (-1.2,-1.1) to [out=60,in=-60] (-1.2,1.1) -- (-1.2,1.2) -- (0,0) -- (-1.2,-1.2) -- cycle;
\draw[dotted, colorR, fill=colorR, opacity=0.3] (1.2,1.2) -- (1.2,1.1) to [out=-120,in=120] (1.2,-1.1) -- (1.2,-1.2) -- (0,0) -- cycle;
\draw[thick, colorL] (-1.2,1.1) to [out=-60,in=60] (-1.2,-1.1);
\draw[thick, colorR] (1.2,1.1) to [out=-120,in=120] (1.2,-1.1);
\draw[thick, colorL] (-0.88,0) -- (0,0);
\draw[thick, colorR] (0.88,0) -- (0,0);
\draw[colorR!50!black, ->] (0.9,-0.6) to [out=120,in=-90] ( 0.7,0) to[out=90,in=-120] (0.9,0.6);
\draw[colorL!50!black, ->] (-0.9,-0.6) to [out=60,in=-90] (-0.7,0) to[out=90,in=-60] (-0.9,0.6);
\node[above] at (0.45,0) {$\Sigma_R$};
\node[above] at (-0.45,0) {$\Sigma_L$};
\end{tikzpicture}
\centering
\caption{BTZ black hole in Kruskal coordinates. Blue and orange regions are the right and left wedges bounded by two asymptotic boundaries. \label{fig:btz}}
\end{subfigure}
\qquad\qquad
\begin{subfigure}[t]{0.4\textwidth}
\begin{tikzpicture}[scale=2.0]
\draw (-1.2,-1.2) -- ( 1.2,1.2);
\draw ( 1.2,-1.2) -- (-1.2,1.2);
\draw[->] (-1.2,0) -- (1.2,0);
\draw[->] (0,-1.2) -- (0,1.2);
\node[above] at (1.2,0) {$X$};
\node[right] at (0,1.2) {$T$};
\draw[dotted, colorL, fill=colorL, opacity=0.3] (-1.2,-1.1) to [out=60,in=-60] (-1.2,1.1) -- (-1.2,1.2) -- (0,0) -- (-1.2,-1.2) -- cycle;
\draw[dotted, colorR, fill=colorR, opacity=0.3] (1.2,1.2) -- (1.2,1.1) to [out=-120,in=120] (1.2,-1.1) -- (1.2,-1.2) -- (0,0) -- cycle;
\draw[thick, colorL] (-1.2,1.1) to [out=-60,in=60] (-1.2,-1.1);
\draw[thick, colorR] (1.2,1.1) to [out=-120,in=120] (1.2,-1.1);
\draw[thick, colorL] (-0.88,0) -- (0,0);
\draw[thick, colorR] (0.88,0) -- (0,0);
\draw[green, fill=green] (0.66,0.22) circle (0.02);
\draw[green, fill=green] (-0.66,-0.22) circle (0.02);
\draw[green] (-0.6,-0.2) -- (0.6,0.2);
\draw[<->] (-0.6,-0.15) to [out=45, in=160] (0.6,0.25);
\draw[white, fill=white] (-0.32,0.58) rectangle (0.32,0.37);
\node[above] at (0,0.3) {identify};
\draw[colorR!50!black, ->] (0.9,-0.6) to [out=120,in=-90] ( 0.7,0) to[out=90,in=-120] (0.9,0.6);
\draw[colorL!50!black, <-] (-0.9,-0.6) to [out=60,in=-90] (-0.7,0) to[out=90,in=-60] (-0.9,0.6);
\end{tikzpicture}
\centering
\caption{Geon is the BTZ spacetime with the antipodal identification between the two wedges as shown by the arrow.\label{fig:geon}}
\end{subfigure}
\caption{The BTZ black hole and the geon.\label{fig:btz_geon}}
\end{figure}
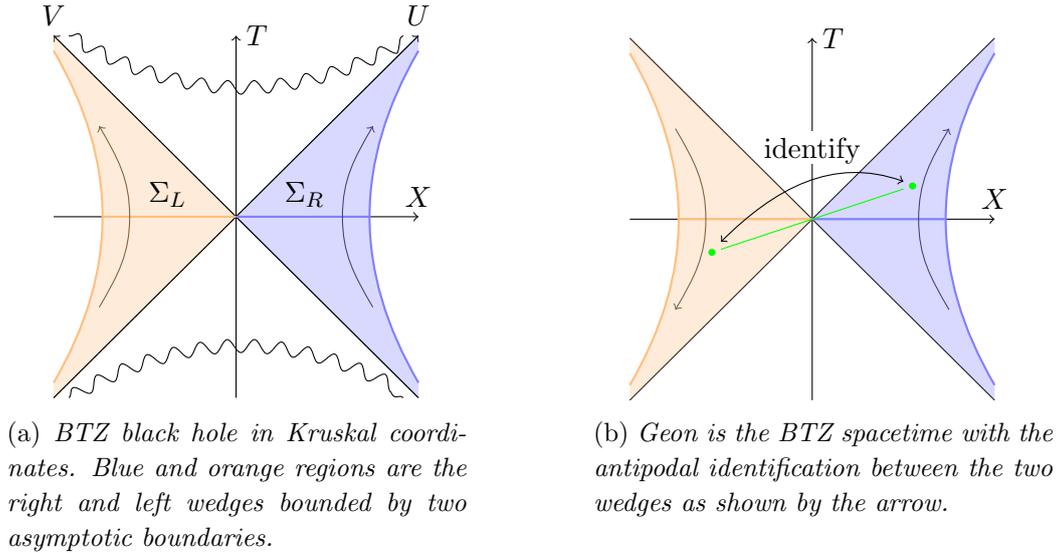

By the term \emph{geon} we will refer to the BTZ black hole with the antipodal $\Z_2$ identification between the left and right wedges. The antipodal map, $\theta$, is the isometry of the BTZ spacetime and in the Kruskal coordinates $(T,X,\varphi)$ is given by
\begin{align}
\theta(T, X, \varphi) = (-T, -X, \varphi + \pi),
\end{align}
The antipodal map swaps the two wedges and inverts the direction of time, as shown in figure \ref{fig:btz_geon}. The geon is the quotient spacetime, $BTZ /\!\! \sim$, where two points $x_L \sim x_R$ are identified if $x_L = \theta x_R$. As the antipodal map maps the boundaries to each other, the resulting spacetime possesses a single asymptotic boundary. 

The motivation behind such a definition is that the geon is smooth and can serve as a toy model of a quantum black hole, where physics beyond the horizon is determined in terms of the physics outside of the black hole. Such a model was extensively studied in \cite{Louko:1998hc,tHooft:2016qoo,tHooft:2016rrl,Betzios:2016yaq,Betzios:2020xuj}.

In this paper we are interested in the structure of the Hilbert space of the geon and its holography dual. The analysis of the Hilbert space was first carried out in \cite{Sanchez:1986qn}, where it was pointed out that the standard Fock quantization fails due to zero-norm states. In the context of holography the analysis was carried in \cite{Louko:1998hc,Guica:2014dfa}. In both papers the authors assume that the Hilbert space is that of the BTZ black hole and in particular it splits into the tensor product of the boundary Hilbert spaces. As we will see this is not the case and we will discuss the relation between this paper and the previous results. To see it clearly, we will treat the geon as a wormhole.

\subsection{BTZ black hole} \label{sec:BTZ}

In this section we want review basic facts about the scalar field in the BTZ black hole. The BTZ black hole is an example of the 2-sided asymptotically AdS spacetime, with two boundary components, left and right, which we label by $I = L, R$. Thus, given the full bulk Hilbert space $\H$, we must be able to identify two boundary Hilbert spaces, $\H_L$ and $\H_R$. Since our analysis here does not involve quantum gravity, the Hilbert space $\H$ of the system does split into the tensor product: $\H \cong \H_L \otimes \H_R$.

\paragraph{Geometry and the modes.} The BTZ metric in a single Schwarzschild wedge reads
\begin{equation} \label{BTZ}
\D s^2 = - (\rho^2 - \p^2) \D t^2 + \frac{L^2 \D \rho^2}{\rho^2 - \p^2} + \rho^2 \D \j^2, \qquad \rho > \p,
\end{equation}
where $\p$ is the Schwarzschild radius. The solution to the Klein-Gordon equation in Schwarzschild coordinates takes form
\begin{align} \label{BTZ_phi}
\phi_{\omega n}(t, \rho, \varphi) = c^{BTZ}_{\omega n} e^{-\I \omega t + \I n \varphi} R_{\omega n}(\rho),
\end{align}
where the explicit form of the radial part $R_{\omega n}$ of the wave function is given in \eqref{BTZmodeR}. These are normalizable modes satisfying
\begin{align}
R_{\omega n} = \left( \frac{\rho}{\p} \right)^{-\Delta} \left[ 1 + O(\rho^{-2}) \right]
\end{align}
near the boundary. The constants $c^{BTZ}_{\omega n}$ are then chosen in such a way that the modes $\phi_{\omega n}$ are properly normalized with respect to the Klein-Gordon product and read
\begin{align} \label{cBTZ}
c_{\omega n}^{BTZ} & = \frac{1}{\sqrt{4 \pi \omega \p} \, \Gamma(\Delta)} \left| \frac{\Gamma \left( \frac{\Delta}{2} + \hat{n} + \hat{\omega} \right) \Gamma \left( \frac{\Delta}{2} - \hat{n} + \hat{\omega} \right)}{\Gamma(2 \hat{\omega})} \right|.
\end{align}

Since the eternal black hole has two wedges, left and right, we define two sets of modes, $\phi^L_{\omega n}$ and $\phi^R_{\omega n}$, each supported in their respective wedges and functionally equal to $\phi_{\omega n}$. The mode decomposition of the field operator reads,
\begin{align} \label{Phi_BTZ}
\op{\Phi}_{BTZ} & = \int_0^{\infty} \frac{\D \omega}{2 \pi} \sum_{n=-\infty}^{\infty} \left( \phi_{\omega n}^{L} \op{\a}^L_{\omega n} + \phi_{\omega n}^R \op{\a}^R_{\omega n} + \phi_{\omega n}^{L \ast} \op{\a}^{L \dagger}_{\omega n} + \phi_{\omega n}^{R \ast} \op{\a}^{R \dagger}_{\omega n} \right).
\end{align}
The creation-annihilation operators $\op{\a}^{L,R \dagger}_{\omega n}, \op{\a}^{L,R}_{\omega n}$ satisfy the canonical commutation relations matching the Klein-Gordon scalar product as in \eqref{ccr_smeared}, 
\begin{align} \label{ccr_BTZ}
\left[ \op{\a}^I_{\omega n}, \op{\a}^{I' \dagger}_{\omega' n'} \right] = (\phi_{\omega n}^I, \phi_{\omega' n'}^{I'}) = 2 \pi \delta(\omega - \omega') \delta^{I I'} \delta_{n n'}.
\end{align}
The two sets of creation-annihilation operators $\op{\a}^{L \dagger}_{\omega n}, \op{\a}^{L}_{\omega n}$ and $\op{\a}^{R \dagger}_{\omega n}, \op{\a}^{R}_{\omega n}$ span two Hilbert spaces, $\H_0^L$ and $\H_0^R$ respectively. The Schwarzschild vacuum state $| 0 \>$ is defined by $\op{\a}^{I}_{\omega n} | 0 \> = 0$ for all $I, \omega, n$ and we can think of the space of 1-particle states $\Hone_{0}$ as spanned by $\op{\a}^{I \dagger}_{\omega n} | 0 \>$. The full Fock space is $\H_0 = Sym(\Hone_{0})$.  The split of the Cauchy surface into $\Sigma = \Sigma_L \cup \Sigma_R$ means that the total Hilbert space $\H_0$ splits as $\H_0 \cong \H_0^L \otimes \H_0^R$.

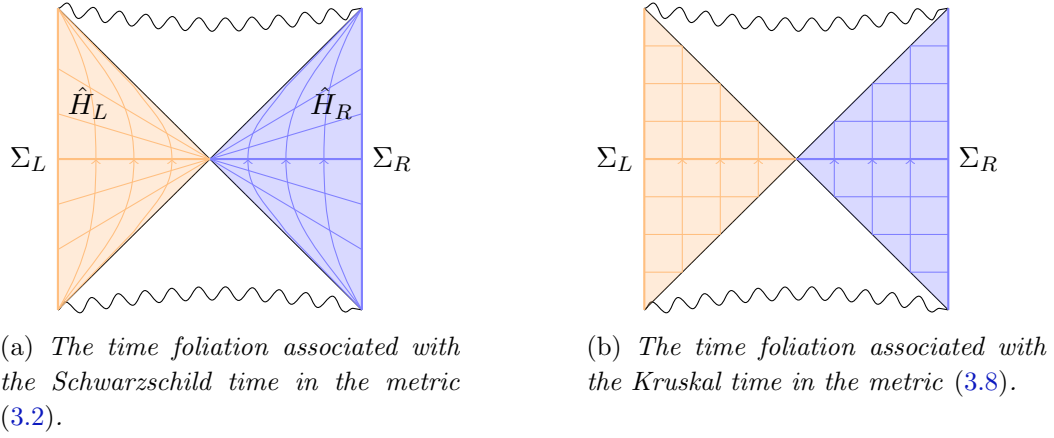
\begin{figure}[ht]
\begin{subfigure}[t]{0.4\textwidth}
\begin{tikzpicture}[scale=2.0]
\draw (-1,-1) -- ( 1,1);
\draw ( 1,-1) -- (-1,1);
\draw[snake it] (-1,1) to [out=-10,in=-170] (1,1);
\draw[snake it] (-1,-1) to [out=10,in=170] (1,-1);
\draw[dotted, colorL, fill=colorL, opacity=0.3] (-1,-1) -- (-1,1) -- (0,0) -- cycle;
\draw[dotted, colorR, fill=colorR, opacity=0.3] (1,1) -- (1,-1) -- (0,0) -- cycle;
\draw[thick, colorL] (-1,1) -- (-1,-1);
\draw[thick, colorR] ( 1,1) -- ( 1,-1);
\draw[thick, colorL] (-1,0) -- (0,0);
\draw[thick, colorR] (1,0) -- (0,0);
\draw[colorR, ->] (1,-1) to [out=125,in=-90] (0.5,0);
\draw[colorR] (0.5,0) to [out=90,in=-125] (1,1);
\draw[colorR, ->] (1,-1) to (0.5,-0.48) to [out=125,in=-90] (0.25,0);
\draw[colorR] (0.25,0) to [out=90,in=-125] (0.5,0.48) to (1,1);
\draw[colorR, ->] (1,-1) to [out=120,in=-90] (0.75,0);
\draw[colorR] (0.75,0) to [out=90,in=-120] (1,1);
\draw[colorL] (-1,1) to [out=-55,in=90] (-0.5,0);
\draw[colorL, <-] (-0.5,0) to [out=-90,in=55] (-1,-1);
\draw[colorL] (-1,1) to (-0.5,0.48) to [out=-55,in=90] (-0.25,0);
\draw[colorL, <-] (-0.25,0) to [out=-90,in=55] (-0.5,-0.48) to (-1,-1);
\draw[colorL] (-1,1) to [out=-60,in=90] (-0.75,0);
\draw[colorL, <-] (-0.75,0) to [out=-90,in=60] (-1,-1);
\draw[colorR] (0,0) -- (1,0.3);
\draw[colorR] (0,0) -- (1,0.6);
\draw[colorR] (0,0) -- (1,-0.3);
\draw[colorR] (0,0) -- (1,-0.6);
\draw[colorL] (0,0) -- (-1,0.3);
\draw[colorL] (0,0) -- (-1,0.6);
\draw[colorL] (0,0) -- (-1,-0.3);
\draw[colorL] (0,0) -- (-1,-0.6);
\node[above] at (0.8,0.2) {$\op{H}_R$};
\node[above] at (-0.8,0.2) {$\op{H}_L$};
\node[right] at (1,0) {$\Sigma_R$};
\node[left] at (-1,0) {$\Sigma_L$};
\end{tikzpicture}
\caption{The time foliation associated with the Schwarzschild time in the metric \eqref{BTZ}.\label{fig:btz_sch}}
\end{subfigure}
\qquad\qquad
\begin{subfigure}[t]{0.4\textwidth}
\begin{tikzpicture}[scale=2.0]
\draw (-1,-1) -- ( 1,1);
\draw ( 1,-1) -- (-1,1);
\draw[snake it] (-1,1) to [out=-10,in=-170] (1,1);
\draw[snake it] (-1,-1) to [out=10,in=170] (1,-1);
\draw[dotted, colorL, fill=colorL, opacity=0.3] (-1,-1) -- (-1,1) -- (0,0) -- cycle;
\draw[dotted, colorR, fill=colorR, opacity=0.3] (1,1) -- (1,-1) -- (0,0) -- cycle;
\draw[thick, colorL] (-1,1) -- (-1,-1);
\draw[thick, colorR] ( 1,1) -- ( 1,-1);
\draw[thick, colorL] (-1,0) -- (0,0);
\draw[thick, colorR] (1,0) -- (0,0);
\draw[colorR] (0.25,0.25) -- (1,0.25);
\draw[colorR] (0.5,0.5) -- (1,0.5);
\draw[colorR] (0.75,0.75) -- (1,0.75);
\draw[colorR] (0.25,-0.25) -- (1,-0.25);
\draw[colorR] (0.5,-0.5) -- (1,-0.5);
\draw[colorR] (0.75,-0.75) -- (1,-0.75);
\draw[colorL] (-0.25,0.25) -- (-1,0.25);
\draw[colorL] (-0.5,0.5) -- (-1,0.5);
\draw[colorL] (-0.75,0.75) -- (-1,0.75);
\draw[colorL] (-0.25,-0.25) -- (-1,-0.25);
\draw[colorL] (-0.5,-0.5) -- (-1,-0.5);
\draw[colorL] (-0.75,-0.75) -- (-1,-0.75);
\draw[colorR, ->] (0.25,-0.25) -- (0.25,0);
\draw[colorR, ->] (0.5,-0.5) -- (0.5,0);
\draw[colorR, ->] (0.75,-0.75) -- (0.75,0);
\draw[colorR] (0.25,0.25) -- (0.25,0);
\draw[colorR] (0.5,0.5) -- (0.5,0);
\draw[colorR] (0.75,0.75) -- (0.75,0);
\draw[colorL] (-0.25,0.25) -- (-0.25,0);
\draw[colorL] (-0.5,0.5) -- (-0.5,0);
\draw[colorL] (-0.75,0.75) -- (-0.75,0);
\draw[colorL, ->] (-0.25,-0.25) -- (-0.25,0);
\draw[colorL, ->] (-0.5,-0.5) -- (-0.5,0);
\draw[colorL, ->] (-0.75,-0.75) -- (-0.75,0);
\node[right] at (1,0) {$\Sigma_R$};
\node[left] at (-1,0) {$\Sigma_L$};
\end{tikzpicture}
\caption{The time foliation associated with the Kruskal time in the metric \eqref{BTZ_Kruskal}.\label{fig:btz_kru}}
\end{subfigure}
\centering
\caption{Two foliations of the BTZ black hole.\label{fig:bru_fol}}
\end{figure}

The left and right wedges can be stitched together to form the maximally extended spacetime. In Kruskal coordinates $(T, X, \varphi)$, which cover the entirety of the spacetime, the metric takes form
\begin{equation} \label{BTZ_Kruskal}
\D s^2 = \frac{4 L^2}{(1 + T^2 - X^2)^2} \left[ - \D T^2 + \D X^2 + \frac{\rho_h^2}{4 L^2} (1 - T^2 + X^2)^2 \D \varphi^2 \right].
\end{equation}
The suitable coordinate transformations are listed in appendix \ref{sec:BTZ_in_Kruskal}. The two asymptotic boundaries are located at $X = \pm \sqrt{1 + T^2}$ while the future and past singularities are at $T = \pm \sqrt{1 + X^2}$. We choose the coordinate change in such a way that the directions of the Schwarzschild times $t$ in both wedges agree with the direction of the Kruskal time $T$ on the Cauchy surface $\Sigma_L \cup \Sigma_R$ as shown in figure \ref{fig:btz_kru} on page \pageref{fig:btz_kru}.

\paragraph{The vacuum.} We use $| 0 \>$ to distinguish the Schwarzschild vacuum, defined with respect to the annihilation operators $\op{\a}_{\omega n}^{L,R}$, from the Kruskal vacuum $|\Omega\>$, a.k.a. the thermofield double state or the Euclidean vacuum. The Schwarzschild vacuum $| 0 \>$ is associated with the foliation by constant-$t$ slices in each wedge in the Schwarzschild coordinates $(t, \rho, \varphi)$. Since the metric \eqref{BTZ} is time-independent, the Schwarzschild vacuum is the ground state for each constant $t$ slice. On the other hand the Kruskal vacuum is the vacuum with respect to the Kruskal time $T$ at the $T = 0$ slice only. Since the BTZ metric \eqref{BTZ_Kruskal} in the Kruskal coordinates is time-dependent, the Kruskal vacuum $| \Omega \>$ evolves with time.

The Kruskal vacuum can be found by investigating which combinations of the modes $\phi_{\omega n}^{L, R}$ and $\phi_{\omega n}^{L, R \ast}$ are analytic in Kruskal coordinates, \cite{Unruh:1976db}. With $| \Omega \> = \bigotimes_{\omega, n} | \Omega \>_{\omega n}$ we can look at the form of the Kruskal vacuum mode by mode. With our conventions the negative frequency Kruskal modes $\chi^{R,L}_{\omega n}$ are
\begin{align} \label{chi_modes}
\chi^{R}_{\omega n} & = \frac{\phi^R_{\omega n} + e^{-\frac{\beta \omega}{2}} \phi^{L \ast}_{\omega,-n}}{\sqrt{1 - e^{-\beta \omega}}}, &
\chi^{L}_{\omega n} & = \frac{\phi^L_{\omega n} + e^{-\frac{\beta \omega}{2}} \phi^{R \ast}_{\omega,-n}}{\sqrt{1 - e^{-\beta \omega}}},
\end{align}
where
\begin{align} \label{temp}
\beta = \frac{2 \pi L}{\p}
\end{align}
is the inverse temperature associated with the black hole, see appendix \ref{sec:BTZ_in_Kruskal} for details. The transformation induces the Bogoliubov transformation of the creation-annihilation coefficients
\begin{align} \label{Bogo_BTZ}
\op{\b}^R_{\omega n} & = \frac{\op{\a}^{R}_{\omega n} - e^{-\frac{\beta \omega}{2}} \op{\a}^{L \dagger}_{\omega, -n}}{\sqrt{1 - e^{-\beta \omega}}}, &
\op{\b}^L_{\omega n} & = \frac{\op{\a}^{L}_{\omega n} - e^{-\frac{\beta \omega}{2}} \op{\a}^{R \dagger}_{\omega, -n}}{\sqrt{1 - e^{-\beta \omega}}}.
\end{align}
The Kruskal vacuum $| \Omega \>$ is the vacuum with respect to $\op{\b}^L_{\omega n}$ and $\op{\b}^R_{\omega n}$, \textit{i.e.}, it obeys $\op{\b}^L_{\omega n} | \Omega \> = \op{\b}^R_{\omega n} | \Omega \> = 0$ for all $\omega, n$. From this it follows that $| \Omega \>$ is a 2-particle squeezed state from the point of view of the Schwarzschild vacuum $| 0 \>$ and it can be written as
\begin{align}
| \Omega \>_{\omega n} & = \sqrt{1 - e^{-\beta \omega}} \exp \left[ e^{-\frac{\beta \omega}{2}} \op{\a}_{\omega,-n}^{L \dagger} \op{\a}_{\omega n}^{R \dagger} \right] | 0 \> \nn\\
& = \sqrt{1 - e^{-\beta \omega}} \sum_{j=0}^{\infty} e^{-\frac{\beta \omega j}{2}} | j \>_{\omega,-n}^L \otimes | j \>_{\omega n}^R. \label{TFD}
\end{align}

As before, we can define the 1-particle Hilbert space $\Hone_{\Omega}$ as spanned by $\op{\b}^{R \dagger}_{\omega n} | \Omega \>$ and $\op{\b}^{L \dagger}_{\omega n} | \Omega \>$ and the full Hilbert space $\H_{\Omega} = Sym(\Hone_{\Omega})$. The Bogoliubov transformation \eqref{Bogo_BTZ} induces the unitary isomorphism between $\H_0$ and $\H_{\Omega}$ given in appendix \ref{sec:2-particle_squeeze}.

\paragraph{Going to the boundary.} After the review of the standard material we can move to the boundary. One would certainly like to say that the two pairs of creation-annihilation operators, $\op{\a}^{L \dagger}_{\omega n}, \op{\a}^L_{\omega n}$ and $\op{\a}^{R \dagger}_{\omega n}, \op{\a}^R_{\omega n}$ are the left and right boundary operators. Clearly, 1-particle states $\Hone_0$ split into the direct sum $\Hone_0 = \H_0^{(1) L} \oplus \H_0^{(1) R}$ and so the full Hilbert space splits into the tensor product, $\H_0 = \H_0^{L} \otimes \H_0^{R}$. This is a valid statement, but notice that the same statement is true for $\H_{\Omega}$ and the Kruskal operators $\op{\b}^{L,R \dagger}_{\omega n}, \op{\b}^{L,R}_{\omega n}$; the 1-particle states split $\H_{\Omega}^{(1)} = \H_{\Omega}^{(1) L} \oplus \H_{\Omega}^{(1) R}$ and the total Hilbert space is $\H_{\Omega} = \H_{\Omega}^{L} \otimes \H_{\Omega}^{R}$.

As we can see the problem is not to define \emph{some} split of the full Hilbert space $\H_{BTZ}$, but to split it into the actual boundary Hilbert spaces $\H_{L,R}$. As defined in section \ref{sec:holo} these Hilbert spaces can be identified with the tensor factors in $\H_{BTZ}$ associated with initial conditions living on $\Sigma_L$ and $\Sigma_R$. In this way we identify $\H^0_L$ and $\H^0_R$ as the boundary Hilbert spaces and the total Hilbert space splits,
\begin{align}
& \H_L = \H_L^0, && \H_R = \H_R^0, && \H_{BTZ} \cong \H_L \otimes \H_R. 
\end{align}
This statement does not depend on the choice of the vacuum. The Hilbert space splits into the tensor product of the boundary Hilbert spaces regardless of the vacuum state selected.

Since both the BTZ metric \eqref{BTZ} and the Rindler-AdS metric \eqref{Rindler_AdS} near the boundary approach the same expression,
\begin{align}
\D s^2_{BTZ}, \D s^2_{RAdS} \overset{\rho \rightarrow \infty}{\longrightarrow} - \rho^2 \D t^2 + \frac{L^2 \D \rho^2}{\rho^2} + \rho^2 \D \varphi^2
\end{align}
we identify the boundary Hilbert spaces $\H_{L,R}$ as the Rindler-AdS Hilbert spaces from section \ref{sec:AdS3}, $\H_{L,R} = \H_{RAdS}$. As the BTZ modes $\phi^{R,L}_{\omega n}$ approach the boundary, they approximate the Rindler-AdS mode $\phi^{RAdS}_{\omega n}$ in \eqref{Rind_phi}, up to normalization. The normalizations of the Rindler-AdS and the BTZ modes differ, since they are normalized with respect to the Klein-Gordon norm in their corresponding bulks. Thus, up to the rescaling by $c^{BTZ}_{\omega n}/c^{RAdS}_{\omega n}$, we can regard the creation-annihilation operators $\op{\a}^{L \dagger}_{\omega n}, \op{\a}^L_{\omega n}$, $\op{\a}^{R \dagger}_{\omega n}, \op{\a}^R_{\omega n}$ as the pair of the independent creation-annihilation operators in two copies of the boundary theory. 

Analogously to \eqref{Obnd}, let us define boundary limits of any bulk quantity $\psi(t, \theta, \varphi)$ as
\begin{align} \label{bnd_lim}
\limRL \psi(t, \varphi) & = \lim_{\rho_{R,L} \rightarrow \infty} \left( \frac{\rho_{R,L}}{\p} \right)^{\Delta} \psi(\rho_{R,L}, t, \varphi).
\end{align}
In particular the boundary limits of the bulk field $\op{\Phi}_{BTZ}$ decomposed in the Schwarzschild modes read
\begin{align} \label{OLR}
\O_{L,R}(t, \varphi) & = \lim_{L,R} \op{\Phi}_{BTZ} = \int_0^{\infty} \frac{\D \omega}{2 \pi} \sum_{n=-\infty}^{\infty} \left( \varphi_{\omega n} \op{\a}^{L,R}_{\omega n} + \varphi_{\omega n}^{\ast} \op{\a}^{L,R \dagger}_{\omega n} \right),
\end{align}
where the boundary mode is
\begin{align} \label{bnd_mode}
\varphi_{\omega n}(t, \varphi) & = \limR \phi^R_{\omega n} = c^{BTZ}_{\omega n} e^{-\I \omega t + \I n \varphi}.
\end{align}
We can regard these operators as the boundary operators composed with the creation-annihilation operators associated with the boundary Hilbert spaces.

The subtle difference between the boundary limit of a bulk operator and the actual boundary operators would manifest itself if we considered the field decomposition in terms of modes other than $\phi_{\omega n}^{L, R}$. Indeed, consider the field operator $\op{\Phi}_{BTZ}$ expressed in terms of the Kruskal modes \eqref{chi_modes}. The right boundary limit of the negative frequency mode $\chi^R_{\omega n}$ is proportional to the negative frequency mode $\varphi_{\omega n}$. On the other hand the right boundary limit of the negative frequency mode $\chi^L_{\omega n}$ is proportional to the positive frequency mode $\varphi^{\ast}_{\omega, -n}$. This simply means that the associated creation-annihilation operators $\op{\beta}^{L, R \dagger}_{\omega n}$ and $\op{\beta}^{L, R}_{\omega n}$ do not decouple and create excitations on both boundaries. Note that it does not mean that the Hilbert space does not factorize into the tensor product $\H \cong \H_{\Omega}^L \otimes \H_{\Omega}^R$, it simply means that $\H_{\Omega}^L$ and $\H_{\Omega}^R$ are not the boundary Hilbert spaces. In particular the vacuum state with respect to $\op{\beta}^{L, R}_{\omega n}$, \textit{i.e.}, the Kruskal vacuum $| \Omega \>$ is an entangled state between the two boundaries.

\paragraph{Correlation functions and temperature.} Let us follow the notation of \cite{Papadodimas:2012aq} and define the Fourier transforms of the left and right operators $\O_{L}$ and $\O_{R}$ defined in \eqref{OLR},
\begin{align} \label{OLR_mom}
\O^{L, R}_{\varpi n} & = \int_{-\infty}^{\infty} \D t \int_0^{2 \pi} \D \varphi \, e^{\I \varpi t - \I n \varphi} \O_{L,R}(t, \varphi) \nn\\
& = c_{\varpi n}^{BTZ} \left( \op{\a}^{L,R}_{\varpi n} \bs{1}_{\varpi > 0} + \op{\a}^{L,R \dagger}_{-\varpi, -n} \bs{1}_{\varpi < 0} \right),
\end{align}
where $\bs{1}_{\varpi > 0} = 1$ for $\varpi > 0$ and zero otherwise. Note that unlike anywhere else in this paper, here $\varpi$ can take both positive and negative values. We want to consider 2-point functions of the operators $\O^{L}_{\varpi n}$ and $\O^{R}_{\varpi n}$ both on the Schwarzschild vacuum $| 0 \>$ and on the Kruskal vacuum $| \Omega \>$. We will compare these correlators to the 2-point functions in the geon background proposed by various authors and these derived in this paper. We can drop the momentum-conserving delta functions in the 2-point functions and define $G^{RR}_{BTZ|\psi\>}$ and $G^{LR}_{BTZ|\psi\>}$ as
\begin{align}
\< \psi | \O^{R}_{\varpi n} \O^{R}_{\varpi' n'} | \psi \> & = 2 \pi \delta(\varpi + \varpi') \delta_{n + n', 0} \times G^{RR}_{BTZ|\psi\>}(\varpi, n), \label{GdefRR} \\
\< \psi | \O^{L}_{\varpi n} \O^{R}_{\varpi' n'} | \psi \> & = 2 \pi \delta(\varpi - \varpi') \delta_{n + n', 0} \times G^{LR}_{BTZ|\psi\>}(\varpi, n), \label{GdefLR}
\end{align}
where $| \psi \>$ stands for $| 0 \>$ or $| \Omega \>$ or any other sufficiently regular state.

The 2-point functions in the Schwarzschild vacuum read
\begin{align}
G^{RR}_{BTZ|0\>}(\varpi, n) & = | c_{\varpi n}^{BTZ} |^2, & G^{LR}_{BTZ|0\>}(\varpi, n) & = 0,
\end{align}
where the normalization constant is given in \eqref{cBTZ}. The cross-boundary correlator vanishes. On the other hand on the Kruskal vacuum we find
\begin{align}
G^{RR}_{BTZ|\Omega\>}(\varpi, n) & = | c_{\varpi n}^{BTZ} |^2 \left[ \frac{e^{\beta \varpi}}{e^{\beta \varpi} - 1} \bs{1}_{\varpi > 0} + \frac{1}{e^{\beta |\varpi|} - 1} \bs{1}_{\varpi < 0} \right], \\
G^{LR}_{BTZ|\Omega\>}(\varpi, n) & = | c_{\varpi n}^{BTZ} |^2 \frac{e^{\frac{\beta |\varpi|}{2}}}{e^{\beta |\varpi|} - 1}
\end{align}
where the inverse temperature is $\beta = 2 \pi L / \p$. The fact that the cross-boundary correlator is non-vanishing is due to the entanglement between the left and right wedge carried by the Kruskal vacuum. Furthermore the $G^{RR}_{BTZ|\Omega\>}$ correlator satisfies the KMS relation
\begin{align}
G^{RR}_{BTZ|\Omega\>}(-\varpi, n) = e^{-\beta \varpi} G^{RR}_{BTZ}(\varpi, n) \text{ for } \varpi > 0.
\end{align}

By taking $\omega' = \omega$ we can evaluate the expectation values of the Schwarzschild number operators in the Kruskal vacuum state. To do it define
\begin{align}
& \op{N}_{L,R} = \int_{0}^{\infty} \frac{\D \omega}{2 \pi} \sum_{n=-\infty}^{\infty} \op{N}^{L,R}_{\omega n}, && \op{N}^{L,R}_{\omega n} = \op{\a}^{L, R \dagger}_{\omega n} \op{\a}^{L, R}_{\omega n},
\end{align}
which measure the number of excited particles with respect to the Schwarzschild vacuum. In the Kruskal state $| \Omega \>$ the expectation value of, say, the right number operators reads
\begin{align}
\< \Omega | \op{N}^{L,R}_{\omega n} | \Omega \> = \frac{\pi}{e^{\beta \omega} - 1} \times \delta(0).
\end{align}
This means that from the point of view of the right boundary QFT the Kruskal vacuum $| \Omega \>$ is the thermal state with the inverse temperature given by \eqref{temp}.

\subsection{The geon}

The antipodal map, $\theta$, is the involutive isometry of the BTZ spacetime, which, in Kruskal coordinates $(T,X,\varphi)$, is given by
\begin{align}
\theta(T, X, \varphi) = (-T, -X, \varphi + \pi).
\end{align}
The geon is the quotient spacetime, $BTZ /\!\! \sim$, where two points $x_L \sim x_R$ are identified if $x_L = \theta x_R$. In particular the geon has a single boundary.

In this paper we want to treat the geon as the full BTZ geometry with the antipodal identification imposed on the level of the bulk field rather than geometry. We can keep the spacetime unfolded, but equivalently require that the values of the scalar field $\Phi$ in the BTZ black hole are related at two points mapped to each other by the antipodal map. In other words if $x_L = \theta x_R$, then $\Phi(x_L) = \pm \Phi(x_R)$. Thus our geon has two asymptotic boundaries and has the geometry of the BTZ black hole. The values of the field in the two wedges, however, are related in a seemingly non-local fashion.

\paragraph{The modes.} For the classical scalar field this means that we are looking for the solutions $\Phi^{(\pm)}$ to the Klein-Gordon equation, which satisfy 
\begin{align} \label{theta_Phi_rel}
\Phi^{(\pm)} \circ \theta = \pm \Phi^{(\pm)}.
\end{align}
With the relation between the Schwarzschild coordinates and the Kruskal coordinates presented in appendix \ref{sec:BTZ_in_Kruskal}, the antipodal map reads
\begin{align}
\theta(t, \rho_R, \varphi) = (-t, \rho_L, \varphi),
\end{align}
where by $\rho_L$ and $\rho_R$ we denote the fact that the two points related by $\theta$ have the same value of the radial variable $\rho = \rho_L = \rho_R$, but live in the opposite wedges. Since the antipodal map reverses the relative direction of time between the two wedges (see figure \ref{fig:btz_geon}) as the quantum operator, $\op{\Theta}$, the antipodal map becomes the anti-unitary involution, $\op{\Theta}^2 = \bs{1}$. From its action on the bulk field operator, $\op{\Theta} \, \op{\Phi}_{BTZ} \, \op{\Theta} = \op{\Phi}_{BTZ} \circ \theta$, we find
\begin{align}
\op{\Theta} \, \phi_{\omega n}^{R} \, \op{\Theta} & = \phi^{R \ast}_{\omega n}, & \op{\Theta} \, \op{\a}^{R}_{\omega n} \, \op{\Theta} & = (-1)^n \op{\a}^{L}_{\omega,-n}.
\end{align}

Let us introduce the following geon modes
\begin{align} \label{geon_modes}
\psi_{\omega n}^{(\pm)} & = \frac{1}{\sqrt{2}} \left[ 
\phi^R_{\omega n} \pm (-1)^n \phi^{L \ast}_{\omega,-n} \right]
\end{align}
and the corresponding operators,
\begin{align}
\op{\a}_{\omega n}^{(\pm)} & = \frac{1}{\sqrt{2}} 
\left[ \op{\a}^R_{\omega n} \pm (-1)^n \op{\a}^{L \dagger}_{\omega,-n} \right]. \label{geon_data}
\end{align}
These modes and operators have the specified parity under $\theta$,
\begin{align}
& \psi_{\omega n}^{(\pm)} \circ \theta = \pm \psi_{\omega n}^{(\pm)}, && \op{\Theta} \, \op{\a}_{\omega n}^{(\pm)} \, \op{\Theta} = \pm \op{\a}_{\omega n}^{(\pm) \dagger}.
\end{align}
Therefore one can use these modes to split the BTZ field operator $\op{\Phi}_{BTZ} = \op{\Phi}^{(+)} + \op{\Phi}^{(-)}$ into two operators of fixed parity under $\op{\Theta}$,
\begin{align} \label{PhiGeon}
\op{\Phi}^{(\pm)} & = \frac{1}{2} \left[ \op{\Phi}_{BTZ} \pm \op{\Theta} \, \op{\Phi}_{BTZ} \op{\Theta} \right] \nn\\
& = \int_0^{\infty} \frac{\D \omega}{2 \pi} \sum_{n=-\infty}^{\infty} \left( \psi_{\omega n}^{(\pm)} \op{\a}^{(\pm)}_{\omega n} + \psi_{\omega n}^{(\pm) \ast} \op{\a}^{(\pm) \dagger}_{\omega n} \right).
\end{align}
From now on and in accordance with \cite{Sanchez:1986qn,Louko:1998hc,Guica:2014dfa} we will concentrate on $\op{\Phi}^{(+)}$ only.

\paragraph{Summary of \cite{Sanchez:1986qn}.} Our analysis so far led us to the problem of the quantization of the scalar field with the mode decomposition
\begin{align} \label{PhiGeonP}
\op{\Phi}^{(+)} & = \int_0^{\infty} \frac{\D \omega}{2 \pi} \sum_{n=-\infty}^{\infty} \left( \psi_{\omega n}^{(+)} \op{\a}^{(+)}_{\omega n} + \psi_{\omega n}^{(+) \ast} \op{\a}^{(+) \dagger}_{\omega n} \right),
\end{align}
where the modes and the creation-annihilation operators are given by \eqref{geon_modes} and \eqref{geon_data}. 

The quantization problem was first discussed in \cite{Sanchez:1986qn}. It was noticed there that, from the point of view of the BTZ black hole, the geon modes, $\psi_{\omega n}^{(\pm)}$, have the vanishing Klein-Gordon norm, $\| \psi_{\omega n}^{(\pm)} \|_{BTZ} = 0$. Equivalently, the operators $\op{\a}^{(+)}_{\omega n}, \op{\a}^{(+) \dagger}_{\omega n}$ are not creation-annihilation operators, but rather they commute between themselves,
\begin{align} \label{comm_rel_1}
& \left[ \op{\a}^{(\pm)}_{\omega n}, \op{\a}^{(\pm) \dagger}_{\omega' n'} \right] = 0, && \left[ \op{\a}^{(\pm)}_{\omega n}, \op{\a}^{(\mp) \dagger}_{\omega' n'} \right] = 2 \pi \delta(\omega - \omega') \delta_{n n'}.
\end{align}

In \cite{Sanchez:1986qn} the authors suggest to keep all the states of the BTZ black hole, \textit{i.e.}, to keep the full factorizable Hilbert space, $\H_{BTZ} = \H_L \otimes \H_R$. In particular one has access to both the Schwarzschild and Kruskal vacua. They argued, however, that one should only consider \eqref{PhiGeonP} as the fundamental field operator on the geon background. In particular, given any state $| \psi \> \in \H_L \otimes \H_R$, the 2-point function is given by equation (16) of \cite{Sanchez:1986qn} and reads
\begin{align} \label{Sanchez2pt}
\< \psi | \op{\Phi}^{(+)}(x) \op{\Phi}^{(+)}(y) | \psi \> = \frac{1}{4} \left[ G(x, y) + G(\theta x, y) + G(x, \theta y) + G(\theta x, \theta y) \right], 
\end{align}
where where $G(x, y)$ is the 2-point function of $\op{\Phi}_{BTZ}$ in the BTZ background. In particular, one can choose two points $x$ and $y$ in two distinct BTZ wedges in such a way that $\theta x$ is in the future cone of $y$. In such a case $[\op{\Phi}^{(+)}(x), \op{\Phi}^{(+)}(y) ] \neq 0$, even though $x$ and $y$ are spacelike-separated. This non-locality can be regarded as a feature, but, as we will see, the treatment of the geon as a wormhole leads to perfectly causal quantum field theory.

\subsection{Geon as the wormhole} \label{sec:geonwh}

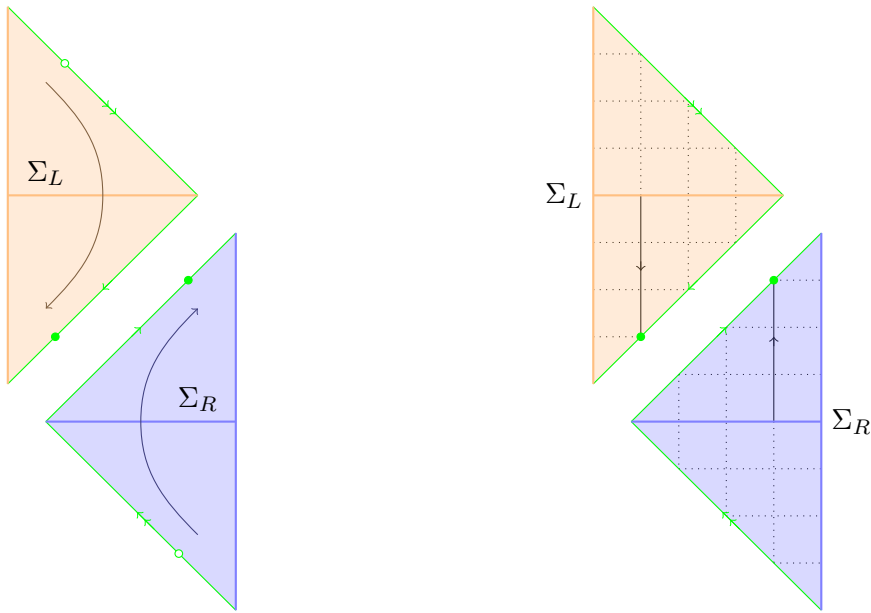
\begin{figure}[ht]
\centering
\begin{subfigure}[t]{0.4\textwidth}
\begin{tikzpicture}[scale=2.5]
\draw[white, colorL, fill=colorL, opacity=0.3] (-1,-1) -- (-1,1) -- (0,0) -- cycle;
\draw[white, colorR, fill=colorR, opacity=0.3] (0.2,-0.2) -- (0.2,-2.2) -- (-0.8,-1.2) -- cycle;
\draw[green] (-1,-1) -- ( 0,0);
\draw[green] (-0.8,-1.2) -- (0.2,-0.2);
\draw[green] (-1,1) -- ( 0,0);
\draw[green] (-0.8,-1.2) -- (0.2,-2.2);
\draw[->, green] (-0.45,-0.45) -- (-0.5,-0.5);
\draw[->, green] (-0.35,-0.75) -- (-0.3,-0.7);
\draw[->, green] (-0.52,0.52) -- (-0.47,0.47);
\draw[->, green] (-0.48,0.48) -- (-0.43,0.43);
\draw[->, green] (-0.27,-1.73) -- (-0.32,-1.68);
\draw[->, green] (-0.23,-1.77) -- (-0.28,-1.72);
\draw[green, fill=green] (-0.75,-0.75) circle (0.02);
\draw[green, fill=green] (-0.05,-0.45) circle (0.02);
\draw[green, fill=white] (-0.7,0.7) circle (0.02);
\draw[green, fill=white] (-0.1,-1.9) circle (0.02);
\draw[thick, colorL] (-1,1) -- (-1,-1);
\draw[thick, colorR] (0.2,-0.2) -- (0.2,-2.2);
\draw[thick, colorL] (-1,0) -- (0,0);
\draw[thick, colorR] (-0.8,-1.2) -- (0.2,-1.2);
\draw[colorR!50!black, ->] (0,-1.8) to [out=135,in=-90] (-0.3,-1.2) to[out=90,in=-135] (0,-0.6);
\draw[colorL!50!black, <-] (-0.8,-0.6) to [out=45,in=-90] (-0.5,0) to[out=90,in=-45] (-0.8,0.6);
\node[above] at (0,-1.2) {$\Sigma_R$};
\node[above] at (-0.8,0) {$\Sigma_L$};
\end{tikzpicture}
\centering
\caption{The Penrose diagram of the geon-wormhole. The two wedges are glued along the horizons according to the directions specified by the green arrows. In particular the pairs of the corresponding points are mapped to each other.\label{fig:geon_glue}}
\end{subfigure}
\qquad\qquad
\begin{subfigure}[t]{0.4\textwidth}
\begin{tikzpicture}[scale=2.5]
\draw (-0.05,-1.2) -- (-0.05,-0.45);
\draw[->] (-0.05,-0.8) -- (-0.05,-0.75);
\draw (-0.75,-0.75) -- (-0.75,0);
\draw[<-] (-0.75,-0.4) -- (-0.75,-0.35);
\draw[dotted] (-0.75,0) -- (-0.75,0.75);
\draw[dotted] (-0.5,-0.5) -- (-0.5,0.5);
\draw[dotted] (-0.25,-0.25) -- (-0.25,0.25);
\draw[dotted] (-1,0.25) -- (-0.25,0.25);
\draw[dotted] (-1,0.5) -- (-0.5,0.5);
\draw[dotted] (-1,0.75) -- (-0.75,0.75);
\draw[dotted] (-1,-0.25) -- (-0.25,-0.25);
\draw[dotted] (-1,-0.5) -- (-0.5,-0.5);
\draw[dotted] (-1,-0.75) -- (-0.75,-0.75);
\draw[dotted] (-0.55,-0.95) -- (0.2,-0.95);
\draw[dotted] (-0.3,-0.7) -- (0.2,-0.7);
\draw[dotted] (-0.05,-0.45) -- (0.2,-0.45);
\draw[dotted] (-0.55,-1.45) -- (0.2,-1.45);
\draw[dotted] (-0.3,-1.7) -- (0.2,-1.7);
\draw[dotted] (-0.05,-1.95) -- (0.2,-1.95);
\draw[dotted] (-0.05,-1.2) -- (-0.05,-1.95);
\draw[dotted] (-0.3,-0.7) -- (-0.3,-1.7);
\draw[dotted] (-0.55,-0.95) -- (-0.55,-1.45);
\draw[white, colorL, fill=colorL, opacity=0.3] (-1,-1) -- (-1,1) -- (0,0) -- cycle;
\draw[white, colorR, fill=colorR, opacity=0.3] (0.2,-0.2) -- (0.2,-2.2) -- (-0.8,-1.2) -- cycle;
\draw[green] (-1,-1) -- ( 0,0);
\draw[green] (-0.8,-1.2) -- (0.2,-0.2);
\draw[green] (-1,1) -- ( 0,0);
\draw[green] (-0.8,-1.2) -- (0.2,-2.2);
\draw[->, green] (-0.45,-0.45) -- (-0.5,-0.5);
\draw[->, green] (-0.35,-0.75) -- (-0.3,-0.7);
\draw[->, green] (-0.52,0.52) -- (-0.47,0.47);
\draw[->, green] (-0.48,0.48) -- (-0.43,0.43);
\draw[->, green] (-0.27,-1.73) -- (-0.32,-1.68);
\draw[->, green] (-0.23,-1.77) -- (-0.28,-1.72);
\draw[green, fill=green] (-0.75,-0.75) circle (0.02);
\draw[green, fill=green] (-0.05,-0.45) circle (0.02);
\draw[thick, colorL] (-1,1) -- (-1,-1);
\draw[thick, colorR] (0.2,-0.2) -- (0.2,-2.2);
\draw[thick, colorL] (-1,0) -- (0,0);
\draw[thick, colorR] (-0.8,-1.2) -- (0.2,-1.2);
\node[right] at (0.2,-1.2) {$\Sigma_R$};
\node[left] at (-1,0) {$\Sigma_L$};
\end{tikzpicture}
\centering
\caption{The Kruskal coordinates in the two wedges. The lines shown in the two wedges are related by the antipodal map. Thus, the values of the scalar field along the lines must match.\label{fig:geon_kru}}
\end{subfigure}
\caption{Geon as a wormhole.\label{fig:geon_as_wormhole}}
\end{figure}

The scalar field $\Phi_g$ on the geon background is equivalent to the scalar field on the full BTZ background obeying \eqref{theta_Phi_rel}. We have selected the even case and the geon field $\Phi_g$ satisfies
\begin{align}
\Phi_g \circ \theta = \Phi_g.
\end{align}
Now we want to show that the geon system is equivalent to a wormhole. Indeed, it is enough to introduce the antipodal identification on the horizons only and the evolution extends it to the entirety of the spacetime. To see it, we split the scalar modes $\phi_{\omega n}^R$ and $\phi_{\omega n}^L$ into the sums of two modes,
\begin{align} \label{phi_decomp_hor}
\phi_{\omega n}^{L,R} =\phi^{R,L (1)}_{\omega n} + \phi^{R,L (2)}_{\omega n},
\end{align}
each with the specified behavior on the horizons. The exact expressions are provided in appendix \ref{sec:BTZ_in_Kruskal}. When approaching the horizons the modes behave as
\begin{align}
& \phi^{R(1)}_{\omega n}(U, 0, \varphi) \sim e^{\I n \varphi} (2 |U|)^{-2 \hat{\omega}}, && \phi^{R(2)}_{\omega n}(0, V, \varphi) \sim e^{\I n \varphi} (2 |V|)^{2 \hat{\omega}}, \\
& \phi^{L(1)}_{\omega n}(0, V, \varphi) \sim e^{\I n \varphi} (2 |V|)^{-2 \hat{\omega}}, && \phi^{L(2)}_{\omega n}(U, 0, \varphi) \sim e^{\I n \varphi} (2 |U|)^{2 \hat{\omega}},
\end{align}
where the null Kruskal coordinates $(U,V,\varphi)$ are related to $(T,X,\varphi)$ in \eqref{BTZ_Kruskal} by $U = T + X$ and $V = T - X$. Since the antipodal map reads $\theta(U,V,\varphi) = (-U,-V,\varphi+\pi)$, the values of the modes at the horizons must match. We find
\begin{align}
\phi^{R(1)}_{\omega n}(U, 0, \varphi) & = (-1)^n \phi^{L(2) \ast}_{\omega, -n}(-U, 0, \varphi + \pi), \\
\phi^{R(2)}_{\omega n}(0, V \varphi) & = (-1)^n \phi^{L(1) \ast}_{\omega, -n}(0, -V, \varphi + \pi).
\end{align}
The two horizons constitute a complete Cauchy surface. By prescribing the values of the fields on the horizon subjected to the relations above, there exists a unique bulk field. Thus, the above relations hold in the entirety of spacetime and for all modes and they combine back to the geon modes
\begin{align} \label{g_mode}
\psi_{\omega n} = \sqrt{2} \psi^{(+)}_{\omega n} = \phi_{\omega n}^{R} + (-1)^n \phi^{L \ast}_{\omega,-n}.
\end{align}
We conclude that the scalar field on the geon background is equivalent to the scalar field on the wormhole background presented in figure \ref{fig:geon_as_wormhole}. The antipodal identifications are imposed on the horizons only and the spacetime becomes that of a 2-sided, traversable wormhole. Furthermore, as shown in figure \ref{fig:geon_kru}, the lines of constant $X$ in the Kruskal coordinates become closed timelike curves.

Let $\psi_{\omega n}|_R = \phi^R_{\omega n}$ and $\psi_{\omega n}|_L = (-1)^n \phi^{L \ast}_{\omega,-n}$ denote the restrictions of $\psi_{\omega n}$ to the respective wedges. We included the factor of $\sqrt{2}$ in the definition \eqref{g_mode} as we want the mode to be equal to $\phi_{\omega n}^R$ in the right wedge. We can then think of $\psi_{\omega n}|_L$ as arising from the evolution of the field $\phi^{R}_{\omega n}$ from the right wedge to the left by the Kruskal Hamiltonian $\op{H}_K$, generating translations in the Kruskal time $T$.

First note that both $\Sigma_L$ and $\Sigma_R$ in figure \ref{fig:geon_as_wormhole} are the complete Cauchy surfaces on their own. By fixing the initial data $\Icx_R$ on, say, $\Sigma_R$, the field is uniquely determined in the entire spacetime. In particular we can think about the initial data $\Icx_L$ on $\Sigma_L$ as the evolution of $\Icx_R$. Thus, we have only a single set of creation-annihilation operators $\op{a}^{\dagger}_{\omega n}, \op{a}_{\omega n}$ obeying the standard canonical commutation relations
\begin{align} \label{ccr_a}
\left[ \op{a}_{\omega n}, \op{a}^{\dagger}_{\omega' n'} \right] = 2 \pi \delta(\omega - \omega') \delta_{n n'}
\end{align}
and the field operator takes form
\begin{align} \label{Phi_geon}
\op{\Phi}_{g} = \int_0^{\infty} \frac{\D \omega}{2 \pi} \sum_{n=-\infty}^{\infty} \left( \psi_{\omega n} \op{a}_{\omega n} + \psi^{\ast}_{\omega n} \op{a}^{\dagger}_{\omega n} \right).
\end{align}
In the right wedge $\psi_{\omega n}|_{R} = \phi^R_{\omega n}$ and we can identify $\op{a}^{\dagger}_{\omega n}, \op{a}_{\omega n}$ with the BTZ creation-annihilation operators $\op{\a}^{R \dagger}_{\omega n}, \op{\a}^R_{\omega n}$. The Hilbert space of the geon, $\Hg$, spanned by $\op{a}^{\dagger}_{\omega n}, \op{a}_{\omega n}$ becomes equal to $\H_R$. The vacuum $|0 \>$ with respect to the annihilation operators $\op{a}_{\omega n}$ is the Schwarzschild vacuum of the single wedge.

The second implication is the choice of the negative frequency modes in the left wedge once such a choice has been done in the right wedge. The Klein-Gordon scalar product on $\Sigma_R$ matches the scalar product in the BTZ background restricted to $\Sigma_R$. Since the Klein-Gordon product is preserved in time, this tells us that we must invert the time direction in the left wedge. This means that the normal vector to the constant time slices in \eqref{KG} in the left wedge points `downwards'. It makes $\phi_{\omega n}^{L \ast}$ negative frequency in the left wedge. The Klein-Gordon scalar product for the geon modes can be written equivalently on $\Sigma_R$ or $\Sigma_L$ as follows,
\begin{align} \label{newKG}
(\psi_{\omega n}, \psi_{\omega' n'})_{g} = (\psi_{\omega n}|_R, \psi_{\omega' n'}|_R)_{BTZ} = -(\psi_{\omega n}|_L, \psi_{\omega' n'}|_L)_{BTZ}.
\end{align}

This conclusion agrees with the analysis of \cite{Louko:1998hc}, here derived from the basic principle of the invariance of the Klein-Gordon product under the time evolution. Furthermore, note that from the point of view of the Klein-Gordon product for the BTZ black hole the normalizations have changed. Since in the BTZ black hole the integral in the Klein-Gordon product extends over $\Sigma_L \cup \Sigma_R$, we have schematically $\| \psi_{\omega n} \|^2_{BTZ} = 2\| \psi_{\omega n} \|^2_{g}$. From the point of view of the BTZ black hole one would expect an additional factor of $1/2$ on the right hand side of the commutation relations \eqref{ccr_a}. That would be incorrect.

From the point of view of the left boundary, the geon modes $\psi_{\omega n}$ become equal to $(-1)^n \phi_{\omega, -n}^{L \ast}$ in the left wedge. Thus the operators $\op{a}^{\dagger}_{\omega n}, \op{a}_{\omega n}$ should be identified with $(-1)^n \op{\a}^{L}_{\omega, -n}, (-1)^n \op{\a}^{L \dagger}_{\omega, -n}$. Not only are $\op{\a}^{L \dagger}_{\omega, -n}$ the annihilation operators from the point of view of the wormhole, but there is the operatorial relation between the left and right boundary operators. To distinguish such operators from the original BTZ operators, for which the excitations in the two wedges are independent, we will change notation from $\op{\a}$ to $\op{a}$. Thus, we have the boundary creation-annihilation operators $\op{a}^{L, R \dagger}_{\omega n}, \op{a}^{L, R}_{\omega n}$, which are not independent and satisfy
\begin{align} \label{geon_rel}
& \op{a}_{\omega n} = \op{a}^R_{\omega n} = (-1)^n \op{a}^{L}_{\omega,-n}, && \op{a}^{\dagger}_{\omega n} = \op{a}^{R \dagger}_{\omega n} = (-1)^n \op{a}^{L \dagger}_{\omega,-n}.
\end{align}
These relations induce the isomorphisms between the geon Hilbert space $\H_g$, spanned by $\op{a}^{\dagger}_{\omega n}$ and the boundary Hilbert spaces $\H_L$ and $\H_R$,
\begin{align} \label{geon_iso}
\Hg \cong \H_R \cong \op{\Theta} \H_L.
\end{align}
There is no tensor product.

\paragraph{The geon dual.} With the structure of the Hilbert space for the geon wormhole figured out, we can seek the vacuum state dual to the wormhole in holography. In \cite{Louko:1998hc,Guica:2014dfa}, where the problem was analyzed, the authors always looked at the factorized BTZ Hilbert space $\H_L \otimes \H_R$. However, as we can see     the Hilbert space $\H_{g}$ of the geon is that of a single side of the BTZ black hole. Thus we simply look for the state $| G \> \in \H \cong \H_R$ annihilated by the Kruskal operators \eqref{Bogo_BTZ}, when the operatorial relation \eqref{geon_rel} are substituted. This means that $| G \> = \bigotimes_{\omega n} | G \>_{\omega n}$ is the ground state with respect to the annihilation operators $\op{b}_{\omega n}$ defined as
\begin{align} \label{Bogo_geon}
\op{b}_{\omega n} & = \frac{\op{a}_{\omega n} - e^{-\frac{\beta \omega}{2}} (-1)^n \op{a}^{\dagger}_{\omega n}}{\sqrt{1 - e^{-\beta \omega}}}, & \beta = \frac{2 \pi L}{\p},
\end{align}
where the inverse temperature $\beta$ is the same as for the BTZ black hole \eqref{temp}. The geon state $| G \>$ is a 1-particle squeezed state (see appendix \ref{sec:1-particle}) and can be written as
\begin{align} \label{G}
| G \>_{\omega n} & = \left( 1 - e^{-\beta \omega} \right)^{1/4} \exp \left[ \frac{1}{2} e^{-\frac{\beta \omega}{2}} (-1)^n \op{a}_{\omega n}^{\dagger} \op{a}_{\omega n}^{\dagger} \right] | 0 \> \nn\\
& = \left( 1 - e^{-\beta \omega} \right)^{1/4} \sum_{j=0}^{\infty} (-1)^{nj} e^{-\frac{\beta \omega j}{2}} \sqrt{\frac{(2j-1)!!}{(2j)!!}} | 2j \>_{\omega n}.
\end{align}

We can think about $| G \>$ as a `thermofield single' state as it mimics the thermofield double, but it is the pure state that can be found in the Hilbert space of a single boundary CFT. Since it is annihilated by $\op{b}_{\omega n}$, the expectation value of the Schwarzschild number operator $\op{N}^R_{\omega n} = \op{a}^{R \dagger}_{\omega n} \op{a}^{R}_{\omega n}$ has the thermal distribution,
\begin{align} \label{G_therma}
\< G | \op{N}^R_{\omega n} | G \> = \frac{2 \pi}{e^{\beta \omega} - 1} \times \delta(0).
\end{align}
The analysis of such states from a more general point of view was recently carried out in \cite{Caputa:2022zsr}.

\paragraph{Evolution.} The evolution of the geon field $\op{\Phi}_g$ is driven by the general Hamiltonian \eqref{genH}. With respect to the foliation by constant $t$ slices, this leads to the standard Hamiltonian
\begin{align}
\op{H} = \int_0^{\infty} \frac{\D \omega}{2 \pi} \sum_{n=-\infty}^{\infty} \omega \op{a}_{\omega n}^{\dagger} \op{a}_{\omega n}.
\end{align}
This Hamiltonian evolves the field in both left and right wedges.

By taking the boundary limits we obtain the boundary Hamiltonians, $\op{H}_L, \op{H}_R$. Note that these are not independent operators, due to the relations \eqref{geon_rel}. Thus, despite the fact that these Hamiltonian have the standard form
\begin{align} \label{HLR}
\op{H}_{L,R} = \int_0^{\infty} \frac{\D \omega}{2 \pi} \sum_{n=-\infty}^{\infty} \omega \op{a}_{\omega n}^{L, R \dagger} \op{a}^{L, R}_{\omega n}
\end{align}
they are in fact the same operator, $\op{H}$. We can write a number of rather trivial identities such as
\begin{align} \label{H_comb}
& \op{H}_R - \op{H}_L = 0, && \op{H} = \frac{1}{2} ( \op{H}_L + \op{H}_R ).
\end{align}
Such identities were advocated in a number of papers on wormholes, \textit{e.g., } \cite{Harlow:2018tqv,Maldacena:2018lmt}. It was argued there that the Hamiltonians driving the boundary dynamics of the two sides are related.

We can also check that the evolution from a given point $x$ in, say, the right wedge to its antipodal point $\bar{x} = \theta x$ is given by $\op{\Theta}$. Indeed, let $x = (T, X, \varphi)$ and consider the Kruskal coordinates with the Hamiltonian $\op{H}_K$ driving the evolution with respect to the $T$ coordinate, see figure \ref{fig:geon_kru}. The Hamiltonian $\op{H}_K$ moves up the line until we reach the horizon at $(X, X, \varphi)$. This point is mapped by $\theta$ to $(-X, -X, \varphi + \pi)$ and then evolved back in time by $-\op{H}_K$ to $\bar{x}$. Thus the evolution operator reads
\begin{align}
U_{RL} = e^{\I (X - T) \op{H}_K} \op{\Theta} e^{- \I (X - T) \op{H}_K} = \op{\Theta},
\end{align}
since all geon operators satisfy $[ \op{H}_K, \op{\Theta}] = 0$. In particular we arrive at the natural isomorphism $\H \cong \H_R \cong \op{\Theta} \H_L$ induced by the evolution.

\paragraph{Bulk correlators.} In the wormhole picture the left wedge is both in the future and the past of the right wedge. For this reason the commutator $[ \op{\Phi}_{g}(x), \op{\Phi}_{g}(y) ]$ will be generally non-vanishing for the two points $x$ and $y$ lying in two different wedges. Now, however, this is a perfectly normal situation, since the two points are in each others future. The entire system remains perfectly causal. 

For example, in the Schwarzschild vacuum $| 0 \>$ the correlation functions within the single wedge match the BTZ correlators,
\begin{align} \label{gCorr_0xx}
\< 0 | \op{\Phi}_{g}(x) \op{\Phi}_{g}(y) | 0 \>_{g} = \< 0 | \op{\Phi}_{BTZ}(x) \op{\Phi}_{BTZ}(y) | 0 \>_{BTZ},
\end{align}
provided $x$ and $y$ belong to the same wedge. This is a more reasonable result than the relation \eqref{Sanchez2pt} as advocated in \cite{Sanchez:1986qn}. The vacuum state $| 0 \>$ is associated with the Hamiltonian driving the evolution within the single wedge. Thus, if the two points $x, y$ belong to the wedge, the 2-point function on $| 0 \>$ should be ignorant of any spacetime beyond it and have the BTZ form. On the other hand, contrary to the BTZ case, the cross-boundary correlators on $| 0 \>$ are non-vanishing. We have
\begin{align} \label{gCorr_0xy}
\< 0 | \op{\Phi}_{g}(x) \op{\Phi}_{g}(y) | 0 \>_{g} = \< 0 | \op{\Phi}_{BTZ}(x) \op{\Phi}_{BTZ}(\theta y) | 0 \>_{BTZ},
\end{align}
assuming $x$ and $y$ lie in different wedges.

As discussed in section \ref{sec:polar}, the choice of a different vacuum state, such as the geon state $| G \>$, corresponds to a different foliation. With $| G \>$ associated with the evolution in the global Kruskal time, we expect that the correlators on this state will exhibit strong cross-boundary correlations. Indeed, for any two points $x, y$ we find
\begin{align} \label{gCorrOmega}
& \< G | \op{\Phi}_{g}(x) \op{\Phi}_{g}(y) | G \>_{g} = \frac{1}{2} \left[ G(x, y) + G(\theta x, y) + G(x, \theta y) + G(\theta x, \theta y) \right], 
\end{align}
where
\begin{align}
G(x, y) = \< \Omega | \op{\Phi}_{BTZ}(x) \op{\Phi}_{BTZ}(y) | \Omega \>_{BTZ}
\end{align}
is the 2-point function of the BTZ field in the Kruskal vacuum. This expression matches \eqref{Sanchez2pt}, up to the multiplicative factor, which can be traced back to the normalization issues discussed below equation \eqref{newKG}.

\paragraph{Boundary theory.} We can also rewrite the conclusions in the language of the boundary operators. By taking the left and right boundary limits of the bulk field we obtain two boundary operators,
\begin{align}
\op{\Phi}_R(t, \varphi) & = \int_0^{\infty} \frac{\D \omega}{2 \pi} \sum_{n=-\infty}^{\infty} \left( \varphi_{\omega n} \op{a}_{\omega n} + \varphi^{\ast}_{\omega n} \op{a}^{\dagger}_{\omega n} \right), \label{PhiRg} \\
\op{\Phi}_L(t, \varphi) & = \int_0^{\infty} \frac{\D \omega}{2 \pi} \sum_{n=-\infty}^{\infty} \left( (-1)^n \varphi^{\ast}_{\omega n} \op{a}_{\omega,-n} + (-1)^n \varphi_{\omega n} \op{a}^{\dagger}_{\omega,-n} \right), \label{PhiLg}
\end{align}
where the boundary modes $\varphi_{\omega n}$ are the same as for the BTZ black hole in \eqref{bnd_mode}. When comparing to the mode decomposition of $\O_L$ and $\O_R$ in \eqref{OLR} we once again arrive at the relations \eqref{geon_rel} between the left and right creation-annihilation operators.

The boundary 2-point functions on $| 0 \>$ and $| G \>$ can be calculated directly, or by taking the boundary limits of the bulk correlation functions from the previous paragraph. We define the momentum space versions of $\O_L$ and $\O_R$ analogously to \eqref{OLR_mom}. On the Schwarzschild vacuum equations \eqref{gCorr_0xx} and \eqref{gCorr_0xy} give
\begin{align} \label{CorrGeon0}
& G^{RR}_{g |0\>}(\varpi, n) = G^{RR}_{BTZ |0\>}(\varpi, n), && G^{LR}_{g |0\>}(\varpi, n) = (-1)^n G^{RR}_{BTZ |0\>}(\varpi, n).
\end{align}
In the geon state $| G \>$ the expression \eqref{gCorrOmega} involves both $G^{RR}_{BTZ |\Omega\>}$ and $G^{RL}_{BTZ |\Omega\>}$ and reads
\begin{align} \label{CorrGeonOmega}
\< G | \O^R_{\varpi n} \O^R_{\varpi' n'} | G \>_{g} & = \frac{| c_{\varpi n}^{BTZ} |^2}{e^{\beta |\varpi|} - 1} \delta_{n + n', 0} \times \left[ 2 \pi \delta(\varpi + \varpi') \times \left( e^{\beta \varpi} \bs{1}_{\varpi > 0} + \bs{1}_{\varpi < 0} \right) \right.\nn\\
& \qquad\qquad\qquad\qquad \left. + \, 2 \pi \delta(\varpi - \varpi') \times e^{\frac{\beta |\varpi|}{2}} \right].
\end{align}
To see it, invert the Bogoliubov transformation \eqref{Bogo_geon} and express $\op{a}^{\dagger}_{\omega n}, \op{a}_{\omega n}$ in terms of $\op{b}^{\dagger}_{\omega n}, \op{b}_{\omega n}$.

\paragraph{Summary.} In this subsection we have constructed the Hilbert space $\H_g$ for the geon, together with its dual state $| G \>$ in \eqref{G}. By treating the geon as a wormhole, we could elucidate the structure of its Hilbert space and show that $\H_g \neq \H_L \otimes \H_R$: the factorization into the tensor product of the boundary Hilbert spaces fails. The Hilbert space $\H_g$ is that of a single boundary, $\H_g \cong \H_R \cong \op{\Theta} \H_L$. The geon state $| G \>$ is the squeezed states in the Hilbert space associated with a single boundary. It exhibits the appropriate thermal distribution \eqref{G_therma} as well as correlation functions \eqref{gCorrOmega}.

In the next two subsections we will show how the results of the previous investigations \cite{Louko:1998hc,Guica:2014dfa} can be retrieved from our construction. The key difference is that the authors of these papers assume factorization of the Hilbert space. We will see how and when the results presented in this subsection can be recovered after the assumption about factorization had been made.

\subsection{\texorpdfstring{Semiclassical approximation: approach of \cite{Louko:1998hc}}{Semiclassical approximation: approach of [1]}} \label{sec:geon_semi1}

In the semiclassical approach one treats $\H_L$ and $\H_R$ as two independent Hilbert spaces spanned by their own sets of independent left and right creation-annihilation operators $\op{\a}_{\omega n}^{L \dagger}, \op{\a}_{\omega n}^{L}$ and $\op{\a}_{\omega n}^{R \dagger}, \op{\a}_{\omega n}^{R}$. The semiclassical Hilbert space is assumed to be the tensor product, $\Hsemi = \H_{BTZ} = \H_L \otimes \H_R$. We want to see here how the results of \cite{Louko:1998hc} match with the analysis presented in the previous subsection. As we will see, the authors effectively impose some variant of the relation \eqref{geon_rel} on $\Hsemi$. In this sense they try to decouple the physical states obeying these relations from the unphysical ones and reduce the Hilbert space to a smaller tensor factor.

\paragraph{Summary of \cite{Louko:1998hc}.} The authors of \cite{Louko:1998hc} observed that the antipodal map reverses the direction of time in the left wedge with respect to the global, Kruskal time, see figure \ref{fig:geon_glue} on page \pageref{fig:geon_glue}. Just as discussed in the previous section they realize that one must identify $\phi^{L \ast}_{\omega n}$ as negative frequency modes and $\op{\a}^{L \dagger}_{\omega n}$ as the annihilation operators. The canonical commutation relations between $\op{\a}^{L}_{\omega n}$ and $\op{\a}^{L \dagger}_{\omega n}$ are now `flipped' and read $[ \op{\a}^{L \dagger}_{\omega n}, \op{\a}^{L}_{\omega n} ] = 2 \pi \delta( \omega - \omega') \delta_{n n'}$. In particular instead of \eqref{comm_rel_1} the commutation relations between the operators $\op{\a}^{(\pm)}_{\omega n}$ and $\op{\a}^{(\pm) \dagger}_{\omega n}$ defined in \eqref{geon_data} read
\begin{align} \label{ccr_pm}
\left[\op{\a}_{\omega n}^{(\sigma)}, \op{\a}_{\omega' n'}^{(\sigma') \dagger} \right] = 2 \pi \delta(\omega - \omega') \delta_{n n'} \delta^{\sigma \sigma'}
\end{align}
Now $\op{\a}_{\omega n}^{(\pm) \dagger}$ and $\op{\a}_{\omega n}^{(\pm)}$ are genuine creation-annihilation operators. The operators $\op{\a}_{\omega n}^{(+) \dagger}$ and $\op{\a}_{\omega n}^{(-) \dagger}$ span two Hilbert spaces $\H_{+}$ and $\H_{-}$ associated with the geon states of the specified parity under $\theta$. Thus the transformations \eqref{geon_data} specify the split of the BTZ, semiclassical Hilbert space as $\Hsemi \cong \H_{+} \otimes \H_{-}$. The authors of \cite{Louko:1998hc} use this decomposition to construct the geon state in (3.17) there. This is the usual thermofield double state, but entangling particles between $\H_{+}$ and $\H_{-}$,
\begin{align}
| \Omega_g \>_{\omega n} & = \sqrt{1 - e^{-\beta \omega}} \exp \left[ e^{-\frac{\beta \omega}{2}} \op{\a}_{\omega n}^{(-) \dagger} \op{\a}_{\omega n}^{(+) \dagger} \right] | 0 \> \nn\\
& = \sqrt{1 - e^{-\beta \omega}} \sum_{j=0}^{\infty} e^{-\frac{\beta \omega j}{2}} | j \>^{(+)}_{\omega n} \otimes | j \>^{(-)}_{\omega n}. \label{TFDg}
\end{align}
This is a reasonable proposal, but it fails to capture the fact that the actual Hilbert space of the system does not factorize into the tensor product. Most importantly, relations \eqref{geon_rel} should hold in any physical Hilbert space describing the wormhole. 

\paragraph{Physical and null states and operators.} The main idea behind the semiclassical approximation is to impose in some way the condition \eqref{geon_rel} on the level of the operators $\op{\a}^{L, R \dagger}_{\omega n}, \op{\a}^{L, R}_{\omega n}$ when acting on the semiclassical, factorizable Hilbert space $\Hsemi = \H_L \otimes \H_R$. First idea would be to require that
\begin{align} \label{strong_cond}
\op{\a}_{\omega n}^{(-)} | \psi \> = \op{\a}_{\omega n}^{(-) \dagger} | \psi \> = 0
\end{align}
for all $\omega, n$ on all physical states $| \psi \> \in \Hsemi$. It is easy to see that no state can satisfy these relations.

One way around it is to impose the condition \eqref{strong_cond} \emph{before} the system is quantized. This can be done within the framework of the constrained quantization, see \textit{e.g.}, \cite{Henneaux:1992ig}. We will discuss this approach in section \ref{sec:jt_semi} while here we will concentrate on the consequences of the semiclassical approximation only.

Once it is assumed that the Hilbert space is factorizable, $\Hsemi = \H_L \otimes \H_R$, the condition \eqref{strong_cond} cannot be consistently imposed. A possible resolution is to demand the weak version of \eqref{strong_cond}, namely
\begin{align} \label{beta_cond}
\op{\a}^{(-)}_{\omega n} | \psi \> = 0
\end{align}
for all $\omega, n$. States that obey this condition are \emph{physical}. All other states are \emph{null}. In contrast to the strong condition \eqref{strong_cond} the action of $\op{\a}^{(-) \dagger}_{\omega n}$ on physical states is in general non-vanishing. However, if $| \psi \>$ is physical, then $\op{\a}^{(-) \dagger}_{\omega n} | \psi \>$ is null due to commutation relations \eqref{ccr_pm}. Thus, physical states are those created by the action of the creation operators $\op{\a}^{(+) \dagger}_{\omega n}$ on the vacuum $|0\>$. With $\H_{+}$ and $\H_{-}$ denoting the Hilbert spaces spanned by $\op{\a}^{(+) \dagger}_{\omega n}$ and $\op{\a}^{(-) \dagger}_{\omega n}$ respectively, we have $\Hsemi = \H_L \otimes \H_R \cong \H_{+} \otimes \H_{-}$. We identify $\H_{+}$ as the physical space and any state which does not belong to $\H_{+}$ is null.

An operator $\O$ is physical if it maps $\H_{+}$ into itself. This is equivalent to
\begin{align}
& \left[ \O, \op{\a}^{(-)}_{\omega n} \right] | \psi \> = 0, && \left[ \O, \op{\a}^{(-) \dagger}_{\omega n} \right] | \psi \> = 0
\end{align}
for all $\omega, n$ on all physical states $| \psi \> \in \H_{+}$. This means that an operator $\O$ is physical if, when presented in terms of the creation-annihilation operators, it contains only $\op{\a}^{(+)}_{\omega n}$ and $\op{\a}^{(+) \dagger}_{\omega n}$, while the operators $\op{\a}^{(-)}_{\omega n}$ and $\op{\a}^{(-) \dagger}_{\omega n}$ are absent.

The left and right boundary creation-annihilation operators are on their own unphysical. Therefore the BTZ boundary operators $\O_L$ and $\O_R$ defined in \eqref{OLR} are in general unphysical as well. Furthermore, the left and right Hamiltonians \eqref{HLR} are unphysical as well. The only well-defined, physical operators are those, where the creation-annihilation operators combine into $\op{\a}^{(+)}_{\omega n}$ and $\op{\a}^{(+) \dagger}_{\omega n}$.

\paragraph{Evolution.} We want to make sure that the physical and null states decouple and that the decoupling is preserved by the evolution. This is achieved by the Hamiltonian
\begin{align} \label{Halpha2}
\op{H}_{+} =\int_0^{\infty} \frac{\D \omega}{2 \pi} \sum_{n=-\infty}^{\infty} \omega \op{\a}^{(+) \dagger}_{\omega n} \op{\a}^{(+)}_{\omega n}
\end{align}
from which we obtain
\begin{align}
e^{\I \op{H}_{+} t} \op{\a}_{\omega n}^R e^{-\I \op{H}_{+} t} & = \frac{1}{\sqrt{2}} \left[ e^{- \I \omega t} \op{\a}^{(+)}_{\omega n} + \op{\a}^{(-)}_{\omega n} \right], \label{evol_aR} \\
e^{\I \op{H}_{+} t} \op{\a}_{\omega n}^L e^{-\I \op{H}_{+} t} & = \frac{(-1)^n}{\sqrt{2}} \left[ e^{- \I \omega t} \op{\a}^{(+)}_{\omega, -n} - \op{\a}^{(-)}_{\omega, -n} \right]. \label{evol_aL}
\end{align}
This means that even if we start we an unphysical operator inserted at $t = 0$, its `unphysicality' does not undergo the evolution in time. As an example, consider the left and right boundary operators $\O_L$ and $\O_R$, defined in \eqref{OLR}. Note, however, that these expressions, derived for the BTZ black hole, assume that the two boundaries evolve by $\op{H}_L$ and $\op{H}_R$ separately. To derive their time-dependent versions with respect to the Hamiltonian \eqref{Halpha2} we must consider the two operators at $t = 0$
\begin{align} \label{OLRzero}
\O_{L,R}(t=0, \varphi) & = \int_0^{\infty} \frac{\D \omega}{2 \pi} \sum_{n=-\infty}^{\infty} c_{\omega n}^{BTZ} \left( \op{\a}^{L,R}_{\omega n} + \op{\a}^{L,R \dagger}_{\omega n} \right)
\end{align}
and apply the relations \eqref{evol_aR} and \eqref{evol_aL},
\begin{align}
\O_L(t, \varphi) & = \frac{1}{\sqrt{2}} \int_0^{\infty} \frac{\D \omega}{2 \pi} \sum_{n=-\infty}^{\infty} \left[ (-1)^n \varphi_{\omega n} \op{\a}^{(+)}_{\omega n} + (-1)^n \varphi^{\ast}_{\omega n} \op{\a}^{(+) \dagger}_{\omega n} \right]\nn\\
& \qquad\qquad + \frac{1}{\sqrt{2}} \int_0^{\infty} \frac{\D \omega}{2 \pi}  \sum_{n=-\infty}^{\infty} (-1)^n c^{BTZ}_{\omega n} \left[ \op{\a}^{(-)}_{\omega n} + \op{\a}^{(-) \dagger}_{\omega n} \right], \label{OLsemi} \\
\O_R(t, \varphi) & = \frac{1}{\sqrt{2}} \int_0^{\infty} \frac{\D \omega}{2 \pi}  \sum_{n=-\infty}^{\infty} \left[ \varphi_{\omega n} \op{\a}^{(+)}_{\omega n} + \varphi^{\ast}_{\omega n} \op{\a}^{(+) \dagger}_{\omega n} \right] \nn\\
& \qquad\qquad + \frac{1}{\sqrt{2}} \int_0^{\infty} \frac{\D \omega}{2 \pi}  \sum_{n=-\infty}^{\infty} c^{BTZ}_{\omega n} \left[ \op{\a}^{(-)}_{\omega n} + \op{\a}^{(-) \dagger}_{\omega n} \right]. \label{ORsemi}
\end{align}
The two time-independent terms containing $\op{\a}^{(-)}_{\omega n}, \op{\a}^{(-) \dagger}_{\omega n}$ are the only artifacts of the fact that the operators $\O_L$ and $\O_R$ evaluated at $t = 0$ in \eqref{OLRzero} were unphysical. The Hamiltonian \eqref{Halpha2} effectively projects out the unphysical operators and after even the shortest evolution only physical operators remain. Thus, we can drop such unphysical terms if we stick to correlators at $t > 0$. 

\paragraph{Normalization.} When the time-dependent terms in \eqref{OLsemi} and \eqref{ORsemi} are compared to \eqref{PhiRg} and \eqref{PhiLg} we see the additional factor of $1/\sqrt{2}$. This is the artifact of the normalization issue discussed underneath equation \eqref{newKG}. From the point of view of the BTZ physics the norm of the state $|1\>_{\omega n}^{R} = \op{\a}_{\omega n}^{R \dagger} | 0 \>$ is the same as the norm of the state $|1\>_{\omega n}^{+} = \op{\a}_{\omega n}^{(+) \dagger} | 0 \>$. However, from the point of view of the wormhole physics the first state contains unphysical part due to the action of $\op{\a}_{\omega n}^{(-) \dagger}$. With this unphysical part dropped its norm-squared, $\| |1\>_{\omega n}^{R} \|^2$ should be half of $\| |1\>_{\omega n}^{(+)} \|^2$. This tension permeates the semiclassical approximation. One way out of it is for the isomorphism between the space $\H_{+}$ of physical states in the semiclassical approximation and the actual Hilbert space $\Hg$ of the geon to rescale the states,
\begin{align} \label{rescaling}
\H_{+} \ni | j \>^{(+)}_{\omega n} \: \longmapsto \: 2^{j/2} | j \>_{\omega n} \in \Hg.
\end{align}

\paragraph{The geon state.} In \cite{Louko:1998hc} the authors advocate for the geon state of the form \eqref{TFDg}. This state, however is unphysical. In order to derive the physical geon state in the semiclassical approximation we simply act with the squeezing operator in \eqref{TFD} on $| 0 \>$. The creation operators $\op{\a}_{\omega n}^{L, R \dagger}$ can be expressed in terms of the creation operators $\op{\a}_{\omega n}^{(+) \dagger}$ and $\op{\a}_{\omega n}^{(-) \dagger}$. The semiclassical geon state reads
\begin{align}
| G_{\text{semi}} \> & = \sqrt{1 - e^{-\beta \omega}} \exp \left[ \frac{1}{2} (-1)^n e^{-\frac{\beta \omega}{2}} \left( \op{\a}_{\omega n}^{(+) \dagger} \op{\a}_{\omega n}^{(+) \dagger} - \op{\a}_{\omega n}^{(-) \dagger} \op{\a}_{\omega n}^{(-) \dagger} \right) \right] | 0 \> \nn\\
& = \sqrt{1 - e^{-\beta \omega}} \sum_{j=0}^{\infty} (-1)^{nj} e^{-\frac{\beta \omega j}{2}} \sqrt{\frac{(2j-1)!!}{(2j)!!}} | 2j_{\omega n} \>_{+} \otimes | 0 \>_{-} + \text{null}.
\end{align}
The omitted portions of the state are null, since they contain at least one insertion of $\op{\a}_{\omega n}^{(-) \dagger}$. The physical portion in $\H_{+}$ is equal to the actual geon state $| G \>$ in \eqref{G}, up to normalization,
\begin{align}
| G_{\text{semi}} \> = \left(1 - e^{-\beta \omega} \right)^{1/4} | G \>_{+} \otimes |0\>_{-}.
\end{align}
As discussed above, the difference in normalization stems from the fact that in the semiclassical approximation all the omitted null states have non-vanishing norms, while in the true, physical system these states are simply absent. Since the Hamiltonian \eqref{Halpha2} effectively projects out the unphysical states, we see that the correct semiclassical analysis reproduces the actual physics of the geon as described in section \ref{sec:geonwh}.

\paragraph{Summary.} In section \ref{sec:geonwh} we showed that the Hilbert space $\H_g$ associated with the geon does not split into the tensor product of its boundary Hilbert spaces, but rather satisfies $\H_g \cong \H_R \cong \op{\Theta} \H_L$. In this subsection we showed how the geon physics can be reconstructed within the semiclassical approximation, where the factorization $\Hsemi = \H_L \otimes \H_R$ is assumed. To do it, we split the states in $\Hsemi = \H_L \otimes \H_R \cong \H_{+} \otimes \H_{-}$ into physical and null states. Physical states are those obeying \eqref{beta_cond}, \textit{i.e.}, the states that in $\H_{+}$, which are spanned by the creation operators $\op{\a}^{(+) \dagger}_{\omega n}$ only. By choosing the Hamiltonian \eqref{Halpha2} we can guarantee that the physical and unphysical states remain decoupled under time evolution. Finally, we can take the state \eqref{TFDg} as advocated in \cite{Louko:1998hc} and realize that the true geon state is obtained by throwing out all unphysical degrees of freedom. In this way we recovered the geon state $| G \>$ as in \eqref{G}, up to normalization.

\subsection{\texorpdfstring{Semiclassical approximation: approach of \cite{Guica:2014dfa}}{Semiclassical approximation: approach of [2]}} \label{sec:geon_semi2}

\paragraph{Summary of \cite{Guica:2014dfa}.} In \cite{Guica:2014dfa} the authors undertook a similar approach to the one discussed above, except that they did not reverse the time direction in the left wedge. They assumed the commutation relations \eqref{comm_rel_1} as they stand, with $\op{\a}_{\omega n}^{L,R}$ designated as annihilation operators. Next, they tried enforcing the field decomposition into \eqref{PhiGeonP} by eliminating states created by $\op{\a}^{(-) \dagger}_{\omega n}$. They looked for physical states $| \psi \> \in \H_L \otimes \H_R$ satisfying the strong condition \eqref{strong_cond}, which now takes form
\begin{align} \label{apm_cond}
& \left[ \op{\a}_{\omega n}^R - (-1)^n \op{\a}_{\omega, -n}^{L \dagger} \right] | \psi \> = \left[ \op{\a}_{\omega n}^L - (-1)^n \op{\a}_{\omega, -n}^{R \dagger} \right] | \psi \> = 0
\end{align}
for all $\omega, n$ on all physical states.

The problem with such an approach is that it leads to a number of divergences and inconsistencies, the first one being that the states obeying the condition \eqref{apm_cond} do not exist in $\H_L \otimes \H_R$. To see it, fix $\omega$ and $n$ and consider a general state 
\begin{align}
| I \>_{\omega n} = \sum_{i,j = 0}^{\infty} c_{ij} |i\>_{\omega, -n} |j\>_{\omega n}
\end{align}
for arbitrary $c_{ij}$. By imposing \eqref{apm_cond} one finds a single `state',
\begin{align}
| I \>_{\omega n} = \sum_{j = 0}^{\infty} (-1)^{jn} |j\>_{\omega, -n} |j\>_{\omega n}.
\end{align}
After \cite{Guica:2014dfa} the state $| I \> = \bigotimes_{\omega, n} | I \>_{\omega, n}$ will be called the \emph{cross-cap state}. This is not a true state since it has infinite norm. Nevertheless, it has the structure of the maximally entangled Bell state. If we cut the sum off at some highly excited level $j = J-1$, the normalized state reads
\begin{align}
| I \>_{\omega n} = \frac{1}{J} \sum_{j = 0}^{J-1} (-1)^{jn} |j\>_{\omega, -n} |j\>_{\omega n}.
\end{align}
The reduced density matrix $\rho_R = \Tr_L | I \> \< I |$ traced over the left Hilbert space is just proportional to the identity matrix, $\rho_R = J^{-1} \bs{1}$. Thus, this is the maximally entangled Bell state with the maximal entropy $S = \log J$. With the philosophy of the `ER=EPR' proposal, \cite{Maldacena:2013xja}, this seemed like a reasonable state with a huge entanglement between the two wedges accounting for the information transfer through the wormhole. 

Although non-normalizable, one can still carry out calculations on $| I \>$, up to overall diverging factors. For example, one can readily calculate the 2-point functions of the operators \eqref{OLR_mom}. Up to an infinite multiplicative constant one finds
\begin{align}
G_{|I\>}^{RR}(\varpi, n) & = \frac{J+1}{2} \times |c_{\varpi n}^{BTZ}|^2 \left[ \left(1 + \frac{2}{J} \right) \bs{1}_{\varpi > 0} + \bs{1}_{\varpi < 0} \right], \label{GIRR} \\
G_{|I\>}^{LR}(\varpi, n) & = \frac{J+1}{2} \times |c_{\varpi n}^{BTZ}|^2 (-1)^n \left(1 + \frac{2}{J} \right). \label{GILR}
\end{align}
Even discarding the divergence, these expressions are problematic. They assume the factorized Hamiltonian, $\op{H}_L + \op{H}_R$, on the tensor product $\Hsemi = \H_L \otimes \H_R$. The problem here is that the cross-cap state $| I \>$ obviously is not the ground state for $\op{H}_L + \op{H}_R$ and thus the use of the Hamiltonian $\op{H}_L + \op{H}_R$ to evolve $\O_L$ and $\O_R$ is unjustified. On the other hand the Hamiltonian, for which $| I \>$ would be the ground state is
\begin{align} \label{shH}
\op{H} = \int_0^{\infty} \frac{\D \omega}{2 \pi} \sum_{n=-\infty}^{\infty} \omega \op{\a}^{(+) \dagger}_{\omega n} \op{\a}^{(+)}_{\omega n}.
\end{align}
However, since now the operators $\op{\a}^{(+)}_{\omega n}$ and $\op{\a}^{(+) \dagger}_{\omega n}$ commute, such a Hamiltonian does not generate any evolution.

Finally, the state dual to the geon, denoted by $| \Psi_g \> = \bigotimes_{\omega, n} | \Psi_g \>_{\omega n}$ in (3.2) in \cite{Guica:2014dfa}, was claimed to be the thermofield double state defined as
\begin{align}
| \Psi_g \>_{\omega n} & = e^{-\frac{\beta}{4} (\op{H}_L + \op{H}_R) } | I \>_{\omega n} \nn\\
& = \sum_{j = 0}^{\infty} (-1)^{jn} e^{-\frac{\beta \omega j}{2}} |j\>_{\omega, -n} |j\>_{\omega n}.
\end{align}
This is nothing but the original thermofield double state \eqref{TFD} with an additional insertion of the factor $(-1)^{jn}$, which can be removed by the redefinition of $\op{\a}^L_{\omega n}$. It does not carry more entanglement than the original TFD state. 

On the other hand, from the point of view of the squeezed states, the geon state should have the form \eqref{TFD} with $| I \>$ replacing $| 0 \>$. Unfortunately, since the cross-cap state is the squeezed state of infinite squeezing, the composition law for the squeezing operators \eqref{opS_comp} implies that $| \Psi_g \> = | I \>$, up to a possible infinite normalization. The prescription fails.

\paragraph{The effective description.} We would like to find a procedure to impose the conditions \eqref{apm_cond} approximately. To do it, let $\nu_{\omega n}$ be a set of complex numbers such that $\nu_{\omega, -n} = \nu_{\omega n}$ and consider the modified condition
\begin{align} \label{nu_cond}
& \left[ \op{\a}_{\omega n}^R - \nu_{\omega n} \op{\a}^{L \dagger}_{\omega, -n}  \right] | \psi_{\nu_n} \> = 0, && \left[ \op{\a}_{\omega n}^{L} - \nu_{\omega n} \op{\a}^{R \dagger}_{\omega, -n}  \right] | \psi_{\nu_n} \> = 0.
\end{align}
We will impose this condition for any $\nu_{\omega n}$ such that $| \nu_{\omega n} | < 1$ and take the limit $\nu_{\omega n} \rightarrow (-1)^n$ afterwards.

To do it, define new creation-annihilation operators $\op{\gamma}^{L \dagger}_{\omega n}, \op{\gamma}^{L}_{\omega n}$ and $\op{\gamma}^{R \dagger}_{\omega n}, \op{\gamma}^{R}_{\omega n}$ related to  $\op{\a}^{L \dagger}_{\omega n}, \op{\a}^{L}_{\omega n}$ and $\op{\a}^{R \dagger}_{\omega n}, \op{\a}^{R}_{\omega n}$ by the Bogoliubov transformation
\begin{align}
\op{\gamma}^{L}_{\omega n} & =  \cosh \lambda_{\omega n} \op{\a}^{L}_{\omega n}- \sinh \lambda_{\omega n} \op{\a}_{\omega,-n}^{R \dagger}, \\
\op{\gamma}^{R}_{\omega n} & =  \cosh \lambda_{\omega n} \op{\a}^{R}_{\omega n} - \sinh \lambda_{\omega n} \op{\a}_{\omega,-n}^{L \dagger},
\end{align}
where $\lambda_{\omega n}$ is a set of real parameters with $\lambda_{\omega, -n} = \lambda_{\omega n}$. The conditions \eqref{nu_cond} become
\begin{align} \label{nu_beta_cond}
\op{\gamma}^{L}_{\omega n} | \psi_{\nu_{\omega n}} \> = \op{\gamma}^{R}_{\omega n} | \psi_{\nu_{\omega n}} \> = 0
\end{align}
where $\nu_{\omega n}$ and $\lambda_{\omega n}$ are related as
\begin{align}
\nu_{\omega n} = \tanh \lambda_{\omega n}.
\end{align}
Thus, the original conditions \eqref{apm_cond} correspond to the limit $\lambda_{\omega n} \rightarrow (-1)^n \times \infty$.

\paragraph{Correlators.} The operators  $\op{\gamma}^{L \dagger}_{\omega n}, \op{\gamma}^{L}_{\omega n}$ and $\op{\gamma}^{R \dagger}_{\omega n}, \op{\gamma}^{R}_{\omega n}$ are genuine creation-annihilation operators satisfying the canonical commutation relations,
\begin{align}
& \left[ \op{\gamma}^{I}_{\omega n}, \op{\gamma}^{I' \dagger}_{\omega' n'} \right] = 2 \pi \delta(\omega-\omega') \delta_{n n'} \delta^{I I'}.
\end{align}
There is only a single state in $\H_L \otimes \H_R$, which satisfies \eqref{nu_beta_cond}: the vacuum state $| I_{\nu_{\omega n}} \>$ annihilated by $\op{\gamma}^{L}_{\omega n}$ and $\op{\gamma}^{R}_{\omega n}$. Nevertheless in the limit $\nu_{\omega n} \rightarrow (-1)^{n}$ we recover relations \eqref{apm_cond}. Thus, we can regard \eqref{apm_cond} as the condition for the vacuum state only, since in the limit $\nu_{\omega n} \rightarrow (-1)^n$ the excited states created by $\op{\gamma}_{\omega n}^{L \dagger}$ and $\op{\gamma}_{\omega n}^{R \dagger}$ coincide.

Let us check now that the procedure leads to the correct 2-point functions for the boundary operators \eqref{OLR}. To do it, we need to use the free Hamiltonian, for which $| I_{\nu_{\omega n}} \>$ is the lowest energy state. We consider
\begin{align} \label{Hgamma}
\op{H}_{\nu} & = \int_0^{\infty} \frac{\D \omega}{2 \pi} \sum_{n=-\infty}^{\infty} \omega \left[ \op{\gamma}^{L \dagger}_{\omega n} \op{\gamma}^{L}_{\omega n} + \op{\gamma}^{R \dagger}_{\omega n} \op{\gamma}^{R}_{\omega n} \right],
\end{align}
which is positive-definite and annihilates $| I_{\nu_{\omega n}} \>$. Now 2-point functions of the operators $\O_{L,R}$ evolved by $\op{H}_{\nu}$ and evaluated on $| I_{\nu_{\omega n}} \>$ can be calculated. By employing the definitions \eqref{GdefRR} and \eqref{GdefLR} we find
\begin{align}
G^{RR}_{| I_{\nu_{\omega n}} \>}(\varpi, n) & = | c_{\varpi n}^{BTZ} |^2 \left[ \cosh^2 \lambda_{\varpi n} \bs{1}_{\varpi > 0} + \sinh^2 \lambda_{-\varpi, n} \bs{1}_{\varpi < 0} \right], \\
G^{LR}_{| I_{\nu_{\omega n}} \>}(\varpi, n) & = \frac{1}{2} | c_{\varpi n}^{BTZ} |^2 \sinh(2 \lambda_{|\varpi| n}).
\end{align}
As $|\nu_{\omega n}|$ approaches $1$, the parameter $\lambda_{\omega n}$ diverges. To take the limit we define
\begin{align} \label{nun}
\nu_{\omega n} = (-1)^n \sqrt{1 - \delta}
\end{align}
so that $1 - \nu_{\omega n}^2 = \delta \ll 1$ is the small parameter. We find
\begin{align}
G^{RR}_{|I\>}(\varpi, n) & = \left( \frac{1}{\delta} + O(\delta^0) \right) | c_{\varpi n}^{BTZ} |^2, \\
G^{LR}_{|I\>}(\varpi, n) & = \left( \frac{1}{\delta} + O(\delta^0) \right) (-1)^n | c_{\varpi n}^{BTZ} |^2.
\end{align}
The overall constant is obviously regularization-dependent. Up to regularization, we recovered \eqref{GIRR} and \eqref{GILR}.

\paragraph{Effective interactions.} It is often claimed that holographically the wormholes can be treated as two copies of the boundary system with a cross-boundary coupling. As we have seen this cannot be true in the literal sense, as the Hilbert space associated with the wormhole is not the tensor product of the boundary Hilbert spaces. However, it can be true in semiclassical approximation in the infinite squeezing limit. Indeed, the Bogoliubov transformation between the creation-annihilation operators $\op{\a}_{\omega n}^{L, R \dagger}, \op{\a}_{\omega n}^{L, R}$ and $\op{\gamma}_{\omega n}^{L, R \dagger}, \op{\gamma}_{\omega n}^{L, R}$ can be interpreted as the introduction of the Hamiltonian quadratic in the creation-annihilation operators. This is reminiscent of the textbook Bogoliubov theory of liquid helium, where the excitations in the superfluid state are decribed by the Bogolibov-transformed creation-annihilation operators. This idea was also applied to black holes, \cite{Hochberg:1991tz}, where the excitations on top of the Kruskal vacuum can be obtained by introducing a suitable Hamiltonian between the Scharzschild creation-annihilation operators. 

Using Bogoliubov transformations we can express the Hamiltonian \eqref{Hgamma} in terms of the original, semiclassical operators $\op{\a}_{\omega n}^{L, R \dagger}, \op{\a}_{\omega n}^{L, R}$. With $\nu_{\omega n} = \nu_{|n|}$ depending on $|n|$ only
\begin{align}
\sinh^2 \lambda_{\omega n} & = \frac{\nu_{|n|}^2}{1 - \nu_{|n|}^2}, &
\cosh(2 \lambda_{\omega n}) & = \frac{1 + \nu_{|n|}^2}{1 - \nu_{|n|}^2}, &
\sinh(2 \lambda_{\omega n}) & = \frac{2 \nu_{|n|}}{1 - \nu_{|n|}^2},
\end{align}
and the Hamiltonian reads
\begin{align}
\op{H}_{\nu} & = \frac{1}{1 - \nu_{|n|}^2} \int_0^{\infty} \frac{\D \omega}{2 \pi} \sum_{n=-\infty}^{\infty} \omega \left[ 2 \nu_{|n|}^2 + (1 + \nu_{|n|}^2) \left( \op{\a}_{\omega n}^{L \dagger} \op{\a}_{\omega n}^{L} + \op{\a}_{\omega n}^{R \dagger} \op{\a}_{\omega n}^R \right) \right.\nn\\
& \qquad\qquad \left. - \, 2 \nu_{|n|} \left( \op{\a}_{\omega, -n}^{L \dagger} \op{\a}_{\omega n}^{R \dagger} + \op{\a}_{\omega, -n}^L \op{\a}_{\omega n}^R \right) \right].
\end{align}
As $\nu_{|n|} \rightarrow (-1)^n$ the divergence factorizes in front of the Hamiltonian. Substituting \eqref{nun} we can write
\begin{align}
\op{H}_{\nu} & = \frac{2}{\delta} \left[ V_0 + \op{H}_0 + \op{H}_{\text{int}} \right], 
\end{align}
where $V_0 = \omega$ is a constant that can be discarded, $\op{H}_0$ is the free Hamiltonian in $\H_L \otimes \H_R$ and $\op{H}_{\text{int}}$ is the interaction Hamiltonian,
\begin{align}
\op{H}_0 & = \int_0^{\infty} \frac{\D \omega}{2 \pi} \sum_{n=-\infty}^{\infty} \omega \left( \op{\a}_{\omega n}^{L \dagger} \op{\a}_{\omega n}^{L} + \op{\a}_{\omega n}^{R \dagger} \op{\a}_{\omega n}^R \right), \\
\op{H}_{\text{int}} & = - \int_0^{\infty} \frac{\D \omega}{2 \pi} \sum_{n=-\infty}^{\infty} \omega (-1)^n \left( \op{\a}_{\omega, -n}^{L \dagger} \op{\a}_{\omega n}^{R \dagger} + \op{\a}_{\omega, -n}^L \op{\a}_{\omega n}^R \right). \label{Hint}
\end{align}
We see that in the semiclassical approximation we can model the wormhole by introducing the interaction between the two boundaries. The interaction Hamiltonian $\op{H}_{\text{int}}$ has the beautiful interpretation of coupling the modes created in the right wedge with $\op{\a}_{\omega n}^{R \dagger}$, with the antipodally identified modes created by $(-1)^n \op{\a}_{\omega, -n}^{L \dagger} = \op{\Theta} \op{\a}_{\omega, -n}^{R \dagger} \op{\Theta}$. Furthermore, the interaction is `infinitely strong' effectively projecting out null states out of the tensor product.

\section{\texorpdfstring{The AdS$_2$ wormhole}{The AdS2 wormhole}} \label{sec:jt}

\begin{figure}[ht]
\centering
\begin{subfigure}[t]{0.4\textwidth}
\begin{tikzpicture}[scale=1.3]
\draw[dotted, colorL, fill=colorL, opacity=0.3] (-1,1) -- (0,0) -- (-1,-1) -- cycle;
\draw[dotted, colorR, fill=colorR, opacity=0.3] (1,1) -- (0,0) -- (1,-1) -- cycle;
\draw[thick] (-1,3.2) -- (-1,-3.2);
\draw[thick] ( 1,3.2) -- ( 1,-3.2);
\draw (0.9,3.1) -- (1,3) -- (-1,1) -- (1,-1) -- (-1,-3) -- (-0.9,-3.1);
\draw (-0.9,3.1) -- (-1,3) -- (1,1) -- (-1,-1) -- (1,-3) -- (0.9,-3.1);
\draw[thick,colorL] (-1,0) -- (0,0);
\draw[thick,colorR] (1,0) -- (0,0);
\node[below] at (1,-3.2) {$\frac{\pi}{2}$};
\node[below] at (-1,-3.2) {$\theta = -\frac{\pi}{2}$};
\draw[dotted] (0,3.1) -- (0,-3.1);
\node[below] at (0,-3.2) {$0$};
\node[colorR!50!black, above] at (0.5,0) {$\Sigma'_R$};
\node[colorL!50!black, above] at (-0.5,0) {$\Sigma'_L$};
\node[right] at (1,0) {$\Sigma_0$};
\draw[thick, colorR] (1,1) -- (1,-1);
\draw[thick, colorL] (-1,1) -- (-1,-1);
\end{tikzpicture}
\centering
\caption{Two independent Rindler wedges and their associated Cauchy surfaces, $\Sigma'_L$ and $\Sigma'_R$, which together form the complete Cauchy surface $\Sigma_0$. By specifying initial data on $\Sigma_0$ the dynamics determines the field in the entirety of the spacetime. Equivalently, one can specify independent boundary data on the orange and blue segments of the boundary. Then the bulk field is uniquely determined as well.\label{fig:AdS2left}}
\end{subfigure}
\qquad\qquad
\begin{subfigure}[t]{0.4\textwidth}
\centering
\includegraphics[width=0.5\textwidth]{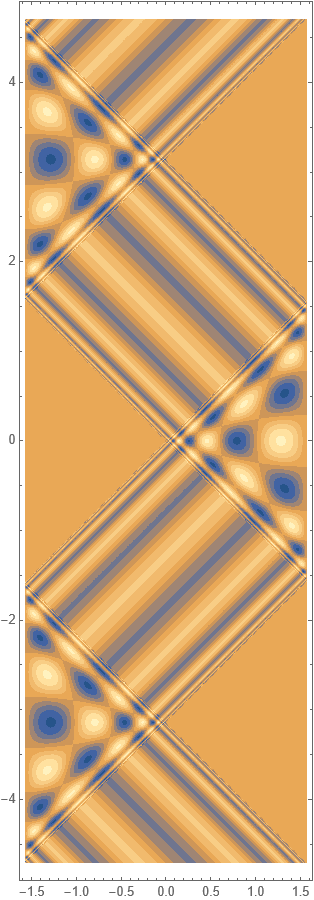}
\caption{A graphical representation of a wave in AdS$_2$ background. The field vanishes in the left wedge (the orange regions) and equals to the Schwarzschild mode $\phi_{\omega}$ in \eqref{phi1} in the right wedge for $\omega = 1.8$. Furthermore we set $L = 3$ and $\p = 1$. The blue regions denote the positive values of the bulk field, while yellow denote negative values. \label{fig:AdS2right}}
\end{subfigure}
\caption{The JT wormhole is the AdS$_2$ spacetime: an infinitely long strip with the metric \eqref{gAdS2} and two asymptotic boundaries (left and right).\label{fig:AdS2}}
\end{figure}
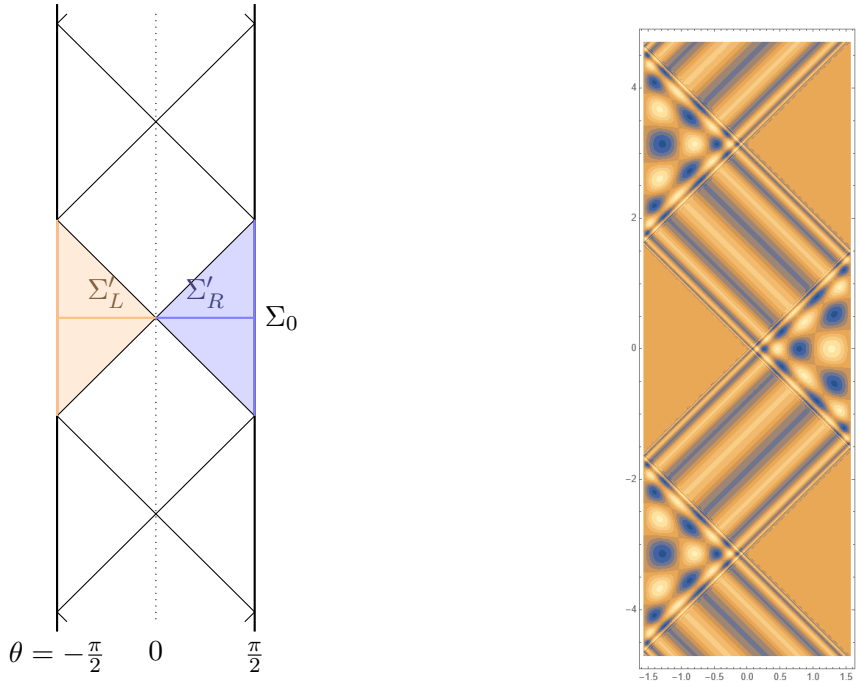

We consider the Jackiw–Teitelboim (JT) action with matter,
\begin{align}
S = \int \D^2 x \sqrt{-g} \left[  \Phi \left( R + \frac{2}{L^2} \right) + \frac{1}{2} \partial_\mu X \partial^\mu X + \frac{1}{2} m_{\Delta}^2 X^2 \right],
\end{align}
where the mass is $L^2 m_{\Delta}^2 = \Delta(\Delta - 1)$. With the dilaton $\Phi$ serving as the Lagrange multiplier, it fixes the background to satisfy $R = -2/L^2$, which corresponds locally to AdS$_2$ spacetime.

The global two-dimensional AdS spacetime, \textit{i.e.}, the JT wormhole, has the metric
\begin{align} \label{gAdS2}
\D s^2 = \frac{L^2}{\cos^2 \theta} ( - \D \tau^2 + \D \theta^2).
\end{align}
Here the radial variable $\theta \in ( -\tfrac{\pi}{2}, \frac{\pi}{2} )$ and, unlike in higher dimensions, the spacetime has two boundaries, at $\theta = -\tfrac{\pi}{2}, \frac{\pi}{2}$. With $\tau \in \R$, the spacetime is a wormhole, as the null rays cross the bulk from one boundary to the opposite one in time equal to $\pi$. The Penrose diagram of the JT wormhole is presented in figure \ref{fig:AdS2left}. Typically, the discussion of the JT wormhole begins with the pair of Schwarzschild wedges as presented in the figure. It is then believed, \cite{Maldacena:2018lmt}, that a suitable coupling between the two boundaries can open and sustain the wormhole. 

In this paper we assume that the wormhole has already been opened and the two boundaries are in causal contact for eternity. Since we consider a scalar field on the fixed background, the boundary data on the orange $\Sigma'_L$ and blue $\Sigma'_R$ portions of the boundary in figure \ref{fig:AdS2left} are independent. The total Hilbert space does factorize into the tensor product of the Hilbert spaces associated with the two subregions of the boundaries, $\H \cong \H_{L'} \otimes \H_{R'}$. 

On the other hand, when the full boundaries are considered, the factorization fails. The left and right boundary Hilbert spaces $\H_L$ and $\H_R$ will be associated with the whole boundaries. As the boundaries are in the causal contact for eternity, $\H \cong \H_L \cong \H_R$, as we will discuss now. This was already observed in \cite{Arias:2010xg}.

\subsection{Matter field solution}

With the background fixed, we consider the matter field $X$ in the fixed AdS$_{2}$ background \eqref{gAdS2}. The Klein-Gordon equation reads
\begin{align} \label{KG_JT}
\left[ - \frac{\partial^2}{\partial \tau^2} + \frac{\partial^2}{\partial \theta^2} - \frac{\Delta(\Delta - 1)}{\cos^2 \theta} \right] X(\tau, \theta) = 0.
\end{align}
The negative frequency solution with respect to the global time can be expressed in terms of the hypergeometric function as
\begin{align} \label{chik}
\chi_{n}(\tau, \theta) & = c_{n} e^{-\I \omega_n \tau} \cos^{\Delta} \theta \, {}_2 F_1 \left( \frac{\Delta - \omega_n}{2}, \frac{\Delta + \omega_n}{2}; \Delta + \frac{1}{2}; \cos^2 \theta \right).
\end{align}
This solution exhibits the correct asymptotics, $\chi_n \sim \cos^{\Delta} \theta$ at both boundaries at $\theta = \pm \tfrac{\pi}{2}$. Smoothness at $\theta = 0$ results in the quantization of frequencies,
\begin{align}
\omega_n = \Delta + n, \quad n = 0,1,2,\ldots
\end{align}
It will be more convenient to split the modes into functions manifestly odd and even under the parity map
\begin{align} \label{parity}
p(\tau, \theta) = (\tau, -\theta).
\end{align}
The modes can written as
\begin{align}
\chi_{2k}(\tau, \theta) & = c_{2k} e^{-\I (\Delta + 2 k) \tau} \cos^{\Delta} \theta \, P_k^{(-\tfrac{1}{2}, \Delta - \tfrac{1}{2})}(\cos (2 \theta)), \label{chiP} \\
\chi_{2k+1}(\tau, \theta) & = c_{2k + 1} e^{-\I (\Delta + 2 k + 1) \tau} \cos^{\Delta} \theta \, \sin \theta \, P_k^{(\tfrac{1}{2}, \Delta - \tfrac{1}{2})}(\cos (2 \theta)) \label{chiM}
\end{align}
where $\chi_{2k}$ are even while $\chi_{2k+1}$ are odd under $p$. Here $k=0,1,2,\ldots$ and $P^{(\alpha \beta)}_k$ denotes Jacobi polynomials. The normalization constants
\begin{align} \label{cAdS2}
c_{2k} & = \frac{(-1)^k}{\sqrt{2}} \sqrt{\frac{k! \, \Gamma(\Delta + k)}{\Gamma(k + \tfrac{1}{2}) \Gamma(\Delta + k + \tfrac{1}{2})}}, & c_{2k+1} & = \frac{(-1)^k}{\sqrt{2}} \sqrt{\frac{k! \, \Gamma(\Delta + k + 1)}{\Gamma(k + \tfrac{3}{2}) \Gamma(\Delta + k + \tfrac{1}{2})}}
\end{align}
are such that the modes are normalized to $1$ in the Klein-Gordon scalar product \eqref{KG}. 

The quantum field operator reads
\begin{align} \label{Chi}
\op{X}(\tau, \theta) & = \sum_{n=0}^{\infty} \left( \chi_{n} \op{a}_{n} + \chi^{\ast}_{n} \op{a}^{\dagger}_{n} \right),
\end{align}
where the creation-annihilation operators satisfy the canonical commutation relations
\begin{align}
\left[ \op{a}_n, \op{a}_{n'}^{\dagger} \right] = (\chi_n, \chi_{n'}) = \delta_{n n'}.
\end{align}
The vacuum state $| \Omega \>$ satisfies $\op{a}_n | \Omega \> = 0$ for all $n$ and the 1-particle states $\Hone$ and the total Hilbert space $\H$ is constructed in the usual way. The normally-ordered Hamiltonian of the bulk theory reads
\begin{align} \label{JT_Ham}
\op{H} = \sum_{n=0}^{\infty} \omega_n \op{a}_n^{\dagger} \op{a}_n, \quad \omega_n = \Delta + n.
\end{align}
Since the metric \eqref{gAdS2} is time-independent, the vacuum state $| \Omega \>$ is preserved by the evolution.

\subsection{Going to the boundary}

AdS$_{2}$ possesses two asymptotic boundaries, we can take both left and right boundary limits of the modes \eqref{chiP} and \eqref{chiM}. Being right boundary-centric we define the boundary modes $\varphi_n$ as
\begin{align} \label{varphi_JT}
\varphi_{n} & = \lim_{\theta \rightarrow \frac{\pi}{2}} \cos^{-\Delta} \theta \chi_{n} = \tilde{c}_{n} e^{-\I (\Delta + n) \tau}.
\end{align}
The constants $\tilde{c}_{n}$ follow from \eqref{cAdS2} and read
\begin{align}
\tilde{c}_{n} & = \frac{1}{2^{\Delta} \Gamma(\Delta + \tfrac{1}{2})} \sqrt{\frac{\Gamma(2 \Delta + n)}{n!}}.
\end{align}
We see that the boundary modes are simply $\varphi_n \sim e^{-\I (\Delta + n) \tau}$ for $n=0,1,2,\ldots$ Unlike in higher dimensional cases, see equation \eqref{omegaQ}, the integer $n$ here can be both even and odd. From the boundary perspective this corresponds to the fact that the boundary theory carries the representation of the diffeomorphism group, \cite{Cadoni:1999ja}. A boundary operator $\O$ together with all its descendants $\partial_{\tau}^n \O$  furnishes a representation of this group. Unlike for conformal symmetry, all integral $n$'s are allowed, both even and odd ones.

We can define the boundary values $X_{L,R}$ of the bulk field $X$ by taking the boundary limits,
\begin{align}
X_R & = \lim_{\theta \rightarrow \frac{\pi}{2}} \left[ \cos^{-\Delta} \theta \, X \right] = \sum_{n=0}^{\infty} \left[ \varphi_{n} a_{n} + \varphi^{\ast}_{n} a^{\ast}_{n} \right], \label{XR} \\
X_L & = \lim_{\theta \rightarrow -\frac{\pi}{2}} \left[ \cos^{-\Delta} \theta \, X \right] = \sum_{n=0}^{\infty} \left[ (-1)^n \varphi_{n} a_{n} + (-1)^n \varphi^{\ast}_{n} a^{\ast}_{n} \right]. \label{XL}
\end{align}
The two boundaries are in causal contact for eternity or, in other words, the wormhole is open for all times. Unlike in the BTZ black hole case, the bulk field $X$ is uniquely specified by prescribing the boundary value of the field on a single boundary. The boundary value on the opposite boundary is then uniquely fixed. Indeed, given the boundary value $X_R$ of the field on the right boundary, the expressions above imply that
\begin{align}
X_R (\tau + k \pi ) = e^{-\I k \pi \Delta} X_L(\tau), \quad k = \pm 1, \pm 3, \pm 5, \ldots
\end{align}
for any odd integer $k$. In particular, the boundary values on a single boundary are periodic with the period of $2\pi$,
\begin{align} \label{period}
X_R(\tau + 2\pi) = e^{-2 \pi \I \Delta} X_R(\tau).
\end{align}
The periodic structure of a bulk solution can be seen in figure \ref{fig:AdS2right}. This means that all modes are wormhole modes propagating between the boundaries. 

\paragraph{Boundary Hilbert spaces and the operators.} We define the boundary operators in terms of the boundary modes \eqref{varphi_JT} and with their own sets of creation-annihilation operators $\op{a}^{L,R \dagger}_n, \op{a}^{L,R}_n$,
\begin{align}
\op{X}_{L,R} & = \sum_{n=0}^{\infty} \left[ \varphi_{n} \op{a}^{L,R}_{n} + \varphi^{\ast}_{n} \op{a}^{L,R \dagger}_{n} \right], \label{qOLR}
\end{align}
which span two Hilbert spaces $\H_L$ and $\H_R$. On the other hand from \eqref{XL} and \eqref{XR} we must identify 
\begin{align} \label{a_rel}
& \op{a}_n = \op{a}^R_n = (-1)^n \op{a}^L_n, && \op{a}^{\dagger}_n = \op{a}^{R \dagger}_n = (-1)^n \op{a}^{L \dagger}_n.
\end{align}
As the two boundaries are in causal contact for eternity, their operators and Hilbert spaces are related to each other. Let $\op{P}$ denote the operator realizing the parity operation \eqref{parity} on the quantum level, \textit{i.e.}, 
\begin{align}
& \op{P} \op{X}(\tau, \theta) \op{P} = \op{X}(\tau, -\theta), && \op{P} \op{a}_n \op{P} = (-1)^n \op{a}_n. 
\end{align}
In particular $\op{P} \op{X}_R \op{P} = \op{X}_L$ and thus we identify 
\begin{align} \label{HLR_iso}
\H \cong \H_R \cong \op{P} \H_L.
\end{align}

\paragraph{Boundary correlation functions.} In \cite{Maldacena:2018lmt} a cross-boundary correlator $\< \O_L \O_R \>$ in the JT wormhole was calculated by introducing a coupling between the two dual theories living on the two boundaries. We can now easily check that the correlation functions reported there are reproduced by our bulk theory. To do it, we replace $\O_R$ and $\O_L$ by $\op{X}_R$ and $\op{X}_L$ and carry out the calculations in the full theory. Indeed, for $\tau_R > \tau'_R > \tau_R - 2 \pi$ the Wightman function follows from \eqref{XR} and \eqref{XL},
\begin{align} \label{2pt_JT}
G_{RR}(\tau_R - \tau'_R) & = \< \Omega | \op{X}_R(\tau_R) \op{X}_R(\tau'_R) | \Omega \> = \frac{\Gamma(\Delta)}{2^{2\Delta + 1} \sqrt{\pi} \Gamma(\Delta + \tfrac{1}{2})} \frac{e^{-\I \pi \Delta}}{\sin^{2 \Delta} \left( \frac{\tau_R - \tau'_R}{2} \right)}.
\end{align}
This is indeed the correct boundary 2-point function, \cite{Maldacena:2018lmt}. Identical result holds for $G_{LL}(\tau_L - \tau'_L) = \< \Omega | \op{X}_L(\tau_L) \op{X}_L(\tau'_L) | \Omega \>$.

We can also calculate the cross-boundary correlator. For $\tau_R + \pi > \tau_L > \tau_R - \pi$ we have
\begin{align}
G_{LR}(\tau_R - \tau_L) & = \< \Omega | \op{X}_L(\tau_L) \op{X}_R(\tau_R) | \Omega \> = \frac{\Gamma(\Delta)}{2^{2\Delta + 1} \sqrt{\pi} \Gamma(\Delta + \tfrac{1}{2})} \frac{1}{\cos^{2 \Delta} \left( \frac{\tau_R - \tau_L}{2} \right)}. \label{2ptLR}
\end{align}
We have derived the correct cross-boundary correlator of \cite{Maldacena:2018lmt} in the free field theory! At no point any interaction between the two boundaries was introduced.

Let us conclude this section by yet another check. By $G_{\Delta}(x, x')$ denote the bulk-to-bulk propagator satisfying
\begin{align}
( - \Box + m^2 ) G_{\Delta}(x, x') = \frac{1}{\sqrt{-g}} \delta(x - x').
\end{align}
In AdS the propagator depends only on the invariant distance $\xi$ between the two points,
\begin{align}
\xi = \frac{\cos \theta \, \cos \theta'}{\cos (\tau - \tau') - \sin \theta \, \sin \theta'}.
\end{align}
The propagator reads
\begin{align}
G_{\Delta}(\xi) = \frac{\Gamma(\Delta)}{2^{\Delta + 1} \sqrt{\pi} \Gamma(\Delta + \tfrac{1}{2})} \, \xi^{\Delta} \, {}_2 F_1 \left( \frac{\Delta}{2}, \frac{\Delta+1}{2}; \Delta + \frac{1}{2}; \xi^2 \right).
\end{align}
By taking the suitable boundary limits we can recalculate the 2-point functions both within a single boundary as well as between the two boundaries. It is easy to verify that
\begin{align}
G_{RR}(\tau_R - \tau'_R) & = \lim_{\theta \rightarrow \tfrac{\pi}{2}} \lim_{\theta' \rightarrow \tfrac{\pi}{2}} \cos^{-\Delta} \theta \cos^{-\Delta} \theta' \, G_{\Delta}(\xi), \\
G_{LR}(\tau_R - \tau_L) & = \lim_{\theta \rightarrow -\tfrac{\pi}{2}} \lim_{\theta' \rightarrow \tfrac{\pi}{2}} \cos^{-\Delta} \theta \cos^{-\Delta} \theta' \, G_{\Delta}(\xi).
\end{align}
Once again, the cross-boundary correlators are obtained from the free theory.

\paragraph{Boundary Hamiltonians.} In the course of the discussion we will also need the boundary Hamiltonians $\op{H}_L$ and $\op{H}_R$. With the total Hamiltonian \eqref{JT_Ham} the substitutions \eqref{a_rel} lead to
\begin{align} \label{jtHLR}
\op{H}_{L,R} & = \sum_{n=0}^{\infty} \omega_n \op{a}_n^{L,R \dagger} \op{a}_n^{L,R}, \quad \omega_n = \Delta + n.
\end{align}
Just as in case of the geon-wormhole, \eqref{HLR}, the two Hamiltonians are in fact identical and the relations \eqref{H_comb} hold. In particular the total Hamiltonian on the wormhole modes is the average rather than a sum over each boundary,
\begin{align} \label{HasAve}
\op{H} = \frac{1}{2} \left( \op{H}_L + \op{H}_R \right).
\end{align}
These Hamiltonians correctly evolve the boundary states, \textit{i.e.}, on the level of boundary operators,
\begin{align}
\op{X}_{L,R}(\tau) = e^{\I \tau \op{H}_{L,R}} \op{X}_{L,R}(0) e^{-\I \tau \op{H}_{L,R}}.
\end{align}
And just as in case of the geon-wormhole the evolution by time $\tau = \pi$ is given by the conjugation by $\op{P}$,
\begin{align}
\op{X}(\tau + \pi) = e^{\I \pi \op{H}} \op{X}(\tau) e^{-\I \pi \op{H}} = e^{-\I \pi \Delta} \op{P} \op{X}(\tau) \op{P}.
\end{align}

\subsection{The wormhole state} \label{sec:wh_state}

Consider the two wedges, orange and blue, presented in figure \ref{fig:AdS2left}. In the context of JT holography one usually thinks about the dual theories being associated with the blue and orange segments of the boundary. As far as the bulk matter is concerned, independent boundary data can be specified on the two segments, similarly to the case of the BTZ black hole. 

We can change coordinates from global AdS to Schwarzschild coordinates, which cover the two wedges. By the appropriate substitution, see appendix \ref{app:RAdS2}, one can bring the metric \eqref{gAdS2} into the Schwarzschild form
\begin{align}
\D s^2 = - (\rho^2 - \p^2) \D t^2 + \frac{L^2 \D \rho^2}{\rho^2 - \p^2}.
\end{align}
This metric covers a single wedge, say the blue wedge in figure \ref{fig:AdS2left}. The bulk field can be decomposed into modes as
\begin{align}
X(t, \rho) = \int_0^{\infty} \frac{\D \omega}{2 \pi} \left[ \alpha^L_{\omega} \phi^L_{\omega} + \alpha^{L \ast}_{\omega} \phi^{L \ast}_{\omega} + \alpha^R_{\omega} \phi^R_{\omega} + \alpha^{R \ast}_{\omega} \phi^{R \ast}_{\omega} \right],
\end{align}
where the negative frequency modes $\phi_{\omega}^{L,R}$ with respect to $t$ are given in \eqref{phi_omega_AdS2}. This is a complete basis of states and their continuation to the entire AdS$_2$ strip can be carried out.

The vacuum state $| 0 \>$ with respect to the annihilation operators $\alpha^L_{\omega}$ and $\alpha^R_{\omega}$ is an element of $\H$. As we will show now, from the point of view of the Schwarzschild wedges, the ground state $| \Omega \>$ is the thermofield double state at the inverse temperature $\beta = 2 L \pi / \p$. We will carry out the analysis for the case of $\Delta = 1$. In this case the Klein-Gordon equation \eqref{KG_JT} simplifies to the simple massless wave equation and the modes \eqref{chiP} and \eqref{chiM} can be collectively written as
\begin{align} \label{chi1}
\chi^{[1]}_{m}(\tau, \theta) & = \frac{1}{\sqrt{\pi m}} e^{-\I m \tau} \sin \left[ m \left( \theta - \tfrac{\pi}{2} \right) \right],
\end{align}
where $m=1,2,3,\ldots$. This matches \eqref{chiP} and \eqref{chiM} with $m = n+1$ and up to the factor of $(-1)^{(n+1)/2}$ when $n$ is odd and $-(-1)^{n/2}$ when $n$ is even, which we dropped. The $\theta$-dependence of these modes describes standing waves stretching between the two boundaries.

The simplification also occurs for the Schwarzschild modes \eqref{phi_omega_AdS2}. Let $\phi^{[1]}_{\omega}$ denote the Schwarzschild mode $\phi^R_{\omega}$ for $\Delta = 1$ in the right wedge and zero in the left wedge. In the right wedge \eqref{phi_omega_AdS2} simplifies to
\begin{align} \label{phi1}
\phi^{[1]}_{\omega} = \frac{1}{\sqrt{2 \omega}} e^{- \I \omega t} \sin \left[ \frac{L \omega}{2 \p} \log \left( \frac{\rho + \p}{\rho - \p} \right) \right].
\end{align}
In figure \ref{fig:AdS2right} we present the plot of $\phi^{[1]}_{\omega}$ for $\omega = 1.8$.

Now it is possible to calculate the Bogoliubov coefficients between $\phi^{[1]}_{\omega}$ and the global modes $\chi^{[1]}_m$. The integrals require proper regularization, as they should be understood in the distributional sense, see appendix \ref{sec:bogo} for details. After the integration is carried out one finds the following Bogolibov coefficients,
\begin{align}
\mu^R_{\omega m}  = ( \phi_{\omega}^{[1]}, \chi_m^{[1]} ) & = - \I^m \sqrt{\frac{2 \pi m}{\omega}} \frac{e^{\frac{\beta \omega}{4}}}{\sinh \left( \frac{\beta \omega}{2} \right)} \, {}_3 F_2 \left( -m, m, \tfrac{1}{2} + \hat{\omega}; \tfrac{1}{2}, 1; 1 \right), \label{bogoA} \\
- \nu^R_{\omega m}  = ( \phi_{\omega}^{[1]}, (\chi_m^{[1]})^{\ast} ) & = e^{-\frac{\beta \omega}{2}} \mu^R_{\omega m}, \label{bogoB}
\end{align}
where
\begin{align} \label{tempR}
\beta = \frac{2 \pi L}{\p}
\end{align}
is the inverse temperature of the Rindler horizon of the two wedges. The details of the calculations are presented in appendix \eqref{sec:bogo}. 

\begin{figure}[ht]
\begin{subfigure}[t]{0.4\textwidth}
\includegraphics[width=1.0\textwidth]{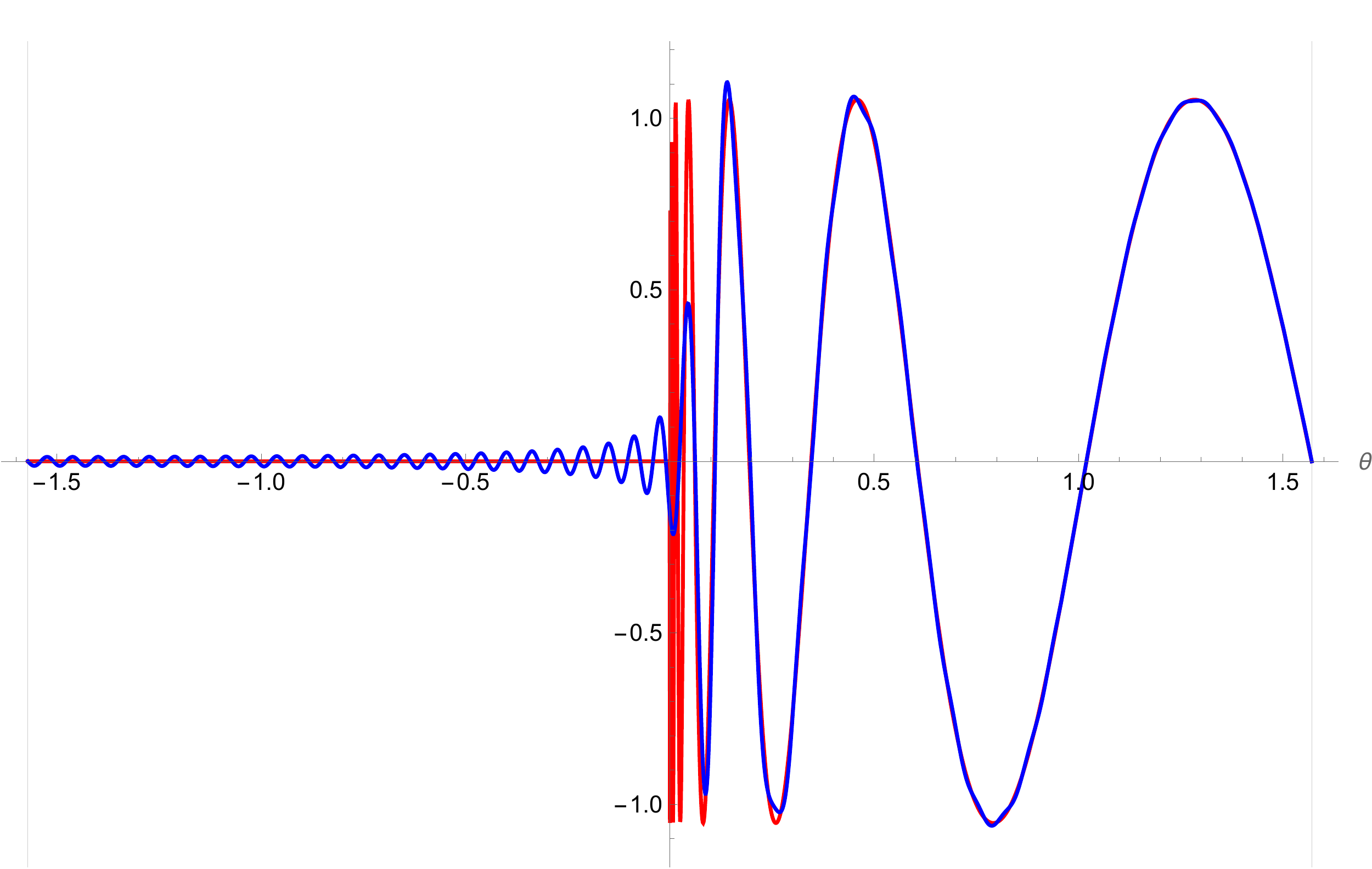}
\centering
\caption{The modes evaluated at the Cauchy surface $\tau = 0$.\label{fig:approx}}
\end{subfigure}
\qquad\qquad
\begin{subfigure}[t]{0.4\textwidth}
\includegraphics[width=1.0\textwidth]{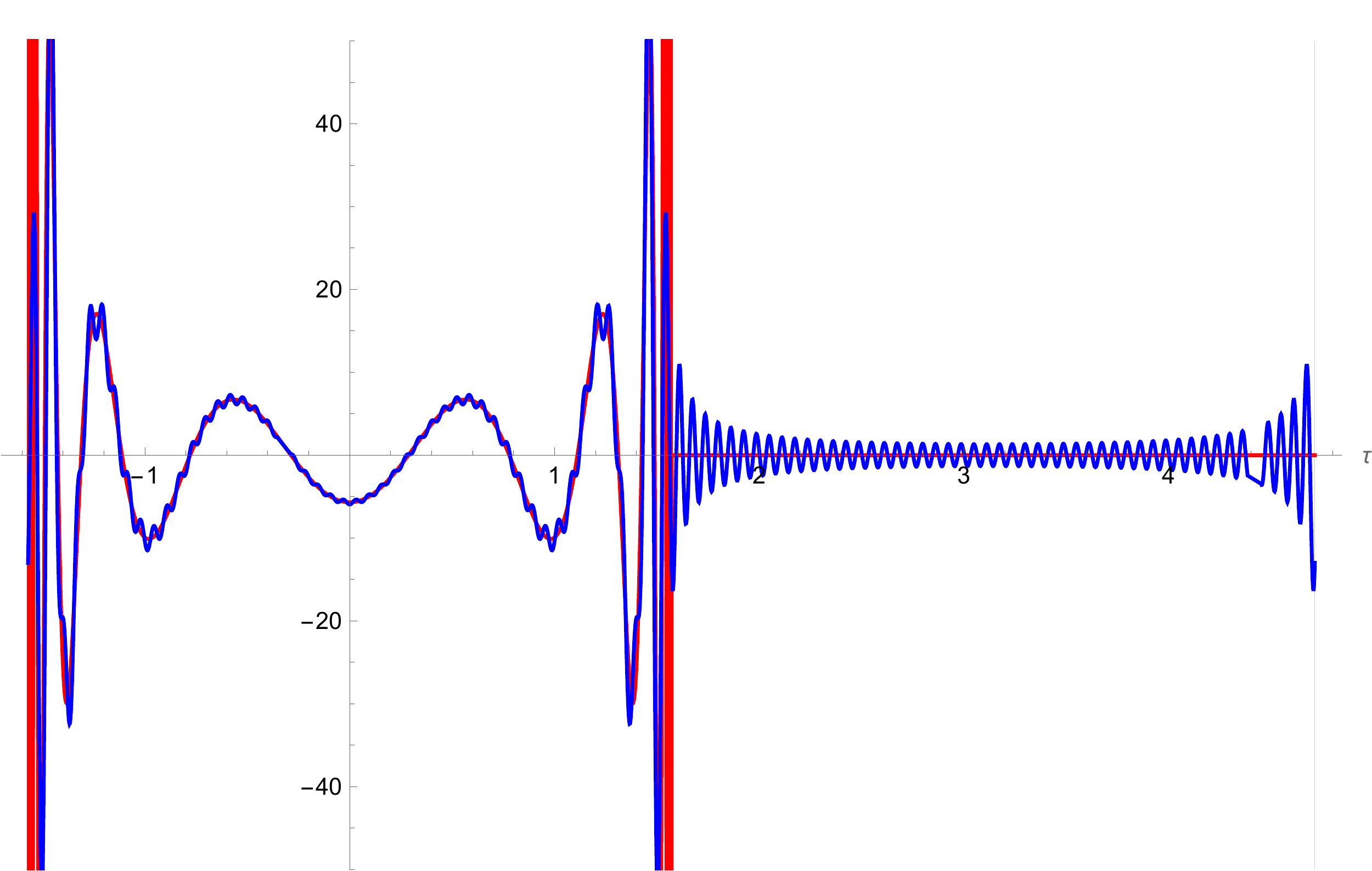}
\centering
\caption{The boundary values of the modes evaluated at the right boundary at $\theta = \pi/2$. The global time $\tau$ ranges from $-\pi/2$ to $3 \pi/2$. Outside this region the picture repeats periodically. \label{fig:approx_bnd}}
\end{subfigure}
\centering
\caption{Comparison of the Schwarzschild mode $\phi^{[1]}_{\omega}$ in \eqref{phi1} (in red) and the composition of 100 global modes $\chi^{[1]}_{m}$ in \eqref{chi1} with the appropriate Bogoliubov coefficients \eqref{bogoA} and \eqref{bogoB} (in blue). We set $L = 3$, $\p = 1$ and $\omega = 1.8$.\label{fig:plots}}
\end{figure}

The relation \eqref{bogoB} between the Bogoliubov coefficients implies that the wormhole state $| \Omega \>$ is \emph{exactly} the thermofield double state from the point of view of the Schwarzschild wedges. Indeed, the same Bogoliubov coefficients follow for the left wedge. Thus, \eqref{bogoB} implies that $| \Omega \>$ is annihilated by the combinations $\op{\a}^R_{\omega} - e^{-\beta \omega/2} \op{\a}^{L \dagger}_{\omega}$ and $\op{\a}^L_{\omega} - e^{-\beta \omega/2} \op{\a}^{R \dagger}_{\omega}$. This, by definition, means that $| \Omega \>$ is the thermofield double state. In particular the expectation value of the Schwarzschild number operator $\op{N}^R_{\omega} = \op{\alpha}^{R \dagger}_{\omega} \op{\alpha}^{R}_{\omega}$ reads
\begin{align} \label{NR}
\< \Omega | \op{N}^R_{\omega} | \Omega \> = \sum_{m=1}^{\infty} |\nu_{\omega m}|^2 = \frac{\pi}{e^{\beta \omega} - 1} \times \delta(0).
\end{align}

In \cite{Maldacena:2018lmt} it was argued that the wormhole state $| \Omega \>$ must be very similar to the thermofield double state. As we can see here, $| \Omega \>$ \emph{is} the thermofield double. The difference between two disconnected Rindler wedges in figure \ref{fig:AdS2left} and the full wormhole lies in the structure of the Hilbert space $\H$ the states live in. For two disconnected boundaries independent boundary data can be specified on both boundaries. In case of the wormhole only periodic boundary data obeying the relation \eqref{period} are allowed.

\subsection{Semiclassical approximation} \label{sec:jt_semi}

In the semiclassical approach one treats $\H_L$ and $\H_R$ as two independent Hilbert spaces spanned by their own sets of left and right creation-annihilation operators $\op{\a}_n^{L \dagger}, \op{\a}_n^{L}$ and $\op{\a}_n^{R \dagger}, \op{\a}_n^{R}$. Unlike the operators $\op{a}_n^{L \dagger}, \op{a}_n^{L}$ and $\op{a}_n^{R \dagger}, \op{a}_n^{R}$ these are regarded as independent and they do not satisfy relation \eqref{a_rel}. This is the essence of the semiclassical approximation, where one works on the tensor product $\Hsemi = \H_L \otimes \H_R$ of the two Hilbert spaces.

The analysis of the semiclassical approximation for the JT wormhole goes analogously to the analysis of the geon-wormhole in section \ref{sec:geon_semi1}. In the semiclassical approximation we want to identify the total Hilbert space $\H$ inside the tensor product $\H_L \otimes \H_R$. Analogously to \eqref{geon_data} define the operators
\begin{align} \label{alpha_beta_op}
& \op{\a}^{(\pm)}_n = \frac{1}{\sqrt{2}} \left[ \op{\a}_n^R \pm (-1)^n \op{\a}_n^{L} \right].
\end{align}
These are creation-annihilation operators satisfying 
\begin{align} \label{comm_ab}
[ \op{\a}^{(\sigma)}_n, \op{\a}_{n'}^{(\sigma') \dagger} ] = \delta_{n n'} \delta^{\sigma \sigma'}.
\end{align}
The idea of the semiclassical approximation is to mimic the relation \eqref{a_rel} and construct the Hilbert space of physical states by imposing $\op{\a}^{(-)}_n = \op{\a}_n^{(-) \dagger} = 0$.

\paragraph{Weak constraint.} As was the case for the geon-wormhole, the condition $\op{\a}^{(-)}_n = \op{\a}_n^{(-) \dagger} = 0$ is too strong to be imposed on the semiclassical Hilbert space $\Hsemi = \H_L \otimes \H_R$. Instead, we  can impose the weak condition
\begin{align} \label{beta_condA}
\op{\a}^{(-)}_{n} | \psi \> = 0
\end{align}
for all $\omega, n$. States that obey this condition are \emph{physical}, while all other states are \emph{null}. If $| \psi \>$ is physical, then $\op{\a}^{(-) \dagger}_{n} | \psi \>$ is null due to commutation relations \eqref{comm_ab}. Thus, physical states furnish the Hilbert space $\H_{+}$ generated by the action of the creation operators $\op{\a}^{(+) \dagger}_{n}$ on the vacuum $|0\>$. We have $\Hsemi = \H_L \otimes \H_R \cong \H_{+} \otimes \H_{-}$.

An operator $\O$ is physical if it maps $\H_{+}$ into itself. The left and right creation-annihilation operators $\op{\a}_n^{L \dagger}, \op{\a}_n^{L}$ and $\op{\a}_n^{R \dagger}, \op{\a}_n^{R}$ are unphysical on their own. Only the combinations that can be rewritten purely in terms of $\op{\a}^{(+)}_{n}$ and $\op{\a}^{(+) \dagger}_{n}$ are physical. Thus, for example, the left and right Hamiltonians $\op{H}_{L,R}$ are unphysical. The physical Hamiltonian $\op{H}_{+}$ reads
\begin{align} \label{Halpha}
\op{H}_{+} = \sum_{n=0}^{\infty} \omega_n \op{\a}_n^{(+) \dagger} \op{\a}_n^{(+)}, \quad \omega_n = \Delta + n
\end{align}
analogously to the case of the geon-wormhole in section \ref{sec:geon_semi1}. In particular the time-dependent boundary operators $\O_{L,R}$ must be evolved with this Hamiltonian. We can define the operator at $t = 0$ by their definitions \eqref{OLR}, which gives
\begin{align}
\O_R(\tau) & = \frac{1}{\sqrt{2}} \sum_{n=0}^{\infty} \tilde{c}_n \left[ \op{\a}_n^{(+)} + \op{\a}_n^{(+) \dagger} \right] + \frac{1}{\sqrt{2}} \sum_{n=0}^{\infty} \tilde{c}_n \left[ \op{\a}_n^{(-)} + \op{\a}_n^{(-) \dagger} \right], \\
\O_L(\tau) & = \frac{1}{\sqrt{2}} \sum_{n=0}^{\infty} (-1)^n \tilde{c}_n \left[ \op{\a}_n^{(+)} + \op{\a}_n^{(+) \dagger} \right] + \frac{1}{\sqrt{2}} \sum_{n=0}^{\infty} (-1)^n \tilde{c}_n \left[ \op{\a}_n^{(-)} + \op{\a}_n^{(-) \dagger} \right].
\end{align}
In each case only the first terms are physical. If we drop the unphysical pieces, the time-evolved operators are
\begin{align}
\O_L(\tau) & = \frac{1}{\sqrt{2}} \sum_{n=0}^{\infty} \left[ (-1)^n \varphi_n(\tau) \op{\a}_n^{(+)} + (-1)^n \varphi^{\ast}_n(\tau) \op{\a}^{(+) \dagger}_n \right], \label{jtOL} \\
\O_R(\tau) & = \frac{1}{\sqrt{2}} \sum_{n=0}^{\infty} \left[ \varphi_n(\tau) \op{\a}^{(+)}_n + \varphi^{\ast}_n(\tau) \op{\a}^{(+) \dagger}_n \right]. \label{jtOR}
\end{align}
Just as for the geon-wormhole, even if we kept the unphysical pieces, they would not be evolved by the Hamiltonian \eqref{Halpha} and effectively decouple. However, by dropping null states, we have effectively changed the normalization of various states. This stems from the fact that in the full, physical theory we have relations \eqref{a_rel} and so all sets of operators $\op{a}_n, \op{a}_n^{\dagger}$, $\op{a}^{L}_n, \op{a}_n^{L \dagger}$ and $\op{a}^{R}_n, \op{a}_n^{R \dagger}$ satisfy canonical commutation relations. Thus, we can either associate $\op{a}_n$ with $\op{\a}_n^R$ and $(-1)^n \op{\a}_n^{L}$ in the semiclassical theory or with $\op{\a}_n^{(+)}$, but not both at the same time. Just as for the geon-wormhole, we will associate $\op{a}_n$ with $\op{\a}_n^R$ and $(-1)^n \op{\a}_n^{L}$, which means that the isomorphism between $\H_{+}$ and the actual Hilbert space $\H$ requires the rescaling as in \eqref{rescaling},
\begin{align} \label{rescaling2}
\H_{+} \ni | j \>^{(+)}_{n} \: \longmapsto \: 2^{j/2} | j \>_{n} \in \H.
\end{align}
This means that in $\H_{+}$ we have the 2-point functions,
\begin{align}
\< \Omega | \O_R(\tau_R) \O_R(\tau'_R) | \Omega \>_{(+)} & = \frac{1}{2} \sum_{n=0}^{\infty} \varphi_n(\tau_R) \varphi^{\ast}_n(\tau'_R) \times {}^{(+)}_n \< 1 | 1 \>_n^{(+)} \nn\\
& = \frac{1}{2} G_{RR}(\tau_R - \tau'_R), \\
\< \Omega | \O_L(\tau_L) \O_R(\tau_R) | \Omega \>_{(+)} & = \frac{1}{2} \sum_{n=0}^{\infty} (-1)^n \varphi_n(\tau_L) \varphi^{\ast}_n(\tau_R) \times {}^{(+)}_n \< 1 | 1 \>_n^{(+)} \nn\\
& = \frac{1}{2} G_{LR}(\tau_L - \tau_R).
\end{align}
The factor of $1/2$ is then removed by \eqref{rescaling2}, when interpreting this result in $\H$. In such a case ${}^{(+)}_n \< 1 | 1 \>_n^{(+)} = 2 {}_n \< 1 | 1 \>_n = 2$, where the norm on the right hand side is in $\H$. Any way, we end up with the correct correlation functions \eqref{2pt_JT} and \eqref{2ptLR}.

\paragraph{Constrained quantization.} The problem with the semiclassical approximation is the fact that we try imposing the condition such as \eqref{beta_condA} only after the system has been quantized. The proper way of imposing the strong condition 
\begin{align} \label{strong_condA}
\op{\a}^{(-)}_n = \op{\a}_n^{(-) \dagger} = 0
\end{align}
is to follow the the constrained quantization, see \textit{e.g.}, \cite{Henneaux:1992ig}. One starts with the set of creation-annihilation operators $\op{\a}^{L \dagger}_n, \op{\a}^{L}_n$ and $\op{\a}^{R \dagger}_n, \op{\a}^{R}_n$ and imposes the constraint. As this is the second-class constraint, it leads to the Dirac brackets,
\begin{align} \label{comm_Dirac}
& [ \op{\a}^{(+)}_n, \op{\a}^{(+) \dagger}_{n'} ]_{D} = \delta_{n n'}, && [ \op{\a}^{(-)}_n, \op{\a}_{n'}^{(-) \dagger} ]_{D} = [ \op{\a}^{(+)}_n, \op{\a}^{(-)}_{n'} ]_{D} = [ \op{\a}^{(+)}_n, \op{\a}_{n'}^{(-) \dagger} ]_{D} = 0.
\end{align}
The constrained system contains a single copy of the Hilbert space, $\H_{+}$, spanned by the creation operators $\op{\a}_{n}^{(+) \dagger}$ defined in \eqref{alpha_beta_op}. The system is equal to the total Hilbert space $\H$ in the sense of section \ref{sec:polar}: the vacuum of $\H_{+}$ is mapped to the vacuum of $\H$ and the algebra of $\op{\a}_{n}^{(+) \dagger}, \op{\a}_n^{(+)}$ to $\op{a}_{n}^{\dagger}, \op{a}_n$ together with the states they create. The Hamiltonian of the constrained system becomes equal to \eqref{Halpha} up to the addition of an arbitrary null operator, which we fix to vanish.

\paragraph{Partition functions.} Since the Hilbert space of the theory is isomorphic to the Hilbert space of a single boundary, the partition function is that of a single boundary as well. The thermal partition function is thus the product
\begin{align} \label{Z}
Z(\beta) & = \Tr e^{-\beta \op{H}} = \prod_{n=0}^{\infty} Z_n(\beta),
\end{align}
where
\begin{align}
Z_n(\beta) = \Tr e^{-\beta \op{H}_n} = \frac{1}{1 - e^{-\beta \omega_n}}.
\end{align}
The thermodynamic entropy is therefore
\begin{align}
S & = - \log Z + \beta \frac{\partial}{\partial \beta} \log Z \nn\\
& = \sum_{n=0}^{\infty} \left[ \log \left( 1 - e^{-\beta \omega_n} \right) - \frac{\beta \omega_n e^{-\beta \omega_n}}{1 - e^{-\beta \omega_n}} \right].
\end{align}

On the other hand in the semi-classical approximation one has the Hamiltonian $\op{H}_{+}$ in \eqref{Halpha}. The correct way of calculating the partition function would be to trace over the physical states only. This yields \eqref{Z}. However, if the trace is taken over unphysical states as well, the Hamiltonian $\op{H}_{+}$ becomes the average of the boundary Hamiltonians $\op{H}_L$ and $\op{H}_R$ as in \eqref{HasAve}. This naive partition function, which overcounts unphysical states, reads
\begin{align}
Z_{\text{naive}}(\beta) = \Tr_{\H_L \otimes \H_R} e^{- \frac{\beta}{2} (\op{H}_L + \op{H}_R)} = \left[ Z \left( \frac{\beta}{2} \right) \right]^2.
\end{align}
Notice that this is the relation advocated in \cite{Saad:2018bqo}, which should hold at the late stage of the black hole evaporation, the so called ramp. Thus, it strongly suggests that the ensamble describing the late stage of the black hole evaporation is dominated by wormholes. Nevertheless, due to the Hilbert space of a wormhole not splitting into the tensor product, the actual partition function is that of the single side, \eqref{Z}.

\section{Summary}

In this paper I carried out the detailed analysis presented in \cite{Bzowski:2021vno}. In section \ref{sec:qh} I argued that the Hilbert space dual to a holographic, traversable wormhole does not split into the tensor product of the boundary Hilbert spaces. The analysis was carried out for the scalar field in the fixed background. The results can be regarded as the $G_N = 0$ or the leading $1/N$ statement in holography, or they can be treated as toy models for non-factorization. Nevertheless, with the gravity turned on, one should only expect that the non-factorization of the Hilbert space becomes even more severe.

In order to present the peculiarities and illusions of the wormholes in a simple set-up, we studied two examples: in section \ref{sec:geon} we analyzed the structure of the geon-wormhole, while in section \ref{sec:jt} the structure of the AdS$_2$ wormhole. The two examples are quite extreme in the sense that the wormholes are open for eternity. In both cases we identified the Hilbert space $\H$ to be isomorphic to the boundary Hilbert spaces $\H_L$ and $\H_R$ separately, $\H \cong \H_L \cong \H_R$. The precise isomorphisms are given in \eqref{geon_iso} and \eqref{HLR_iso}. 

Next, in sections \ref{sec:geonwh} and \ref{sec:wh_state} we identified the wormhole dual states. In case of the geon-wormhole the dual state is the `thermofield single state' given in \eqref{G}. It exhibits the thermal properties of the usual thermofield double, when perceived from the point of view of a single boundary. For the JT wormhole, the global vacuum is the thermofield double state from the point of view of the two independent Rindler wedges. In section \ref{sec:wh_state} we calculated the Bogoliubov coefficients explicitly. We find that the difference between the two disconnected Rindler wedges and the eternal wormhole lies in the structure of the Hilbert space rather than the dual state.

Finally, in sections \ref{sec:geon_semi1} and \ref{sec:jt_semi}  we showed to what extent the semiclassical analysis under the assumption of the tensor product factorization reconstructs the full, physical system $\H$. We showed that the number of peculiarities and illusions emerge as the result of such an assumption. This includes the `illusions' stated in the introduction:
\begin{enumerate}
\item \emph{Illusion of the null states.} As the actual Hilbert space $\H$ is `smaller' than $\H_L \otimes \H_R$, one must remove certain null states from the tensor product in order to describe the actual system. Physical states must obey relations such as \eqref{geon_rel}, \eqref{apm_cond} or \eqref{a_rel}.
\item \emph{Illusion of physical operators.} Only those operator that map physical states to physical states correctly reproduce the algebra of operators on $\H$.
\item \emph{Illusion of entanglement.} If the factorized system $\H_L \otimes \H_R$ was to describe the physical system $\H$, relations such as \eqref{apm_cond} must be imposed by hand. As shown in section \ref{sec:geon_semi2} this results in the illusion that the vacuum state is a highly-entangled, infinite temperature, Bell-like state.
\item \emph{Illusion of interactions.} The difference between the Hamiltonians driving the system and `free' Hamiltonians can be regarded as an interaction. The interaction Hamiltonian in \eqref{Hint} acts as the effective projector on the physical states.
\end{enumerate}
All these illusions are the avatars of the analysis of the wormhole system on the factorizable Hilbert space $\Hsemi = \H_L \otimes \H_R$. They are all naturally embedded in the actual, physical, non-factorizable system on $\H$.

Finally, one can speculate what should be the Hilbert space of the full gravitational theory. If anything, gravity should make the Hilbert space `smaller', leading to more null states from the point of view of semiclassical approximation. As an oversimplified example, consider a pair of wormhole modes, which should be related between the two sides of a wormhole. The key observation is that in order to introduce a coupling we do not need to take the tensor product of their Hilbert spaces. Instead, the modes can be coupled in the way two approximate harmonic oscillators are coupled in the double-well system. The difference between the described system and the factorized system is analogous to the difference between the description of a single electron in the molecule $H_2^{+}$ and the description of two electrons in the atom of helium. The electron in $H_2^{+}$ moves in the double-well potential of the two hydrogen nuclei. For sufficiently small energies the two potential wells around the two hydrogen nuclea are almost decoupled and look like the factorizable system. At higher energies, however, the difference between the truly factorized system and the double-well becomes visible.

\section*{Acknowledgments}

I would like to thank Marika Taylor and Daniel L. Jafferis for valuable discussions. I am very grateful to Ruben Monten and John Gardiner, who participated in the early stages of this work. I am supported by the NCN POLS grant No.~2020/37/K/ST2/02768 financed from the Norwegian Financial Mechanism 2014-2021 \includegraphics[width=12pt]{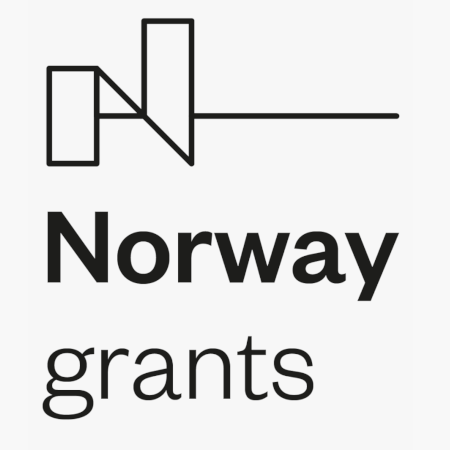} \includegraphics[width=12pt]{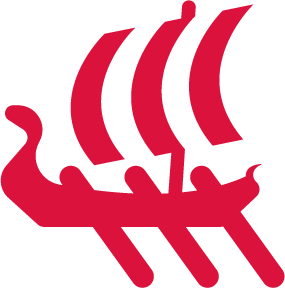}.

\appendix

\section{Useful formulas}

\subsection{Jacobi polynomials} \label{app:Jacobi}

Jacobi polynomials are defined as
\begin{align}
P^{(\alpha, \beta)}_k(x) = \frac{\Gamma(\alpha + k + 1)}{n! \Gamma(\alpha + 1)} \: {}_2 F_1 \left( -k, k + \alpha + \beta + 1, \alpha + 1; \frac{1 - x}{2} \right).
\end{align}
They satisfy
\begin{align}
P_k^{(\alpha, \beta)}(-x) & = (-1)^k P_k^{(\beta, \alpha)}(x), \\
P_k^{(\alpha, \beta)}(1) & = \frac{\Gamma(\alpha + k + 1)}{k! \Gamma(\alpha + 1)}
\end{align}
Jacobi polynomials are orthogonal polynomials, which satisfy
\begin{align} \label{JacobiDef}
0 = (1 - x^2) f''(x) + \left[ \beta - \alpha - (\alpha + \beta + 2) x \right] f'(x) + k(k + \alpha + \beta + 1) f(x)
\end{align}
where $f = P_k^{(\alpha, \beta)}$. The orthogonality properties are
\begin{align}
\int_{-1}^{1} \D x \, w^{(\alpha, \beta)}(x) P_m^{(\alpha, \beta)}(x) P_n^{(\alpha, \beta)}(x) & = \mathcal{N}^{(\alpha, \beta)}_{n} \delta_{mn}, \label{ortho1:Jacobi}\\
\sum_{n=0}^{\infty} \frac{P_{n}^{(\alpha, \beta)}(x) P_{n}^{(\alpha, \beta)}(y)}{\mathcal{N}^{(\alpha, \beta)}_{n}} & = \frac{\delta(x - y)}{w^{(\alpha, \beta)}(y)}, \label{ortho2:Jacobi}
\end{align}
where
\begin{align}
w^{(\alpha, \beta)} & = (1 - x)^{\alpha} (1 + x)^{\beta}, \\
\mathcal{N}^{(\alpha, \beta)}_{n} & = \frac{2^{\alpha + \beta + 1}}{2 n + \alpha + \beta + 1} \frac{\Gamma(n + \alpha + 1) \Gamma(n + \beta + 1)}{n! \Gamma(n + \alpha + \beta + 1)}
\end{align}

\subsection{Hypergeometric function relations}

We can use results of \cite{Wimp_1964} to obtain orthogonality properties for the hypergeometric functions. Using equations (4.13) and (4.14) there we find
\begin{align} \label{ortho1}
	& \int_0^{\infty} \D u \: u^{\xi-\sigma} {}_2 F_1( \xi - \I x, \xi + \I x; 1 + \xi - \sigma; -u ) \times \nn\\
	& \qquad \times {}_2 F_1( 1 - \sigma + \I y, 1 - \sigma - \I y; 1 + \xi - \sigma; -u ) = \mathcal{N}_{\xi\sigma}(x) \, \delta(x - y), \\
	& \int_0^{\infty} \frac{\D u}{\mathcal{N}_{\xi\sigma}(u)} \: {}_2 F_1( \xi - \I u, \xi + \I u; 1 + \xi - \sigma; -x ) \times \nn\\
	& \qquad \times {}_2 F_1( 1 - \sigma + \I u, 1 - \sigma - \I u; 1 + \xi - \sigma; -y ) = x^{\xi - \sigma} \, \delta(x - y),
\end{align}
where
\begin{align}
	\mathcal{N}_{\xi\sigma}(x) & = \frac{\Gamma(\sigma - \I x) \Gamma(\sigma + \I x) \Gamma^2(1 + \xi - \sigma)}{\Gamma(\xi - \I x) \Gamma(\xi + \I x)} \times \frac{\cosh^2(\pi x) - \cos^2 (\pi \sigma)}{x \sinh(2 \pi x)} \nn\\
	& = \frac{2 \pi \Gamma(2 \I x) \Gamma(-2 \I x) \Gamma^2(1 + \xi - \sigma)}{\Gamma(\xi - \I x) \Gamma(\xi + \I x) \Gamma(1 - \I x - \sigma) \Gamma(1 + \I x - \sigma)}.
\end{align}
By substituting 
\begin{align}
\xi & = \frac{\Delta}{2} + \hat{n}, &
\sigma & = 1 - \frac{\Delta}{2} + \hat{n}, & u & = \frac{\rho_h^2}{\rho^2 - \rho_h^2}, \\
x & = - \I \hat{\omega} = \frac{\omega L}{2 \rho_h}, & y & = - \I \hat{\omega}' = \frac{\omega' L}{2 \rho_h}
\end{align}
to the first identity we find
\begin{align}
& \int_{\rho_h}^{\infty} \frac{\rho\,\D \rho}{\rho^2 - \rho_h^2} \: \left( \frac{\rho_h^2}{\rho^2 - \rho_h^2} \right)^\Delta \times {}_2 F_1 \left( \frac{\Delta}{2} + \hat{n} - \hat{\omega}, \frac{\Delta}{2} + \hat{n} + \hat{\omega}; \Delta; \frac{\rho_h^2}{\rho_h^2 - \rho^2} \right) \times \nn\\
& \qquad\qquad\qquad\qquad \times {}_2 F_1 \left( \frac{\Delta}{2} - \hat{n} + \hat{\omega}', \frac{\Delta}{2} - \hat{n} - \hat{\omega}'; \Delta; \frac{\rho_h^2}{\rho_h^2 - \rho^2} \right) = \nn\\
& \qquad = \frac{\p}{L} N_{\omega n} 2 \pi \delta(\omega - \omega'),
\end{align}
where
\begin{align}
N_{\omega n} & = \frac{\Gamma^2 (\Delta) \Gamma(2 \hat{\omega}) \Gamma(-2 \hat{\omega})}{\Gamma \left( \frac{\Delta}{2} + \hat{n} + \hat{\omega} \right) \Gamma \left( \frac{\Delta}{2} - \hat{n} + \hat{\omega} \right) \Gamma \left( \frac{\Delta}{2} + \hat{n} - \hat{\omega} \right) \Gamma \left( \frac{\Delta}{2} - \hat{n} - \hat{\omega} \right)} \nn\\
& = \left| \frac{\Gamma(\Delta) \Gamma(2 \hat{\omega})}{\Gamma \left( \frac{\Delta}{2} + \hat{n} + \hat{\omega} \right) \Gamma \left( \frac{\Delta}{2} - \hat{n} + \hat{\omega} \right)} \right|^{2}.
\end{align}
The hatted variables are
\begin{align}
\hat{x} = \frac{\I x L}{2 \p}.
\end{align}
Note that this constant is real and positive. By substituting 
\begin{align}
\xi & = \frac{\Delta}{2} + \hat{n}, &
\sigma & = 1 - \frac{\Delta}{2} + \hat{n}, & u & = \frac{\omega L}{2 \rho_h}, \\
x & = \frac{\rho_h^2}{\rho^2 - \rho_h^2}, & y & = \frac{\rho_h^2}{\rho'^2 - \rho_h^2}
\end{align}
to the second identity we find
\begin{align}
& \int_{0}^{\infty} \frac{\D \omega}{2 \pi} \frac{1}{N_{\omega n}} \: {}_2 F_1 \left( \frac{\Delta}{2} + \hat{n} - \hat{\omega}, \frac{\Delta}{2} + \hat{n} + \hat{\omega}; \Delta; \frac{\rho_h^2}{\rho_h^2 - \rho^2} \right) \times \nn\\
& \qquad\qquad\qquad\qquad \times {}_2 F_1 \left( \frac{\Delta}{2} - \hat{n} + \hat{\omega}, \frac{\Delta}{2} - \hat{n} - \hat{\omega}; \Delta; \frac{\rho_h^2}{\rho_h^2 - \rho'^2} \right) = \nn\\
& \qquad = \frac{\rho_h}{L} \left( \frac{\rho^2 - \rho_h^2}{\rho_h^2} \right)^{\Delta} \frac{\rho^2 - \rho_h^2}{\rho} \: \delta(\rho - \rho').
\end{align}

\section{Scalar field}

\subsection{\texorpdfstring{AdS$_{d+1}$ in global coordinates}{AdS[d+1] in global coordinates}} \label{app:gAdS}

The AdS metric in global coordinates takes form
\begin{equation}
\D s^2 = \frac{L^2}{\cos^2 \theta} \left( - \D \tau^2 + \D \theta^2 + \sin^2 \theta \D \Omega^2_{d-1} \right),
\end{equation}
where $\theta \in (0, \pi/2)$ and $\tau \in \R$. The useful combination is
\begin{align}
\sqrt{\frac{\gamma}{|g_{\tau\tau}|}} = L^{d-1} \tan^{d-1} \theta.
\end{align}
In these coordinates the Klein-Gordon equation for the scalar field of mass $m$ takes form
\begin{align}
0 = \left[ - \frac{\partial^2}{\partial \tau^2} + \frac{\partial^2}{\partial \theta^2} + \frac{d - 1}{\sin \theta \cos \theta} \frac{\partial}{\partial \theta} + \frac{1}{\sin^2 \theta} \Delta_{S^{d-1}} - \frac{L^2 m^2}{\cos^2 \theta} \right] \Phi(\tau, \theta, \Omega),
\end{align}
where $L^2 m^2 = \Delta(\Delta - d)$. We look for a factorized solution of the form
\begin{equation}
\phi_{\omega \ell}(\tau, \theta, \Omega) = c_{\omega \ell} e^{-\I \omega \tau} Y_{\ell}(\Omega) \Theta_{\omega \ell}(\theta),
\end{equation}
for which the equation becomes
\begin{align} \label{eqSepAdS}
0 = \Theta''(\theta) + \frac{(d-1) \Theta'(\theta)}{\sin \theta \cos \theta}  + \left( \omega^2 - \frac{\ell (\ell + d -2)}{\sin^2 \theta} - \frac{\Delta(\Delta - d)}{\cos^2 \theta} \right) \Theta(\theta),
\end{align}
where
\begin{align}
\Delta_{S^{d-1}} Y_{\ell}(\Omega) = - \ell(\ell+d-2) Y_{\ell}(\Omega).
\end{align}
If we substitute
\begin{equation}
\Theta_{\omega \ell}(\theta) = \cos^{\Delta} \theta \sin^{\ell} \theta \chi_{\omega \ell}(\cos(2 \theta)),
\end{equation}
then $\chi_{\omega \ell}$ satisfies the radial equation
\begin{equation} \label{eqchi}
0 = (1 - x^2) \chi''(x) - \left[ (\ell + \Delta + 1) x + (\ell + d - \Delta - 1) \right] \chi'(x) + \frac{1}{4} \left[ \omega^2 - (\ell + \Delta)^2 \right] \chi(x).
\end{equation}
Two independent solutions for $\Theta_{\omega \ell}$ can be expressed in terms of hypergeometric functions and the radial mode reads
\begin{align} \label{solAdS1}
\Theta_{\omega \ell} & = c_1 \cos^{\Delta} \theta \sin^{\ell} \theta {}_2 F_1 \left( \frac{1}{2} (\ell + \Delta - \omega), \frac{1}{2} (\ell + \Delta + \omega); \Delta - \frac{d}{2} + 1; \cos^2 \theta \right) \nn\\
& \qquad + c_2 \cos^{d - \Delta} \theta \sin^{\ell} \theta {}_2 F_1 \left( \frac{1}{2} (\ell + d - \Delta - \omega), \frac{1}{2} (\ell + d - \Delta + \omega); - \Delta + \frac{d}{2} + 1; \cos^2 \theta \right).
\end{align}
In this representation the hypergeometric approach $1$ at the boundary. Thus, we can see that only the first solution exhibits the correct asymptotics: $\sim \cos^{\Delta} \theta$. So we must set $c_2 = 0$. Next we must ensure the regularity in the interior at $\theta = 0$. We carry out another linear transformation and from the two modes one is regular at $\theta = 0$, while the other blows up. This introduces the quantization of frequencies, (also see \cite{Balasubramanian:1998sn, Skenderis:2008dg})
\begin{equation} \label{freqdd}
\omega_{k \ell} = \Delta + \ell + 2 k,
\end{equation}
where $k = 0,1,2,\ldots$ is an arbitrary non-negative integer. With this condition we recognize in \eqref{eqchi} the differential equation \eqref{JacobiDef} defining Jacobi polynomials. All in all we find
\begin{align} \label{PhiAdSd}
\phi_{k \ell}(\tau, \theta, \Omega) =  c_{k \ell} e^{-\I \omega_{k \ell} \tau} Y_{\ell}(\Omega) \cos^\Delta \theta \sin^{\ell} \theta \: P_k^{(\ell + \frac{d}{2} - 1, \Delta - \frac{d}{2})}(\cos(2 \theta)).
\end{align}
With the normalization constant
\begin{align} \label{AdS_norm}
	c_{k \ell} = \sqrt{ \frac{1}{L^{d-1}} \frac{k! \: \Gamma(\Delta + \ell + k)}{\Gamma(\ell + k + \tfrac{d}{2}) \Gamma(\Delta - \tfrac{d}{2} + k + 1)}},
\end{align}
one can use \eqref{ortho1:Jacobi} to see that the modes are orthonormal in the Klein-Gordon norm
\begin{align} \label{KGglob}
(\Phi, \Psi) = - \I L^{d-1} \int_{S^{d-1}} \D \Omega_{d-1} \int_0^{\frac{\pi}{2}} \D \theta \, \tan^{d-1} \theta \left[ \Phi \partial_{\tau} \Psi^{\ast} - \partial_{\tau} \Phi \, \Psi^{\ast} \right]_{\tau = 0},
\end{align}
\textit{i.e.}, $(\phi_{k \ell}, \phi^{\ast}_{k' \ell'}) = \delta_{kk'} \delta_{\ell \ell'}$. We normalize the spherical harmonics such that $\int \D \Omega \sqrt{g} Y_{\ell} Y^{\ast}_{\ell'} = \delta_{\ell \ell'}$.

\subsection{\texorpdfstring{AdS$_3$ in Rindler coordinates}{AdS3 in Rindler coordinates}} \label{app:RAdS}

In Rindler coordinates the AdS metric takes form
\begin{equation}
\D s^2 = - \rho^2 \D t^2 + \frac{L^2 \D \rho^2}{L^2 + \rho^2} + ( L^2 + \rho^2 ) \D \varphi^2,
\end{equation}
with $\rho$ ranging from $0$ to $\infty$. The useful combination is
\begin{align}
\sqrt{\frac{\gamma}{|g_{tt}|}} = \frac{L}{\rho}.
\end{align}
With the mode ansatz
\begin{align}
& \phi_{\omega n}(t, \rho, \varphi) = c_{\omega n} e^{-\I \omega t + \I n \varphi} R_{\omega n}(\rho),
\end{align}
where $n \in \Z$, $\omega > 0$, the Klein-Gordon equation is
\begin{align}
0 = (L^2 + \rho^2) R''_{\omega n}(\rho) + \frac{L^2 + 3 \rho^2}{\rho} R'_{\omega n}(\rho) + L^2 \left( - m^2 + \frac{\omega^2}{\rho^2} - \frac{n^2}{L^2 + \rho^2} \right) R_{\omega n}(\rho),
\end{align}
where $m^2 L^2 = \Delta (\Delta - 2)$. The mode functions are
\begin{align} \label{RAdSmode}
R_{\omega n}(\rho) & = \left( \frac{L}{\rho} \right)^{\Delta} \left( \frac{\rho^2 + L^2}{\rho^2} \right)^{\frac{\I n}{2}} {}_2 F_1 \left( \frac{\I n}{2} + \frac{\I \omega}{2} + \frac{\Delta}{2}, \frac{\I n}{2} - \frac{\I \omega}{2} + \frac{\Delta}{2}; \Delta; - \frac{L^2}{\rho^2} \right).
\end{align}
From a linear transformation of hypergeometric function one can see that the solution is real, $R_{\omega n}^{\ast} = R_{\omega n} = R_{\omega, -n}$. These are normalizable modes and near the boundary $\rho \rightarrow \infty$ we have
\begin{align}
R_{\omega n} = \left( \frac{\rho}{L} \right)^{-\Delta} \left[ 1 + O(\rho^{-2}) \right].
\end{align}
The normalization constant is the same as for the BTZ black hole with $\p = L$,
\begin{align}
c_{\omega n} & = \left. \frac{1}{\sqrt{4 \pi \omega \p N_{\omega n}}} \right|_{\p = L} \nn\\
& = \sqrt{ \frac{1}{4 \pi \omega L} \frac{\Gamma \left( \frac{\Delta}{2} + \frac{\I n}{2} + \frac{\I \omega}{2} \right) \Gamma \left( \frac{\Delta}{2} - \frac{\I n}{2} + \frac{\I \omega}{2} \right) \Gamma \left( \frac{\Delta}{2} + \frac{\I n}{2} - \frac{\I \omega}{2} \right) \Gamma \left( \frac{\Delta}{2} - \frac{\I n}{2} - \frac{\I \omega}{2} \right)}{\Gamma^2 (\Delta) \Gamma(\I \omega) \Gamma(- \I \omega)}}.
\end{align}
Note that $N_{\omega n}$ is real and positive, so we can choose the positive root and $c_{\omega n} = c_{\omega, -n} = c_{-\omega, n} = c_{\omega n}^{\ast}$. The normalization is such that with
\begin{align}
\left[ a_{\omega n}, a_{\omega' n'}^{+} \right] & = 2 \pi  \delta(\omega - \omega') \delta_{nn'}
\end{align}
the canonical commutation relations hold,
\begin{align}
\left[ \Phi(0, \rho, \varphi), \Pi(0, \rho', \varphi') \right] & = \I \, \delta(\rho - \rho') \delta(\varphi - \varphi').
\end{align}
The Klein-Gordon scalar product is
or equivalently,
\begin{align}
 ( \phi_{\omega n}, \phi_{\omega' n'}) & = 2 \pi \delta(\omega - \omega') \delta_{n n'}.
\end{align}

\subsection{BTZ in Schwarzschild coordinates}

In the left and right wedges the metric in Schwarzschild coordinates reads,
\begin{equation}
\D s^2 = - (\rho^2 - \rho_h^2) \D t^2 + \frac{L^2 \D \rho^2}{\rho^2 - \rho_h^2} + \rho^2 \D \varphi^2, \qquad \rho > \rho_h.
\end{equation}
A useful combination is
\begin{align}
\sqrt{\frac{\gamma}{|g_{tt}|}} = \frac{L |\rho|}{\rho^2 - \rho_h^2}.
\end{align}
With the mode ansatz
\begin{align}
& \phi_{\omega n}(t, \rho, \varphi) = c_{\omega n} e^{-\I \omega t + \I n \varphi} R_{\omega n}(\rho),
\end{align}
where $n \in \Z$, $\omega > 0$, the Klein-Gordon equation is
\begin{align}
0 = - (\rho^2 - \p^2) R''_{\omega n}(\rho) + \frac{\p^2 - 3 \rho^2}{\rho} R'_{\omega n}(\rho) + L^2 \left( m^2 + \frac{n^2}{\rho^2} + \frac{\omega^2}{\p^2 - \rho^2} \right) R_{\omega n}(\rho),
\end{align}
where $m^2 L^2 = \Delta (\Delta - 2)$. The mode functions are
\begin{align} \label{BTZmodeR}
R_{\omega n}(\rho) & = \left( \frac{\rho^2}{\rho^2_h} \right)^{\hat{n}} \left( \frac{\rho^2_h}{\rho^2 - \rho_h^2} \right)^{\hat{n} + \frac{\Delta}{2}} {}_2 F_1 \left( \frac{\Delta}{2} + \hat{n} - \hat{\omega}, \frac{\Delta}{2} + \hat{n} + \hat{\omega}; \Delta; \frac{\rho_h^2}{\rho_h^2 - \rho^2} \right),
\end{align}
where the hatted variables are
\begin{align}
\hat{x} = \frac{\I x L}{2 \p}.
\end{align}
From a linear transformation of hypergeometric function one can see that the solution is real, $R_{\omega n}^{\ast} = R_{\omega n} = R_{\omega, -n}$. These are normalizable modes and near the boundary $\rho \rightarrow \infty$ we have
\begin{align}
R_{\omega n} = \left( \frac{\rho}{\rho_h} \right)^{-\Delta} \left[ 1 + O(\rho^{-2}) \right].
\end{align}
Consider a general mode decomposition
\begin{align}
\Phi = \sum_I \sum_{n=-\infty}^{\infty} \int_0^{\infty} \frac{\D \omega}{2 \pi} \left[ a_{\omega n}^I \phi^I_{\omega n} + a_{\omega n}^{I +} \phi^{I \ast}_{\omega n} \right],
\end{align}
where the outer sum either goes over $I = L,R$ or a single $I = R$, depending on whether we analyze one or two wedges. The normalization constant then reads
\begin{align}
c_{\omega n} & = \frac{1}{\sqrt{4 \pi \omega \p N_{\omega n}}} \nn\\
& = \sqrt{ \frac{1}{4 \pi \omega \p} \frac{\Gamma \left( \frac{\Delta}{2} + \hat{n} + \hat{\omega} \right) \Gamma \left( \frac{\Delta}{2} - \hat{n} + \hat{\omega} \right) \Gamma \left( \frac{\Delta}{2} + \hat{n} - \hat{\omega} \right) \Gamma \left( \frac{\Delta}{2} - \hat{n} - \hat{\omega} \right)}{\Gamma^2 (\Delta) \Gamma(2 \hat{\omega}) \Gamma(-2 \hat{\omega})}}.
\end{align}
Note that $N_{\omega n}$ is real and positive, so we can choose the positive root and $c_{\omega n} = c_{\omega, -n} = c_{-\omega, n} = c_{\omega n}^{\ast}$. The normalization is such that with
\begin{align}
\left[ a_{\omega n}^I, a_{\omega' n'}^{I' \dagger} \right] & = 2 \pi  \delta(\omega - \omega') \delta_{nn'} \delta^{II'},
\end{align}
the canonical commutation relations hold,
\begin{align}
\left[ \Phi(0, \rho, \varphi), \Pi(0, \rho', \varphi') \right] & = \I \, \delta(\rho - \rho') \delta(\varphi - \varphi').
\end{align}
The Klein-Gordon scalar product is
\begin{align}
( \Phi, \Psi ) = - \I \sum_{I = L,R} \int_0^{2 \pi} \D \varphi \int_{\p}^{\infty} \frac{L \rho_I \D \rho_I}{\rho_I^2 - \p^2} \left[ \Phi \partial_t \Psi^{\ast} - \partial_t \Phi \, \Psi^{\ast} \right]_{t=0}.
\end{align}
and the modes satisfy
\begin{align}
( \phi^I_{\omega n}, \phi^{I'}_{\omega' n'}) & = 2 \pi \delta(\omega - \omega') \, \delta_{n n'} \, \delta^{I I'}.
\end{align}

\subsection{\texorpdfstring{AdS$_2$ in Schwarzschild coordinates}{AdS2 in Schwarzschild coordinates}} \label{app:RAdS2}

In the embedding coordinates $T_1, T_2, X$ AdS$_2$ is described as the surface satisfying
\begin{align}
-T_1^2 - T_2^2 + X^2 = - L^2
\end{align}
and with the metric induced from
\begin{align}
\D s^2 = - \D T_1^2 - \D T_2^2 + \D X^2.
\end{align}
With
\begin{align}
T_1 & = \frac{L \cos \tau}{\cos \theta}, \\
T_2 & = \frac{L \sin \tau}{\cos \theta}, \\
X & = L \tan \theta
\end{align}
we obtain the global AdS metric
\begin{align}
\D s^2 = \frac{L^2}{\cos^2 \theta} ( - \D \tau^2 + \D \theta^2 ).
\end{align}
On the other hand the Schwarzschild coordinates $(t ,\rho)$ cover a single Rindler wedge of AdS$_2$. They are related to the embedding coordinates as
\begin{align}
T_1 & = \frac{L \rho}{\p}, \\
T_2 & = \frac{L}{\p} \sqrt{\rho^2 - \p^2} \sinh \left( \frac{\p t}{L} \right), \\
X & = \frac{L}{\p} \sqrt{\rho^2 - \p^2} \cosh \left( \frac{\p t}{L} \right)
\end{align}
for arbitrary $\p > 0$. The induced metric reads
\begin{align}
\D s^2 = - (\rho^2 - \p^2) \D t^2 + \frac{L^2}{\rho^2 - \p^2} \D \rho^2.
\end{align}
The useful combination is
\begin{align}
\sqrt{\frac{\gamma}{|g_{tt}|}} = \frac{L}{\rho^2 - \p^2}.
\end{align}

The mode ansatz is
\begin{align} \label{phi_omega_AdS2}
\phi_\omega(t, \rho) = c_{\omega} e^{-\I \omega t} R_{\omega}(\rho)
\end{align}
and the Klein-Gordon equation becomes
\begin{align}
0 = - (\rho^2 - \p^2) R''_{\omega}(\rho) - 2 \rho R'_{\omega}(\rho) + L^2 \left( m^2 + \frac{\omega^2}{\p^2 - \rho^2} \right) R_{\omega}(\rho).
\end{align}
The solution can be neatly written in terms of Legendre functions
\begin{align}
R_{\omega}(\rho) = c^P_{\omega} P_{\Delta - 1}^{2 \hat{\omega}} \left( \frac{\rho}{\p} \right) + c^Q_{\omega} Q_{\Delta - 1}^{2 \hat{\omega}} \left( \frac{\rho}{\p} \right),
\end{align}
where $\hat{x} = \I L \omega / (2 \p)$ and $c^P_{\omega}, c^Q_{\omega}$ are arbitrary constants. Unfortunately, both Legendre functions contain sources and vevs and one needs to take a suitable linear transformation to decouple them. Eventually, the mode function with the appropriate boundary conditions reads
\begin{align}
R_{\omega}(\rho) = \left( \frac{\p^2}{\rho^2 - \p^2} \right)^{\frac{\Delta}{2}} {}_2 F_1 \left( \frac{\Delta}{2} - \hat{\omega}, \frac{\Delta}{2} + \hat{\omega}; \Delta + \frac{1}{2}; \frac{\p^2}{\p^2 - \rho^2} \right).
\end{align}

\subsection{\texorpdfstring{The Bogoliubov transformation for $\Delta = 1$}{The Bogoliubov transformation for Delta = 1}} \label{sec:bogo}

Let $\phi^{[1]}_{\omega}$ denote the Schwarzschild mode $\phi^R_{\omega}$ for $\Delta = 1$ in the right wedge and zero in the left wedge. The normalized solution equals
\begin{align} \label{phiSch}
\phi^{[1]}_{\omega}(t, \rho) = \sqrt{\frac{2}{\omega}} e^{-\I \omega t} \sin \left[ \frac{L \omega}{2 \p} \log \left( \frac{\rho + \p}{\rho - \p} \right) \right].
\end{align}
We want to find the Bogoliubov coefficients between the Schwarzschild basis $\{ \phi^{[1]}_{\omega} \}_{\omega > 0}$ and the basis of modes in global coordinates. To do it, we rewrite the mode in the global coordinates where it becomes
\begin{align} \label{app_phi1}
\phi^{[1]}_{\omega}(\tau, \theta) = \frac{\I}{\sqrt{2 \omega}} \left[ \tan^{2 \hat{\omega}} \left( \frac{\theta - \tau}{2} \right) - \tan^{- 2 \hat{\omega}} \left( \frac{\theta + \tau}{2} \right) \right] \bs{1}_{\theta > 0},
\end{align}
where $\bs{1}_{\theta > 0} = 1$ for $\theta > 0$ and $0$ otherwise. Furthermore, $\hat{\omega} = \I L \omega/(2 \p)$. The global positive frequency modes $\chi^{[1]}_m$ are given by \eqref{chiP} and \eqref{chiM} and for $\Delta = 1$ they read
\begin{align}
\chi^{[1]}_{m}(\tau, \theta) & = \frac{1}{\sqrt{\pi m}} e^{-\I m \tau} \sin \left[ m \left( \theta - \tfrac{\pi}{2} \right) \right],
\end{align}
where $m=1,2,3,\ldots$. The Klein-Gordon scalar product \eqref{KG} between $\phi^{[1]}_{\omega}$ and $\chi^{[1]}_m$ becomes
\begin{align} \label{app_KG}
( \phi^{[1]}_{\omega}, \chi^{[1]}_m ) = \int_{0}^{\frac{\pi}{2}} \D \theta \, \phi^{1}_{\omega}(0, \theta) \chi^1_m(0, \theta) \left[ m \sigma + \frac{L \omega}{\p \sin \theta} \right],
\end{align}
with $\sigma = 1$. The scalar product $( \phi^{[1]}_{\omega}, \chi^{[1] \ast}_m)$ is identical but with $\sigma = -1$.

The first issue is that the integral involving the $1/\sin \theta$ term diverges at $\theta = 0$. This is a manifestation of the fact that, from the point of view of the global AdS we have to treat the Schwarzschild modes in the distributional sense, see \textit{e.g.}, \cite{Morrison:2014jha}. The positivity properties of the modes imply that \eqref{app_phi1} should be understood as
\begin{align}
\phi^{[1]}_{\omega}(\tau, \theta) = \frac{\I}{\sqrt{2 \omega}} \left[ \tan^{\frac{L \omega}{\p} (\I + \epsilon)} \left( \frac{\theta - \tau}{2} \right) - \tan^{\frac{L \omega}{\p} (-\I + \epsilon)} \left( \frac{\theta + \tau}{2} \right) \right] \Theta(\theta).
\end{align}
Now the regulated version of \eqref{app_KG} can be written down and the integrals calculated. The calculations require two integrals,
\begin{align}
\int_0^{\frac{\pi}{2}} \D \theta \, \tan^{2a} \frac{\theta}{2} \, \sin (n \theta) \sin^{2b} \theta & = \int_0^1 \D x \, (1-x)^{b+a} (1+x)^{b-a} U_{n-1}(x), \label{cheb_intU} \\
\int_0^{\frac{\pi}{2}} \D \theta \, \tan^{2a} \frac{\theta}{2} \, \cos(n \theta) \sin^{2b} \theta & = \int_0^1 \D x \, (1-x)^{b+a-\frac{1}{2}} (1+x)^{b-a-\frac{1}{2}} T_{n}(x), \label{cheb_intT}
\end{align}
where $U_n$ and $T_n$ are Chebyshev polynomials. This can be done with the help of entries 2.18.2.1 and 2.18.1.3 of \cite{Prudnikov2},
\begin{align}
\int_{-1}^1 \D x \, (1-x)^{\alpha} (1+x)^{\beta} U_{n}(x) & = 2^{\a + \b + 1} (1 + n) \frac{\Gamma(1 + \a) \Gamma(1 + \b)}{\Gamma(\a + \b + 2)} \times\nn\\
& \qquad \times {}_3 F_2 \left( \a+1, -n, n+2; \tfrac{3}{2}, \a+\b+2; 1 \right), \\
\int_{-1}^1 \D x \, (1-x)^{\alpha} (1+x)^{\beta} T_{n}(x) & = 2^{\a + \b + 1} \frac{\Gamma(1 + \a) \Gamma(1 + \b)}{\Gamma(\a+\b+2)} \times\nn\\
& \qquad \times {}_3 F_2 \left( \a+1, -n, n; \tfrac{1}{2}, \a+\b+2; 1 \right).
\end{align}
The integrals \eqref{cheb_intU} occur only for even $n$'s, while \eqref{cheb_intT} for odd $n$'s only. The parity of Chebyshev polynomials implies then that the two terms in \eqref{app_phi1} combine in such a way that the above expressions, involving integration from $-1$ to $1$, can be used.

When put together the finite $\epsilon \rightarrow 0$ limit exists and one finds
\begin{align}
( \phi_{\omega}^{[1]}, \chi_m^{[1]} ) & = \left\{ \begin{array}{ll} \frac{\I L}{\p} (-1)^{\frac{m}{2}} \sqrt{\frac{\pi \omega m}{2}} \left[ \frac{\sigma m \: {}_3 F_2 \left( 1 - m, 1 + m, 1 + \hat{\omega}; 2, \tfrac{3}{2}; 1 \right)}{\sinh \left( \frac{L \pi \omega}{2 \p} \right)} \right. & \\
\qquad\qquad \left. + \frac{{}_3 F_2 \left( 1 - m, 1 + m, \tfrac{1}{2} + \hat{\omega}; 1, \tfrac{3}{2}; 1 \right)}{\cosh \left( \frac{L \pi \omega}{2 \p} \right)} \right] & \text{ if } m \text{ is even,} \\
\frac{L}{\p} (-1)^{\frac{m + 1}{2}} \sqrt{\frac{\pi \omega m}{2}} \left[ \frac{m \: {}_3 F_2 \left( 1 - m, 1 + m, 1 + \hat{\omega}; 2, \tfrac{3}{2}; 1 \right)}{\sinh \left( \frac{L \pi \omega}{2 \p} \right)} \right. & \\
\qquad\qquad \left. - \frac{\sigma \: {}_3 F_2 \left( - m, m, \tfrac{1}{2} + \hat{\omega}; 1, \tfrac{1}{2}; 1 \right)}{2 \hat{\omega} \cosh \left( \frac{L \pi \omega}{2 \p} \right)} \right] & \text{ if } m \text{ is odd}.
\end{array} \right.
\end{align}
with $\sigma = 1$ and with the identical expressions for $( \phi_{\omega}^{[1]}, \chi_m^{[1] \ast} )$ but with $\sigma = -1$ instead. Finally, we can use the Corollaries 3.3.5 and 3.3.4 (Sheppard's identity) of \cite{andrews_askey_roy_1999},
\begin{align}
{}_3 F_2 \left( \begin{array}{c} a,b,c \\[-6pt] d,e \end{array}; 1 \right) & = \frac{\Gamma(e) \Gamma(d+e-a-b-c)}{\Gamma(e-a) \Gamma(d+e-b-c)} {}_3 F_2 \left( \begin{array}{c} a,d-b,d-c \\[-6pt] d,d+e-b-c \end{array}; 1\right), \\
{}_3 F_2 \left( \begin{array}{c} -n,a,b \\[-6pt] d,e \end{array}; 1 \right) & = \frac{(d-a)_n (e-a)_n}{(d)_n (e)_n} {}_3 F_2 \left( \begin{array}{c} -n, a, a+b-n-d-e+1 \\[-6pt] a-n-d+1, a-n-e+1 \end{array}; 1\right)
\end{align}
to find the relations
\begin{align}
& {}_3 F_2 \left( -m, m, x + \tfrac{1}{2}; \tfrac{1}{2}, 1; 1 \right) = \nn\\
& \qquad = -2 x \, {}_3 F_2 \left( 1-m, 1+m, x + \tfrac{1}{2}; \tfrac{3}{2}, 1; 1 \right) \nn\\
& \qquad = -2 m x \, {}_3 F_2 \left( 1-m, 1+m, x + 1; \tfrac{3}{2}, 2; 1 \right).
\end{align}
When substituted to the expressions above we end up with \eqref{bogoA} and \eqref{bogoB},
\begin{align}
\mu^R_{\omega m}  = ( \phi_{\omega}^{[1]}, \chi_m^{[1]} ) & = - \I^m \sqrt{\frac{2 \pi m}{\omega}} \frac{e^{\frac{\beta \omega}{4}}}{\sinh \left( \frac{\beta \omega}{2} \right)} \, {}_3 F_2 \left( -m, m, \tfrac{1}{2} + \hat{\omega}; \tfrac{1}{2}, 1; 1 \right), \\
- \nu^R_{\omega m}  = ( \phi_{\omega}^{[1]}, \chi_m^{[1] \ast} ) & = e^{-\frac{\beta \omega}{2}} \mu^R_{\omega m},
\end{align}
where
\begin{align}
\beta = \frac{2 \pi L}{\p}.
\end{align}

One can independently verify the normalization. The Bogoliubov coefficients can be rewritten as
\begin{align}
\mu_{\omega m}^R = - \sqrt{\frac{2 m}{\omega}} \frac{\pi}{\Gamma( \tfrac{1}{2} + m)} \frac{e^{\frac{\beta \omega}{4}}}{\sinh \left( \frac{\beta \omega}{2} \right)} \, p_m \left( \frac{\beta \omega}{4 \pi}; \frac{1}{2}, 0, 0, \frac{1}{2} \right),
\end{align}
where $p_n$ denote continuous Hahn polynomials,
\begin{align} \label{Hahn_pol}
p_n(x; a,b,c,d) = \I^n \frac{(a+c)_n (a+d)_n}{n!} \, {}_3 F_2 \left( -n, n+a+b+c+d-1, a+\I x; a+c, a+d; 1 \right).
\end{align}
For fixed $a,b,c,d > 0$ the polynomials are orthogonal satisfying
\begin{align} \label{Hahn_ortho}
& \int_{-\infty}^{\infty} \frac{\D x}{2 \pi} \Gamma(a + \I x) \Gamma(b + \I x) \Gamma(c - \I x) \Gamma(d - \I x) p_m(x) p_n(x) = \nn\\
& \qquad = \frac{\Gamma(n+a+c)\Gamma(n+a+d) \Gamma(n+b+c) \Gamma(n+b+d)}{(2n+a+b+c+d-1) \Gamma(n+a+b+c+d-1)} \frac{\delta_{mn}}{n!}.
\end{align}
Equation (6.10.10) of \cite{andrews_askey_roy_1999} seems to be missing the factorial.

We can use \eqref{Hahn_ortho} to check the orthogonality relation of the Bogoliubov coefficients. The full orthogonality relation requires the analogous Bogoliubov coefficients $\mu_{\omega m}^L, \nu_{\omega m}^L$ with respect to the modes in the left wedge. These are equal to $\mu_{\omega m}^R, \nu_{\omega m}^R$ and thus
\begin{align}
& \sum_{I=L,R} \int_0^{\infty} \frac{\D \omega}{2 \pi} \left[ \mu^I_{\omega m} \mu^{I \ast}_{\omega n} - \nu^I_{\omega m} \nu^{I \ast}_{\omega n} \right] = \nn\\
& \qquad = 2 \int_0^{\infty} \frac{\D \omega}{2 \pi} ( 1 - e^{-\beta \omega} ) \mu^R_{\omega m} \mu^{R \ast}_{\omega n} \nn\\
& \qquad = \delta_{mn}.
\end{align}
By writing $\mu^R_{\omega m} = \sum_{n=1}^{\infty} \mu^R_{\omega n} \delta_{mn}$ and substituting the second line above for the Kronecker's delta we confirm the orthogonality relations
\begin{align}
\sum_{m=1}^{\infty} \mu^R_{\omega m} \mu^{R \ast}_{\omega' m} & = \frac{\pi \delta(\omega - \omega')}{1 - e^{- \beta \omega}}, & \sum_{m=1}^{\infty} \nu^R_{\omega m} \nu^{R \ast}_{\omega' m} & = \frac{\pi \delta(\omega - \omega')}{e^{\beta \omega} - 1},
\end{align}
from which the expectation value of the number operator in \eqref{NR} follows.

\subsection{BTZ in Kruskal coordinates} \label{sec:BTZ_in_Kruskal}

We use conventions
\begin{align}
& U = T + X, && V = T - X,
\end{align}
so the right wedge corresponds to $U > 0$ and $V < 0$. In the right wedge, with $\rho > \p > 0$ and $t$ flowing up, the Kruskal coordinates are given by
\begin{align}
& U = e^{\frac{\p t}{L}} \sqrt{\frac{\rho - \p}{\rho + \p}}, && V = -e^{-\frac{\p t}{L}} \sqrt{\frac{\rho - \p}{\rho + \p}}.
\end{align}
We choose the time in the left wedge to flow up as well, which gives
\begin{align}
& U = - e^{- \frac{\p t}{L}} \sqrt{\frac{\rho - \p}{\rho + \p}}, && V = e^{\frac{\p t}{L}} \sqrt{\frac{\rho - \p}{\rho + \p}}
\end{align}
in the left wedge. In particular in both wedges
\begin{align}
\frac{\rho^2}{\p^2} & = \left( \frac{1 - U V}{1 + U V} \right)^2, & \frac{\rho^2 - \p^2}{\p^2} & = \frac{-4 U V}{(1 + U V)^2},
\end{align}
while
\begin{align}
e^{-\I \omega t} = \left( \frac{U}{-V} \right)^{\mp\hat{\omega}}
\end{align}
with minus sign in the right wedge and plus in the left.

Using linear transformation of the hypergeometric function the modes \eqref{BTZmodeR} can be expressed in terms of the modes with specified behavior at the horizons. With
\begin{align}
R^{hor}_{\omega n}(\rho) & = \left( \frac{\rho^2}{\rho^2_h} \right)^{\hat{n}} \left( \frac{\rho^2_h}{\rho^2 - \rho_h^2} \right)^{\hat{\omega}} {}_2 F_1 \left( \frac{\Delta}{2} + \hat{n} - \hat{\omega}, 1 - \frac{\Delta}{2} + \hat{n} - \hat{\omega}; 1 - 2 \hat{\omega}; \frac{\rho_h^2 - \rho^2}{\rho_h^2} \right)
\end{align}
the decomposition of the modes \eqref{BTZmodeR} reads
\begin{align}
R_{\omega n} & = \gamma_{\omega n} R^{hor}_{\omega n} + \gamma_{\omega, -n}^{\ast} R^{hor \ast}_{\omega, -n},
\end{align}
where
\begin{align}
\gamma_{\omega n} & = \frac{\Gamma(\Delta) \Gamma(2 \hat{\omega})}{\Gamma \left( \frac{\Delta}{2} + \hat{n} + \hat{\omega} \right) \Gamma \left( \frac{\Delta}{2} - \hat{n} + \hat{\omega} \right)}.
\end{align}

Next we define modes 
\begin{align}
& \phi^{R, L (1)}_{\omega n} = e^{-\I \omega t + \I n \varphi} R^{hor}_{\omega n}(\rho_{R,L}), && \phi^{R, L (2)}_{\omega n} = e^{-\I \omega t + \I n \varphi} R^{hor \ast}_{\omega, -n}(\rho_{R,L}).
\end{align}
These modes have the specific behavior at the horizons,
\begin{align}
& \phi^{R(1)}_{\omega n}(U, 0, \varphi) = e^{\I n \varphi} (2 |U|)^{-2 \hat{\omega}}, && \phi^{R(2)}_{\omega n}(0, V, \varphi) = e^{\I n \varphi} (2 |V|)^{2 \hat{\omega}}, \\
& \phi^{L(1)}_{\omega n}(0, V, \varphi) = e^{\I n \varphi} (2 |V|)^{-2 \hat{\omega}}, && \phi^{L(2)}_{\omega n}(U, 0, \varphi) = e^{\I n \varphi} (2 |U|)^{2 \hat{\omega}}.
\end{align}
We also have the following identities,
\begin{align}
& R^{hor \ast}_{\omega, n} = R_{-\omega, -n}, && 
\gamma_{\omega n} = \gamma_{\omega, -n} = \gamma^{\ast}_{-\omega, n}, && |c^{BTZ}_{\omega n}|^2 = \frac{1}{4 \pi \omega \p} \frac{1}{|\gamma_{\omega n}|^2},
\end{align}
although $c^{BTZ}_{\omega n}$ is strictly real, while $\gamma_{\omega n}$ is complex.

It follows that the combinations
\begin{align}
& \phi^{R(1)}_{\omega n} + e^{-\frac{\pi \omega L}{\p}} \phi^{L (2) \ast}_{\omega, -n}, && \phi^{L(1)}_{\omega n} + e^{-\frac{\pi \omega L}{\p}} \phi^{R (2) \ast}_{\omega, -n}
\end{align}
are suitably analytic in the Kruskal coordinates. Thus, following the reasoning of \cite{Unruh:1976db}, the normalized Kruskal modes are
\begin{align}
& \chi^{R}_{\omega n} = \frac{\phi^R_{\omega n} + e^{-\frac{\pi \omega L}{\p}} \phi^{L \ast}_{\omega, -n}}{\sqrt{1 - e^{-\frac{2 \pi \omega L}{\p}}}}, && \chi^{L}_{\omega n} = \frac{\phi^L_{\omega n} + e^{-\frac{\pi \omega L}{\p}} \phi^{R \ast}_{\omega, -n}}{\sqrt{1 - e^{-\frac{2 \pi \omega L}{\p}}}}
\end{align}
and the inverse temperature is
\begin{align}
\beta = \frac{2 \pi L}{\p}.
\end{align}

\section{Squeezed states}

\subsection{1-particle squeezed states} \label{sec:1-particle}

Let $z = r e^{\I \theta}$ and $\op{a}^{\dagger}, \op{a}$ be a pair of creation-annihilation operators. Consider the Bogoliubov operator,
\begin{align}
\mathcal{S}(z) = \exp \left( \frac{1}{2} z \op{a}^{\dagger} \op{a}^{\dagger} - \frac{1}{2} z^{\ast} \op{a} \op{a} \right).
\end{align}
The operator is unitary and $\mathcal{S}^{-1}(z) = \mathcal{S}^{\dagger}(z) = \mathcal{S}(-z)$ and realizes the Bogoliubov transformation,
\begin{align}
\op{b} & = \op{a} \, \cosh r + \op{a}^{\dagger} \, e^{\I \theta} \sinh r = \mathcal{S}^{\dagger}(z) \op{a} \mathcal{S}(z), \\
\op{b}^{\dagger} & = \op{a}^{\dagger} \, \cosh r + \op{a} \, e^{-\I \theta} \sinh r = \mathcal{S}^{\dagger}(z) \op{a} \mathcal{S}(z).
\end{align}
With
\begin{align}
\zeta = e^{\I \theta} \tanh r
\end{align}
its expansion reads
\begin{align}
\mathcal{S}(z) & = \exp \left( \frac{1}{2} \zeta \, \op{a}^{\dagger} \op{a}^{\dagger} \right) \times \left[ \sqrt{\sech r} \, \sum_{n=0}^{\infty} \frac{( \sech r - 1)^n}{n!} ( \op{a}^{\dagger})^n \op{a}^n \right] \times \exp \left( - \frac{1}{2} \zeta^{\ast} \, \op{a} \op{a} \right),
\end{align}
where $\sech r = 1/\cosh r = \sqrt{1 - |\zeta|^2}$. Let $|0_a \>$ and $|0_b\>$ be two vacuua satisfying $\op{a} |0_a \> = \op{b} |0_b \> = 0$. Thus,
\begin{align}
| 0_b \> & = \mathcal{S}^{\dagger}(z) | 0_a \> \nn\\
& = \sqrt{\sech r} \, \exp \left( - \frac{1}{2} \zeta \, \op{a}^{\dagger} \op{a}^{\dagger} \right) | 0_a \> \nn\\
& = \sqrt{\sech r} \, \sum_{n=0}^{\infty} (- \zeta)^n \sqrt{\frac{(2n-1)!!}{(2n)!!}} | 2 n_a \>.
\end{align}

\subsection{2-particle squeezed states} \label{sec:2-particle_squeeze}

Let $z = r e^{\I \theta}$ and consider two commuting pairs of creation-annihilation operators,  $\op{a}^{\dagger}_{R,L}, \op{a}_{R,L}$. Consider the Bogoliubov operator,
\begin{align} \label{opS}
\mathcal{S}(z) = \exp \left( z \op{a}_L^{\dagger} \op{a}_R^{\dagger} - z^{\ast} \op{a}_L \op{a}_R \right).
\end{align}
The operator is unitary, $\mathcal{S}^{-1}(z) = \mathcal{S}^{\dagger}(z) = \mathcal{S}(-z)$ and realizes the Bogoliubov transformation,
\begin{align}
\op{b}_L & = \op{a}_L \, \cosh r + \op{a}_R^{\dagger} \, e^{\I \theta} \sinh r = \mathcal{S}^{\dagger}(z) \op{a}_L \mathcal{S}(z), \\
\op{b}_R & = \op{a}_R \, \cosh r + \op{a}_L^{\dagger} \, e^{\I \theta} \sinh r = \mathcal{S}^{\dagger}(z) \op{a}_R \mathcal{S}(z).
\end{align}
With
\begin{align}
\zeta = e^{\I \theta} \tanh r
\end{align}
its expansion reads
\begin{align}
\mathcal{S}(z) & = \exp \left( \zeta \, \op{a}_L^{\dagger} \op{a}_R^{\dagger} \right) \times \left[ \sqrt{\sech r} \, \sum_{m=0}^{\infty} \frac{( \sech r - 1)^m}{m!} ( \op{a}_L^{\dagger})^m \op{a}_L^m \right] \times \nn\\
& \qquad\qquad \times \left[ \sqrt{\sech r} \sum_{n=0}^{\infty} \frac{( \sech r - 1)^n}{n!} ( \op{a}_R^{\dagger})^n \op{a}_R^n \right] \times \exp \left( - \zeta^{\ast} \, \op{a}_L \op{a}_R \right).
\end{align}
Let $|0_a \>$ and $|0_b\>$ be two vacua satisfying $\op{a}_L |0_a \> = \op{a}_R |0_a \> = 0$ and $\op{b}_L |0_b \> = \op{b}_R |0_b \> = 0$. Thus,
\begin{align}
| 0_b \> & = \mathcal{S}^{\dagger}(z) | 0_a \> \nn\\
& = \sqrt{1 - | \zeta |^2 } \, \exp \left( - \zeta \, \op{a}_L^{\dagger} \op{a}_R^{\dagger} \right) | 0_a \> \nn\\
& = \sqrt{1 - | \zeta |^2 } \, \sum_{n=0}^{\infty} (- \zeta)^n | n_a \>_L | n_a \>_R.
\end{align}
When tracing over the left states one obtains the density matrix,
\begin{align}
\rho_R & = \Tr_L | 0_b \> \< 0_b | = ( 1 - | \zeta |^2 ) \sum_{n=0}^{\infty} | \zeta |^{2n} | n \>\< n |.
\end{align}
The product of two Bogoliubov transformations can be expressed as follows
\begin{align} \label{opS_comp}
\mathcal{S}(z_1) \mathcal{S}(z_2) = \mathcal{S}(z_3) \exp \left[ \frac{1}{2} \log \left( \frac{1 + \zeta_1 \zeta_2^{\ast}}{1 + \zeta_1^{\ast} \zeta_2} \right) \left( \op{a}_L^{\dagger} \op{a}_L + \op{a}_R^{\dagger} \op{a}_R + 1 \right) \right],
\end{align}
where
\begin{align}
\zeta_j = e^{\I \theta_j} \tanh r_j, \quad j = 1,2
\end{align}
and the parameter $z_3 = r_3 e^{\I \theta_3}$ is determined by
\begin{align}
\zeta_3 = e^{\I \theta_3} \tanh r_3 = \frac{\zeta_1 + \zeta_2}{1 + \zeta_1^{\ast} \zeta_2}.
\end{align}
In particular, if $z_1$ and $z_2$ are real,
\begin{align}
\mathcal{S}(z_1) \mathcal{S}(z_2) = \mathcal{S}(z_3), \quad z_1, z_2 \in \R.
\end{align}

\bibliographystyle{JHEP}
\bibliography{wh}

\end{document}